\documentclass[11pt,a4paper]{article}
\pdfoutput=1
\usepackage{fancyhdr}  
\usepackage{jcappub}
\usepackage{bbm}
\usepackage{mathrsfs}
\usepackage{slashed}
\usepackage{caption}
\usepackage{epstopdf}
\usepackage[normalem]{ulem}
\usepackage[bottom]{footmisc}
\usepackage{subcaption}
\usepackage{bbold}
\usepackage{titlesec}
\usepackage{threeparttable}
\usepackage{booktabs}
\usepackage{changepage}
\usepackage[utf8]{inputenc}
\usepackage{dsfont}
\usepackage{grffile}
\usepackage{graphicx}  % needed for figures
\usepackage{dcolumn}   % needed for some tables
\usepackage{bm}        % for math
\usepackage{setspace}
\usepackage{amsmath,amssymb,setspace}
\usepackage{array}
\usepackage{booktabs}
\usepackage{indentfirst}
\usepackage{float}
\usepackage{lmodern}
\usepackage{multirow}
\usepackage{soul}
\usepackage[dvipsnames]{xcolor}
\usepackage{extpfeil}
\usepackage{enumitem} % customize \item
\usepackage{braket}
\usepackage{comment}
\makeatletter
\newcommand*\bigcdot{\mathpalette\bigcdot@{.5}}
\newcommand*\bigcdot@[2]{\mathbin{\vcenter{\hbox{\scalebox{#2}{$\m@th#1\bullet$}}}}}
\makeatother
\newcommand{\mpl}{M_{\text{Pl}}}

\newcommand{\calR}{{\mathcal{R}}}
\newcommand{\AR}{\mathcal{A}_\mathcal{R}}
\newcommand{\calP}{{\cal P}}
\newcommand{\be}{\begin{equation}}
\newcommand{\ee}{\end{equation}}
\newcommand{\nn}{ }

\newcommand*{\dif}{\mathop{}\!\mathrm{d}}

\usepackage{verbatim}
\usepackage{tikz}
\usetikzlibrary{arrows.meta}

\tikzset{%
  >={Latex[width=2mm,length=2mm]},
            base/.style = {rectangle, rounded corners, draw=black,
                           minimum width=4cm, minimum height=1cm,
                           text centered, font=\sffamily},
  activityStarts/.style = {base, fill=blue!30},
       startstop/.style = {base, fill=red!30},
    Blockmuth/.style = {base, fill=green!30},
         process/.style = {base, minimum width=2.5cm, fill=orange!15,
                           font=\ttfamily},
}
\title{\huge{Primordial Black Hole Formation from Power Spectrum with Finite-width}}

\author[a,b,c]{Shi Pi,}
\author[c,d,e,f]{Misao Sasaki,}
\author[g,h,c,i]{Volodymyr Takhistov,}
\author[c,a,j]{Jianing Wang}

\affiliation[a]{CAS Key Laboratory of Theoretical Physics, Institute of Theoretical Physics, Chinese Academy of Sciences, Beijing 100190, China}
\affiliation[b]{Center for High Energy Physics, Peking University, Beijing 100871, China}
\affiliation[c]{Kavli Institute for the Physics and Mathematics of the Universe (WPI), UTIAS, The University of Tokyo, Kashiwa, Chiba 277-8583, Japan}
\affiliation[d]{Asia Pacific Center
for Theoretical Physics, Pohang 37673, Korea}
\affiliation[e]{Center for Gravitational Physics and Quantum Information,
Yukawa Institute for Theoretical Physics, Kyoto University, Kyoto 606-8502, Japan}
\affiliation[f]{Leung Center for Cosmology and Particle Astrophysics,
National Taiwan University, Taipei 10617}
\affiliation[g]{International Center for Quantum-field Measurement Systems for Studies of the Universe and Particles (QUP,WPI), High Energy Accelerator Research Organization (KEK), Oho 1-1, Tsukuba, Ibaraki 305-0801, Japan}
\affiliation[h]{Theory Center, Institute of Particle and Nuclear Studies (IPNS),
High Energy Accelerator Research Organization (KEK), Tsukuba 305-0801, Japan}
\affiliation[i]{Graduate University for Advanced Studies (SOKENDAI), 
1-1 Oho, Tsukuba, Ibaraki 305-0801, Japan}
\affiliation[j]{School of Physical Sciences, University of Chinese Academy of Sciences, Beijing 100049, China}

\emailAdd{shi.pi@itp.ac.cn}
\emailAdd{misao.sasaki@ipmu.jp} 
\emailAdd{vtakhist@post.kek.jp}
\emailAdd{jianing.wang@ipmu.jp}

\abstract{
Primordial black holes (PBHs) can form from gravitational collapse of large overdensities in the early Universe, giving rise to rich phenomena in astrophysics and cosmology.
We develop a novel, general, and systematic method based on theory of density contrast peaks to calculate the abundance of PBHs for a broad power spectrum of curvature perturbations with Gaussian statistics. We introduce a window function to account for the relevant perturbation scales associated with PBHs of different masses, along with a filter function that removes unphysical contributions from super-horizon-scale overdensities. While some uncertainties remain due to the limited understanding of the nonlinear collapse process, our approach substantially reduces the discrepancy previously observed between peaks theory and the Press–Schechter formalism.}

\begin{document}

\noindent YITP-24-184, KEK-QUP-2024-0028, KEK-TH-2676, KEK-Cosmo-0369  

\maketitle
\flushbottom

%%%%%%%%%%%%%%%%%%%%%%%%%%%%%%%%%%%%%%%%%%%%%%%%%%%%%%%%%%%%%%%%%%%%%%%%

\section{Introduction}\label{s:introduction}

Primordial black holes (PBHs) can form in the early Universe through the gravitational collapse of overdense regions after inflation. Proposed decades ago~\cite{Zeldovich:1967lct,Hawking:1971ei,Carr:1974nx,Meszaros:1974tb,Carr:1975qj,Khlopov:1985jw}, PBHs prominently link a multitude of topics in cosmology, rich astrophysical phenomena and can serve as potential non-particle dark matter (DM) candidates. 
While a variety of PBH formation scenarios have been proposed (see \textit{e.g.}~\cite{Sasaki:2018dmp,Carr:2020gox,Green:2020jor} for review), the most extensively explored\footnote{Although in some scenarios PBH formation has been suggested to occur through drastically different pathways, for example involving scalar field fragmentation into Q-balls~\cite{Cotner:2016cvr,Cotner:2019ykd} or oscillons~\cite{Cotner:2018vug,Cotner:2019ykd}.} involves quantum fluctuations generated during inflation in the early Universe. These fluctuations are stretched to scales beyond the Hubble horizon, where they freeze and evolve into classical density perturbations. These fluctuations are stretched to scales beyond the Hubble horizon, where they freeze and evolve into classical density perturbations. Upon re-entry into the horizon during the decelerated expansion phase, perturbations with sufficiently large amplitudes collapse to form PBHs, with masses approximately equal to the mass contained within the horizon at the time.
 
Both the theoretical predictions and the observational constraints sensitively rely on precise calculation of the PBH abundance. Considering the gravitational collapse of density perturbations, PBH formation relies on the criteria used to determine when perturbations collapse into black holes.
An approximate criterion for the required density contrast of $\delta_\mathrm{cr}\sim 1/3$ based on the Jeans instability as a threshold for gravitational collapse during the radiation-dominated era has been put forth in Ref.~\cite{Carr:1975qj}.
This threshold represents the minimum overdensity required for a perturbation to overcome pressure forces and collapse into PBH. 
Taking inspiration from Press-Schechter formalism~\cite{Press:1973iz} to analyze DM halo formation and considering primordial perturbations to follow Gaussian statistics, 
the abundance of PBHs can be estimated by 
 calculating the fraction of regions where the density contrast exceeds the critical threshold as
\begin{align}
\beta\sim\int_{\delta_\mathrm{cr}}\exp\left(-\dfrac{\delta^2}{2\sigma_\delta^2}\right)\mathrm{d}\delta
\sim\mathrm{erfc}\left(\dfrac{\delta_\mathrm{cr}}{\sqrt2\sigma_\delta}\right),
\end{align}
where $\delta\equiv\delta\rho/\rho$ is the density contrast over the mean energy density $\rho$ and $\sigma_\delta$ is the root-mean-square (RMS) of the density contrast, quantifying the typical amplitude of density fluctuations. This method has been widely used to estimate the abundance of PBHs. We refer to it as the simplest Press-Schechter method based on the density contrast, or just PS$\delta$ for short. 

The compaction function, introduced in Ref.~\cite{Shibata:1999zs}, measures the mass excess within a spherical region relative to its radius and provides a more precise criterion for identifying regions likely to collapse into black holes, particularly in numerical relativity. The compaction function can be roughly expressed as $\mathcal{C} \sim 2G\delta M(R)/R$, where $\delta M(R)$ is the mass excess enclosed within a sphere of radius $R$. This function is derived using the Poisson equation, which relates the density perturbation to the gravitational potential. In its simplest form, the compaction function is interpreted as the volume average of the density contrast~\cite{Shibata:1999zs,Musco:2018rwt}. However, due to the nonlinear relation between the density contrast and the curvature perturbation on a comoving slice~\cite{Harada:2015yda}, the compaction function is more properly described as a quadratic form of the gradient of the curvature perturbation. The compaction function provides a natural framework for extending the Press-Schechter formalism. By integrating the probability density function of the compaction function from a critical threshold to an upper bound, the abundance of primordial black holes (PBHs) can be estimated~\cite{Biagetti:2021eep,Gow:2022jfb,Ferrante:2022mui,Young:2024jsu,Pi:2024jwt}. We refer to this as the Press-Schechter-type formalism of compaction function, or just PSC for short.

In the PS$\delta$ method based on density contrast, the abundance of PBHs is estimated considering the probability of regions exceeding a critical threshold. However, this approach lacks information about the profile of the density field.
The lack of information about density field's profile remains the same in the PSC method.
This limitation is addressed by the peaks theory method, which incorporates the typical profile of the density field as determined by the power spectrum and its higher order statistical moments. In this framework, the number density of over-dense regions is calculated by taking the ensemble average on the comoving volume of the maxima of a Gaussian field.  
The peaks theory method has been applied to calculate the PBH number density using either the comoving curvature perturbation~\cite{Yoo:2018kvb,Atal:2019cdz,Yoo:2019pma,Atal:2019erb,Riccardi:2021rlf,Germani:2019zez,Young:2022phe} or its Laplacian \cite{Yoo:2020dkz,Kitajima:2021fpq} as Gaussian variables. 
For cases with narrow peak in power spectrum the calculation is significantly simplified, but such spectra are challenging to realize in theories and could lead to unrealistic results~\cite{Germani:2023ojx}. 
For power spectra with finite width, PBHs naturally form at different scales, resulting in a range of masses, which makes it necessary to introduce a window function~\cite{Ando:2018qdb,Young:2019osy}. In Ref.~\cite{Yoo:2020dkz}, window functions were considered in analyzing the PBH formation within peaks theory. The smoothing scale $R_s$ in the window function acts as a UV cut-off, which allows them to deal with broad power spectra. To estimate the PBH mass spectrum, they first calculate the mass spectrum for each $R_s$, and take the envelope curve of the mass spectra when varying $R_s$. Although this procedure may have some relevance, the physical interpretation of the enveloped mass function is not clear, considering in particular the fact that the only role of the smoothing scale $R_s$ in their prescription is to single out a unique mass given by the maximum of the mass spectrum determined by $R_s$. Furthermore, the PBH mass was still defined by the size of the over-dense region, which approaches a constant when $R_s\to0$, inconsistent with our intuition. 

In this work we develop a systematic method for determining PBH abundance based on peaks theory. For power spectra with finite width, PBHs naturally form at different scales, resulting in a range of masses. As PBHs form at different scales, under-dense regions can be embedded within over-dense regions with smaller-scale peaks nested within larger-scale peaks, so-called cloud-in-clouds~\cite{Bardeen:1985tr,Bond:1990iw,Peacock:1990zz,Sheth:1999su,Escriva:2023qnq,MoradinezhadDizgah:2019wjf}.
However, a detailed treatment of this cloud-in-cloud problem is beyond the scope of this work.  
Our approach considers that PBH formation is an extremely rare event and that the timescale for gravitational collapse is much shorter than the typical evolutionary timescale of the Hubble horizon.  
This suggests that effects of cloud-in-cloud configurations are likely to be subdominant in the regimes of interest we consider.
A more robust resolution of the cloud-in-cloud problem would require N-body simulations. 
Our approach assumes that PBH formation is an extremely rare event and that the timescale for gravitational collapse is much shorter than the typical evolutionary timescale of the Hubble horizon, ensuring the validity of our estimation. 
The choice of the window function plays a key role in our calculations, as the PBH mass is directly related to the smoothing scale. We select a filtering scale such that the smoothed over-dense region contains the maximum of the compaction function, which serves as the criterion for PBH formation. Introducing an appropriate window function significantly suppresses PBH abundance, as both the power spectrum and its higher-order moments are suppressed. Importantly, we note that the window function should also be included for consistency even in the case of a monochromatic peak in the spectrum.

For our analysis, we focus on the comoving curvature perturbation $\cal R$, of which the power spectrum is enhanced on small scales to have a log-normal bump, but still assuming Gaussian statistics. Such a setup can be realized, \textit{e.g.}, in ultra-slow-roll inflation with a smooth transition to slow-roll~\cite{Cai:2018dkf,Pi:2022ysn,Pi:2024jwt}, as seen in models like the Starobinsky model~\cite{Starobinsky:1992ts,Leach:2001zf,Biagetti:2018pjj,Pi:2022zxs,Cielo:2024poz} and some other potentials \cite{Domenech:2023dxx,Ivanov:1994pa,Ozsoy:2019lyy,Cole:2022xqc,Wang:2024wxq,Caravano:2024tlp,Caravano:2024moy}. The non-Gaussianity arising from the nonlinear relation between $\cal R$ and the density contrast $\delta$ is naturally encoded in the definition of the compaction function $\mathcal{C}$ \cite{Harada:2015yda,Kawasaki:2019mbl,Young:2019yug,DeLuca:2019qsy}.
Since $\mathcal{C}$ is a quadratic function of $\nabla\cal R$, the amplitude of $\mathcal{R}$ itself does not directly affect PBH formation, as it can be absorbed into a redefinition of the scale factor. Hence, as in Refs.~\cite{Yoo:2020dkz,Kitajima:2021fpq}, we adopt the Laplacian of $\cal R$ as the Gaussian variable instead of $\cal R$ itself. 
We then apply peaks theory, incorporating the power spectrum and its moments smoothed over a given scale with an appropriately chosen window function.

The paper is organized as follows. In Sec. \ref{s:Executive Summary} we provide the executive summary of our method, including outline of the procedure, differences with previous analyses and new results. In Sec. \ref{s:PT}, we discuss the algorithm for calculating PBH abundance using peaks theory, applicable for general scenarios.
Specifically, in Sec. \ref{s:Applications and Results}, we analyze in detail the log-normal form that represents a finite-width power spectrum. In Sec. \ref{s:results}, we discuss our results and compare with those obtained from previous methods, highlighting the differences and advantages of our approach. We conclude in Sec. \ref{s:conclusion}.

\section{Executive Summary}\label{s:Executive Summary}

Numerical and analytical studies have demonstrated that the profile of over-densities is crucial for determining the threshold of the compaction function and hence criterion when PBHs form~\cite{Musco:2018rwt,Escriva:2019phb,Young:2019osy}. However, PSC formalism does not contain the description of density profile, thus can not account for this. Choosing a typical profile independently or relying on the mean profile from peaks theory inevitably introduces inconsistencies in the calculation methodology. These uncertainties in the choice of profile and threshold can lead to significant differences in the predicted PBH abundance, motivating the adoption of peaks theory. In this approach, the mean peak profile is rigorously defined by the multiple statistical moments and two-point correlation functions of the Gaussian curvature perturbation field~\cite{Yoo:2018kvb,Atal:2019cdz,Atal:2019erb,Yoo:2020dkz,Kitajima:2021fpq}.  The shapes of the peaks, including their height, ellipticity, and oblateness, can also be approprietly captured~\cite{Bardeen:1985tr}. Since for the rare event of PBH formation non-sphericity can be safely neglected (see, \textit{e.g.},~\cite{Sheth:1999su,Kuhnel:2016exn,Yoo:2020lmg,Escriva:2024aeo})), one can focus on analyzing the height and width of spherical peaks. Additionally, peaks theory provides the number density of maxima for any Gaussian random field based on its multiple moments, which are derived from the power spectrum. These quantities follow Gaussian distributions with variances determined by the power spectrum of the comoving curvature perturbation $\mathcal{R}$, smoothed through an appropriately chosen window function.

We put forth a general framework for analyzing PBH formation from finite-width power spectra within the peaks theory approach. Previous analyses of PBH formation using peaks theory have primarily focused on the monochromatic power spectrum of the comoving curvature perturbation, $\mathcal{P_R}\sim\delta(\ln(k/k_*))$.
In this case, even without introducing a window function, one could compute the PBH abundance by assuming that $k_*$ serves as both the peak scale and the scale determining the PBH mass. 
While such a narrow peak in the power spectrum can arise in some theories including multi-field inflation~\cite{Palma:2020ejf,Fumagalli:2020adf,Boutivas:2022qtl} or based on parametric resonance \cite{Cai:2018tuh,Chen:2019zza,Cai:2019bmk}, simple inflationary models typically predict\footnote{For example, in Starobinsky inflation, a sudden change in the potential slope produces modulated oscillations with a period of $\pi$, where the first oscillatory peak represents the maximum of the power spectrum~\cite{Pi:2022zxs}. The width of this peak, determined by the oscillation period arising from the interplay of positive- and negative-frequency modes of $\cal R$, is approximately $\sim 1/2\pi$ and remains robust against variations in other parameters~\cite{Dalianis:2021iig,Pi:2022zxs}.} a spectrum with a peak of finite width, $\Delta k\sim k$. In our analysis, we advance beyond the monochromatic power spectrum and develop a general method adopting
a log-normal spectrum as basis capturing a wide range of theory predictions~(\textit{e.g.}~\cite{Pi:2020otn})
\be\label{def:PR}
\mathcal{P_R}=\displaystyle\frac{\mathcal{A_R}}{\sqrt{2\pi}\Delta}\exp\left(-\frac{\ln^2(k/k_*)}{2\Delta^2}\right)~,
\ee
where $\Delta$ and $\mathcal{A}_{\mathcal{R}}$ are the width and the {variance} of the curvature perturbation, respectively. That is, $\int \mathrm{d} \ln k ~ \mathcal{P}_{\mathcal{R}}(k)=\mathcal{A}_{\mathcal{R}}$. The log-normal peaked spectrum of Eq.~\eqref{def:PR} also allows to approximate well the smoothed junction point of the broken power-law spectra that can appear in ultra-slow-roll inflation models~\cite{Riccardi:2021rlf,Atal:2021jyo}.  

\begin{figure}[t]
\begin{center}
Table of Notation
\begin{equation*}
\setlength{\arraycolsep}{5pt}
\begin{array}{ccccccccc}
\hline\hline
\text{Notation} & \text{Meaning} & \text{Definition} & \\
\hline 
f_\mathrm{PBH}(M) & \text{PBH mass spectrum} & {\rm Eq.~}\eqref{eq:fPBH} & \\
\mathcal{R} & \text{Curvature perturbation in comoving slice} & {\rm Eq.~}\eqref{eq:Gaufield} & \\
\mathcal{P}_\mathcal{R}(k) & \text{Power spectrum of curvature perturbation} & {\rm Eq.~}\eqref{eq:Rk} & \\
\mu_2 & \text{Height of }-\nabla^2\mathcal{R}\text{ peak} & {\rm Eq.~}\eqref{def:mu2k3} & \\
K_3 & \text{Width of }-\nabla^2\mathcal{R}\text{ peak} & {\rm Eq.~}\eqref{def:mu2k3} & \\
\sigma_n^2 & \text{Multiple moments of Gaussian random field} & {\rm Eq.~}\eqref{eq:staticsPT} & \\
\gamma_n & \sigma_n^2 / (\sigma_{n-1} \sigma_{n+1} )& {\rm Eq.~}\eqref{eq:staticsPT} & \\
R_n & \sqrt{3} \sigma_n / \sigma_{n+1} & {\rm Eq.~}\eqref{eq:staticsPT} & \\
\psi_n & \text{Two point correlation function of Gaussian random field} & {\rm Eq.~}\eqref{eq:staticsPT} & \\
\mu & \text{Rescaled } -\nabla^2\mathcal{R} \text{ peak height: }\mu_2   \sigma_1^2 / \sigma_2^2 & {\rm Eq.~}\eqref{def:muK} & \\
K & \text{Rescaled } -\nabla^2\mathcal{R} \text{ peak width: }K_3  \sqrt{\sigma_2 / \sigma_4} & {\rm Eq.~}\eqref{def:muK} & \\
\hat{\mathcal{R}} & \text{Peak profile of }\mathcal{R} & {\rm Eq.~}\eqref{eq:profilefullbare} & \\
\mathscr{N}_{\text {pk }}\left(\mu, K\right)& \text{ \text{Number density of peaks with height $\mu$ and width $K$}} & {\rm Eq.~}\eqref{eq:npeakk3} & \\
R_s & \text{Smoothing scale of window function} & {\rm Eq.~}\eqref{eq:pswin} & \\
\mathscr{N}_{\mathrm{PBH}}(M) & \text{PBH number density as a function of mass} & {\rm Eq.~}\eqref{def:nPBH}  & \\
\kappa & k R_s & {\rm Eq.~}\eqref{eq:kappa} & \\
\kappa_* & k_* R_s & {\rm Eq.~}\eqref{eq:kappa} & \\
x & k r & {\rm Eq.~}\eqref{eq:x} & \\
x_* & k_* r & {\rm Eq.~}\eqref{eq:x} & \\
x_W & r / R_s & {\rm Eq.~}\eqref{eq:x} & \\
\hline\hline
\end{array}
\end{equation*}
\end{center}
\end{figure}

An important aspect of PBH formation is the consideration of a window function, which defines a smoothing scale and effectively isolates perturbation contributions relevant to that scale. 
The introduction of smoothing scales is necessary to take account of the expansion of the universe and the temporal evolution of perturbations. 
Although all wavelength components of perturbations generated during inflation freeze on super-horizon scales, they re-enter the horizon at different times. 
Shorter-wavelength (i.e. small-scale) perturbations enter earlier, while longer-wavelength perturbations are still outside the horizon. 
If we focus on a given comoving scale $R_s$, we consider the configuration of the curvature perturbation when it is on superhorizon scales, while the perturbations on scales smaller than $R_s$ may have already entered the horizon. 
Thus the window function effectively selects only the large-scale components that are still super-horizon at a given time. 

The choice of different window functions can significantly affect PBH abundance calculations~\cite{Ando:2018qdb, Young:2019osy}. Using an appropriately chosen filtering scale to remove the unphysical contributions from configurations larger than smoothing scale, we further find that PBH abundance calculated via peaks theory is significantly reduced compared to other methods without a window function, even for narrow spectra, but remains higher than estimates from the PS$\delta$ approach.

\begin{figure}
    \centering
    \includegraphics[width=0.9\textwidth]{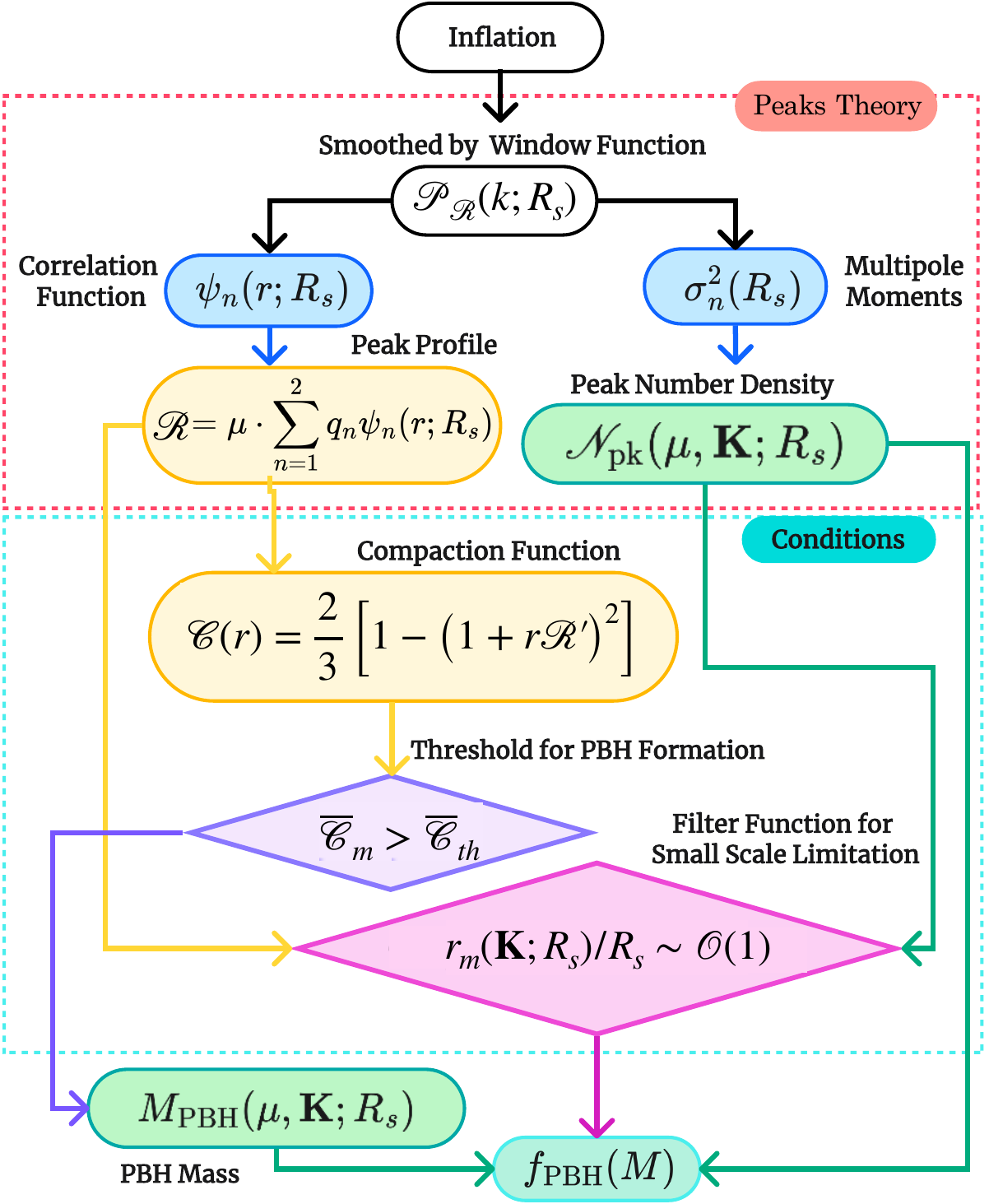}
    \caption{Algorithm flowchart of our method for calculating PBH abundance.}
    \label{fig:flowchart}
\end{figure}

In Fig.~\ref{fig:flowchart} we provide flowchart summarizing the algorithm of the method we put forth. \\

\noindent \underline{Key algorithm steps to evaluate PBH abundance:}
 \begin{enumerate}
     \item Smooth the primordial perturbation power spectrum generated from an inflation model by employing a window function.
     \item Analyze the statistical properties of the Gaussian random field, the peak profile and number density considering peaks theory.
     \item Establish the threshold conditions for the formation of PBHs by considering the compaction function and the scale of the over-density region.  
     \item Determine the abundance of PBHs using the number density and the horizon mass given the evolution of the Universe.
 \end{enumerate}
~\newline

Our approach is based on several key innovations as follows.  
First, we note that the peak profile is defined by the smoothed power spectrum 
$  
\hat{\mathcal{R}} \sim \mu  \sum_{n=1}^2 q_n \psi_n \left(r, R_s\right)~,
$ 
where $\mu$ is the rescaled peak height and $\psi_n$ denotes the weighted two point correlation functions. Then, the PBH mass is determined by the smoothing scale $R_s$ of the window function $M_H\propto R_s^{2}$, instead of the size of the overdense region (i.e. the radius $r_m$ of the compaction function maximum) $M_H\propto r_m^{2}$. 

An important aspect we discuss is the necessity to filter
out large $k$ modes.
As the smoothing scale decreases, the corresponding PBH mass also reduces. However, beyond a certain limit, the maximum of the compaction function falls outside the horizon, preventing PBH formation. 
To take this effect into account, we exclude unwanted small PBHs arising from fluctuations on very small scales, ensuring the total PBH abundance, calculated as an integral over all scales, remains finite. This approach aligns with the findings of Ref.~\cite{Germani:2019zez}.
Namely, the horizon scale, approximately corresponding to the smoothing scale $R_s$, should be larger than the size of the over-dense region, $r_m / R_s \lesssim \mathcal{O}(1)$. Otherwise, if $r_m\gg R_s$, the horizon cannot effectively account for the Laplacian of $\cal R$, i.e., the density contrast of the over-dense region, and no PBH can form.

We note that there is inherent uncertainty in determining the exact value of $r_m/R_s$, reflecting the ambiguity in the precise timing of the horizon re-entry of the over-dense region.
The exact ratio can only be determined through numerical relativity, which is beyond the scope of this work. Here, we impose this condition by requiring that the central mass in the PBH mass spectrum corresponds to $R_s=1/k_*$ (see Eq.~\eqref{eq:MH}), at least in the case of a monochromatic power spectrum. In addition to presenting the full expression for the number density, we also re-analyze all our formulas in the high-peak limit \cite{Green:2004wb}, and find that the difference of PBH abundances between them is only mild. Therefore, for a simple calculation, the high-peak limit can suffice.

The master formula to calculate the PBH mass spectrum $f_{\rm PBH}$ for PBHs of mass $M$ from our method is given by
\begin{equation}
\boxed{f_{\mathrm{PBH}}\left(M\right) = \int \frac{\mathrm{d} R_s}{R_s} \int \mathrm{d} K \int \mathrm{d} \mu ~\delta_{\mathrm{D}}\left(\ln \frac{M}{M\left(\mu, K, R_s \right)}\right) \frac{M\left(\mu, K, R_s \right)\mathscr{N}_{\text {PBH } }} {\rho_\mathrm{DM}}.
\label{eq:fPBH}}
\end{equation}
Here, $\rho_{\mathrm{DM}}$ is the energy density of dark matter, while $\mathscr{N}_{\mathrm{PBH}}$ is comoving number density of PBHs that can be obtained using peaks theory. The $\delta$-function imposes constraint on PBH mass $M$ to be a function of smoothing scale $R_s$, rescaled peak height $\mu$, and rescaled width $K$. All of these quantities will be defined later. Note that the $R_s$-dependence appears also in $\mathscr{N}_{\mathrm{PBH}}$, which originates from considering the peaks 
of size $R_s$.
 
Analyzing consequences of Eq.~\eqref{eq:fPBH}, among our new results are the following. A broader power spectrum produces a wider PBH mass function. However, for a log-normal spectrum as described by Eq.~\eqref{def:PR} with a logarithmic width $\Delta\lesssim1$, the PBH mass function resembles that of a narrow spectrum. This similarity arises from the critical behavior in the infrared (IR) and the rapid exponential decay in the ultraviolet (UV) regimes. Additionally, a broad power spectrum can reduce the overall PBH abundance, as the peak value of the power spectrum decreases when the {variance} $\mathcal{A_R}$ is fixed.
  
We find distinct predictions for PBH formation compared to other approaches. When a Gaussian window function is used, which ensures smoothness in both real and momentum space, our new method based on peaks theory and described by Eq.~\eqref{eq:fPBH} predicts an enhancement in PBH formation compared to the PS$\delta$ formalism, though the increase is not dramatic. For a monochromatic power spectrum, our results indicate that peaks theory without a window function and the PSC (Press-Schechter formalism using the compaction function) yield similar PBH abundances. This similarity arises partly because the latter adopts the threshold of the compaction function from the former.
Notably, both the peaks theory without a window function and the PSC method predict PBH abundances approximately $\mathcal{O}(30)$ orders of magnitude higher than the PS$\delta$ formalism based on density contrast for a characteristic spectrum amplitude  of $\mathcal{A_R}\sim10^{-2}$.  
In contrast, our method provides a significantly milder enhancement, predicting PBH abundances about $\mathcal{O}(10)$ orders of magnitude higher than the PS$\delta$ formalism. This is shown explicitly in Figs. \ref{fig:diffWinfM} and \ref{fig:diffmethods-D041} for comparison.

\section{Peaks Theory Method}\label{s:PT}

We consider the perturbed metric on comoving slices,
\be
\mathrm{d} s_3^2=a(t)^2 e^{2 \mathcal{R}(\boldsymbol{r})}\left(\mathrm{d} r^2+r^2 \mathrm{d} \Omega^2\right),
\label{eq:metric-ud}
\ee
where $\mathcal{R}(\boldsymbol{r})$ is the comoving curvature perturbation \cite{Mukhanov:1988jd,Sasaki:1986hm},  
$\boldsymbol{r}$ is the position vector, $r$ denotes the radial coordinates, and $\mathrm{d} \Omega^2=\mathrm{d} \theta^2+\sin ^2 \theta \mathrm{d} \phi^2$ is an element of a unit two-dimensional sphere. In this work, our discussion is limited to spherical symmetric peaks, such that $\mathcal{R}(\boldsymbol{r})$ reduces to $\mathcal{R}(r)$ for simplicity. This approximation is known to hold for very high peaks, but may not be so good for moderate peaks. 
An extension of our formalism to non-spherical peaks is left for future investigation.  

The inflationary phase of the early Universe naturally generates a randomly distributed curvature perturbation, $\calR$, with its primordial non-Gaussianity primarily determined by its super-horizon evolution~\cite{Motohashi:2017kbs}. In this study, we focus on a Gaussian $\calR$, whose probability distribution function (PDF) is given by 
\be
\mathbb{P}_G(\mathcal{R})=\frac{1}{\sqrt{2 \pi} \sigma_{\mathcal{R}}}
\exp\left(-\frac{\mathcal{R}^2}{2 \sigma_{\mathcal{R}}^2}\right),
\label{eq:Gaufield}
\ee
where $\sigma_{\calR}^2$ is the variance of $\mathcal{R}$, given by its power spectrum
\begin{align}
\sigma_{\calR}^2(R_s)&=\int \frac{\mathrm{d} k}{k} \mathcal{P}_{\mathcal{R}}(k) \widetilde{W}^2\left(k , R_s\right),\\\label{eq:Rk}
\mathcal{P}_{\mathcal{R}}(k)&=\frac{k^3}{2\pi^2}\left|\tilde{\mathcal{R}}\left(\boldsymbol{k}\right)\right|^2.
\end{align}
PBHs form when a quantity characterizing the over-dense region in a given area exceeds a critical threshold. This quantity, designed to describe the compactness of the over-dense region, can be represented by the density contrast~(see Eq.~\eqref{eq:contrast}), the compaction function~(see Eq.~\eqref{eq:coarse-grained}), the curvature perturbation, or similar quantities, all of which are interconnected through the Einstein equations.

\subsection{Random Field Seed Background}
Peaks theory provides the number density and shapes of local maxima of a Gaussian random field, originally developed by Bardeen, Bond, Kaiser, and Szalay (BBKS) to analyze halo formation~\cite{Bardeen:1985tr}. We briefly review this theory in App.~\ref{s:OverviewPT}. We adopt such a Gaussian random field as the initial condition ``seeding'' for PBH formation, which actually occurs later, well inside the Hubble horizon. For type I fluctuations, the formation process itself requires numerically solving the Misner-Sharp equations in comoving gauge, which in turn provides the threshold for the initial Gaussian peaks on super-horizon scales. We assume spherical symmetry for the peaks, neglecting ellipticity and oblateness due to the rarity of high peaks. Furthermore, because of their rareness, we assume that peaks are sparsely distributed in space, with no overlap between peaks of the same scale. 

Distinct Gaussian random fields can be considered for analysis. However, as first noted in Ref.~\cite{Yoo:2020dkz}, the Gaussian random field used in peaks theory for PBH formation should be $\nabla^2\mathcal{R}$ instead of $\mathcal{R}$. This choice arises because PBH formation is inherently a local process, and the long-wavelength components of $\mathcal{R}$ can be absorbed into a redefinition of the scale factor $a$. 
Furthermore, it has an intuitive physical motivation as $\nabla^2\mathcal{R}\sim-\delta\rho/\rho$ at linear order. 
The power spectrum of $\nabla^2\mathcal{R}$ satisfies \cite{Yoo:2018kvb,Kitajima:2021fpq,Yoo:2022mzl}
\be
\mathcal{P}_{\nabla^2 \mathcal{R}}(k)=k^4 \mathcal{P}_{\mathcal{R}}(k)~.
\label{eq:n+2}
\ee 

Around an over-dense region, one can identify the $-\nabla^2 \mathcal{R}$-peak profile with statistic quantities of $\mathcal{R}$-peak as
\be
-\widehat{\nabla^2 \mathcal{R}}(r)=\frac{\mu_2}{1-\gamma_3^2}\left[\psi_2(r)+\frac{R_3^2}{3} \nabla^2 \psi_2(r)-\frac{K_3^2}{\gamma_3} \frac{\sigma_2}{\sigma_4}\left(\gamma_3^2 \psi_2(r)+\frac{R_3^2}{3} \nabla^2  \psi_2(r)\right)\right].
\label{eq:peakprofileDeltazeta}
\ee
Here, $\mu_2$ and $K_3$ are the height and width of the $-\nabla^2 \mathcal{R}$ peak profile, respectively, \textit{i.e.}
\be
\mu_2\equiv\left.-\nabla^2 \mathcal{R}\right|_{r=0}, \quad K_3^2\equiv -\dfrac{1}{\mu_2} \left.\nabla^2\left(-\nabla^2 \mathcal{R}\right)\right|_{r=0} ~.
\label{def:mu2k3}
\ee
Note that both quantities are positive. For a detailed derivation of \eqref{eq:peakprofileDeltazeta} from the peaks theory, see App.~\ref{s:OverviewPT}. The local coordinates are defined around the peaks, with their origins positioned at the maximum points. The hat denotes the profile near the peak, while without the hat represents the general field. 
Here, $\nabla^2\equiv\delta^{ij}\partial_i\partial_j$ is the Laplacian of the background conformal spatial metric. Other statistic quantities from the peaks  theory 
$\sigma_n$, $\psi_n$, $\gamma_n$, $ R_n$ are defined as  
\be
\begin{aligned}
 \sigma_n^2=&~\int \frac{\mathrm{d} k}{k} k^{2 n} \mathcal{P}_\mathcal{R}(k), \\
 \psi_n(r)=&~\frac{1}{\sigma_n^2} \int \frac{\mathrm{d} k}{k} k^{2 n} \frac{\sin (k r)}{k r} \mathcal{P}_\mathcal{R}(k), \\
 \gamma_n=&~\frac{\sigma_n^2}{\sigma_{n-1} \sigma_{n+1}}, \\
 R_n=&~\sqrt{3} \frac{\sigma_n}{\sigma_{n+1}},
 \label{eq:staticsPT}
\end{aligned}
\ee
where $\sigma_n^2$ are the multiple moments. Note that $\gamma_n$ and $R_n$ are only valid for odd $n$.

Using the iteration relation 
\be
\nabla^2 \psi_n(r)=-\frac{\sigma_{n+1}^2}{\sigma_n^2} \psi_{n+1}(r),
\label{eq:recurrence}
\ee
one can solve for the peak profile of $\mathcal{R}$ from the peak profile of $-\nabla^2 \mathcal{R}$ in Eq.~\eqref{eq:peakprofileDeltazeta}
\be
\hat{\mathcal{R}}(r)=\frac{\mu}{1-\gamma_3^2}\left[\psi_1(r)+\frac{R_3^2}{3} \nabla^2 \psi_1(r)-\frac{K^2}{\gamma_3}\left(\gamma_3^2 \psi_1(r)+\frac{R_3^2}{3} \nabla^2 \psi_1(r)\right)\right]+\mathcal{R}_{\infty}.
\label{eq:hatzetaG}
\ee
Here,
\be
\mu\equiv\mu_2 \frac{\sigma_1^2}{\sigma_2^2}, \quad K^2 \equiv K_3^2 \frac{\sigma_2}{\sigma_4}
\label{def:muK}
\ee
are the rescaled dimensionless height and width of $\mathcal{R}$-peak, respectively. 
In Eq.~\eqref{eq:hatzetaG}, $\mathcal{R}_{\infty}\equiv \hat{\mathcal{R}}(r\to \infty)$ is an integration constant, which represents the effect of the long-wavelength perturbation and can always be absorbed in the redefinition of scale factor $a$. Therefore, we set $\mathcal{R}_\infty=0$ in the following.

Considering Eq.~\eqref{eq:recurrence}, the peak profile in Eq.~\eqref{eq:hatzetaG} can be organized as  
\be
\hat{\mathcal{R}} \left(r\right) = \mu \sum_{n=1}^2 q_n (K)\psi_n \left(r\right).
\label{eq:profilefullbare}
\ee
Here
\be
\begin{aligned}
q_1\left(K\right) & =\frac{1-K^2 \gamma_3 }{1-\gamma_3^2} ,\\
q_2\left(K\right) & =\frac{1}{1-\gamma_3^2}\left(-\frac{\sigma_3^2}{\sigma_4^2} \frac{\sigma_2^2}{\sigma_1^2}\right)\left(1-\frac{K^2}{\gamma_3}\right).
\end{aligned}
\label{def:q1q2}
\ee
In case of a monochromatic power spectrum one has $q_2=0$.

Analogously, the number density of $-\nabla^2 \mathcal{R}$ peaks with height $\mu_2$ and width $K_3$ in a comoving volume element can be obtained by relating quantities denoted by $n$ with those denoted by $n+2$ in the results obtained from BBKS as described in App.~\ref{s:OverviewPT}~(see Eq.~\eqref{eq:comNpkvx}, Eq.~\eqref{eq:Npkvx}, Eq.~\eqref{def:f(x)}), yielding  
\be
\mathscr{N}_{\mathrm{pk}}\left(\mu_2, K_3\right) \mathrm{d} \mu_2 \mathrm{~d} K_3=2\left(\frac{1}{6 \pi}\right)^{3 / 2}\mu_2 K_3 \frac{\sigma_4^2}{\sigma_2 \sigma_3^3} f\left(\frac{\mu_2 K_3^2}{\sigma_4}\right) P_1^{(3)}\left(\frac{\mu_2}{\sigma_2}, \frac{\mu_2 K_3^2}{\sigma_4}\right) \mathrm{d} \mu_2 \mathrm{~d} K_3,
\label{eq:npeakmuok1}
\ee
where 
\begin{align}\nonumber
f(x)=&~\frac{x^3-3 x}{2}\left[\operatorname{erf}\left(\sqrt{\frac{5}{2}}x\right)+\operatorname{erf}\left( \frac{1}{2}\sqrt{\frac{5}{2}}x\right)\right] \\\label{eq:f(x)} &+\sqrt{\frac{2}{5 \pi}}\left[\left(\frac{31 x^2}{4}+\frac{8}{5}\right) e^{-5 x^2 / 8}+\left(\frac{x^2}{2}-\frac{8}{5}\right) e^{-5 x^2 / 2}\right].
\end{align}
and
\begin{align}
 \label{def:P1(n)}
P_1^{(n)}(v, x)=&~\frac{1}{2 \pi \sqrt{1-\gamma_n^2}} \exp \left[-\frac{1}{2}\left(v^2+\frac{\left(x-\gamma_n v\right)^2}{1-\gamma_n^2}\right)\right].
\end{align}
The number density of peaks of $-\nabla^2 \mathcal{R}$ in a volume element around $(\mu,K)$ can then be obtained by replacing $\mu_2,K_3$ with $\mu,K$ by considering Eq.~\eqref{def:muK}, resulting in
\be
\mathscr{N}_{\mathrm{pk}}\left(\mu, K\right) \mathrm{d} \mu \mathrm{~d} K=2\left(\frac{1}{6 \pi}\right)^{3 / 2} \frac{\sigma_2^2}{\sigma_1^4} \frac{\sigma_4^3}{\sigma_3^3} \mu K f\left(\frac{\sigma_2}{\sigma_1^2} \mu K^2\right) P_1^{(3)}\left(\frac{\sigma_2}{\sigma_1^2} \mu, \frac{\sigma_2}{\sigma_1^2} \mu K^2\right) \mathrm{d} \mu \mathrm{~d} K.
\label{eq:npeakk3}
\ee

\subsection{Window Function}

An essential consideration for PBH formation is the role of the window function. Generally, the window function is introduced to smooth out small-scale perturbations, preventing unwanted small PBHs from influencing the PBH population at the scales of interest. The PBH mass is determined by the smoothing scale, which spans all relevant physical scales and ultimately generates the PBH mass function.
As a result, the physical interpretation of the PBH mass function inherently depends on the choice of window function, which can lead to some variation in the total PBH abundance. This uncertainty can only be diminished with sufficiently advanced numerical simulations of PBH formation, which are complementary to our analysis.
We thus emphasize the relevance of the window function and its connection to the smoothing scale and PBH mass. Taking into account a window function, the smoothed power spectrum is given by
\be
\mathcal{P}_{\mathcal{R}}\left(k , R_s \right)=\mathcal{P}_{\mathcal{R}}(k) \widetilde{W}^2\left(k , R_s\right).
\label{eq:pswin}
\ee
Here, $\widetilde{W}^2\left(k , R_s\right)$ represents the window function in momentum space, with the smoothing scale denoted by $R_s$ as before. 

Since PBHs form when the perturbation scale $1 / k$ is approximately equal to the horizon scale $R_H$, one might consider that we only need to select these perturbations directly out of the power spectrum. However, imposing such Dirac delta-type ``window function'' is pathological, essentially because the window function would be given by the square root of the Dirac delta function. 

Considering Eq.~\eqref{eq:pswin}, the statistic quantities from peaks theory listed in Eq.~\eqref{eq:staticsPT} are to be replaced using definitions based on the smoothed power spectrum  and become dependent on $R_s$ as
\be
\begin{aligned}
\sigma_n^2(R_s)=&~\int \frac{\mathrm{d} k}{k} k^{2 n} \mathcal{P}_{\mathcal{R}}\left(k, R_s\right),\\
\psi_n(r,R_s)=&~\frac{1}{\sigma_n^2(R_s)} \int \frac{\mathrm{d} k}{k} k^{2n} \frac{\sin (k r)}{k r} \mathcal{P}_{\mathcal{R}}\left(k, R_s\right).
\label{eq:kWenter}
\end{aligned}
\ee
In particular, $R_s$ enters Eqs.~\eqref{eq:staticsPT}, \eqref{eq:profilefullbare}, \eqref{def:q1q2}, and \eqref{eq:npeakk3}, as well as the high-peak limit formulas \eqref{eq:zetaGHPL} and \eqref{eq:npeakHPL} in the following section. Following replacement of $\sigma_n^2$, $\psi_n$ as described by Eq.~\eqref{eq:kWenter}, the dependence on $R_s$ also enters quantities such as $\hat{\mathcal{R}}\left(r,\mu,K,R_s\right)$, $q_n(K,R_s)$, $\mathscr{N}_{\text {pk }}\left(\mu, K,R_s\right)$, $\gamma_3(R_s)$, $\hat{\mathcal{R}}_{\mathrm{HPL}}(r,\mu,R_s)$, and $\mathscr{N}_{\mathrm{pk,HPL}}(\mu,R_s)$, which incorporate the $R_s$ dependence through $\mathcal{P_R}(r,R_s)$.

\subsection{Collapse Threshold}

The PBH formation process is highly nonlinear and detailed description generally requires numerical relativity. Broadly, PBH formation occurs when the compaction function reaches a value of unity at late times, as the over-dense region evolves non-linearly. On super-horizon scales, the formation criterion is tied to the initial conditions and the critical compaction function that corresponds to the collapse of a region to a PBH with zero mass. Numerical studies have shown that the critical compaction function on super horizon scales, $\mathcal{C}_{\rm th}$, varies approximately in the range of 0.41 to 0.67 depending on the density profile~\cite{Musco:2018rwt}. Additionally, numerical calculations suggest a nearly universal threshold for the averaged compaction function~\cite{Escriva:2019phb}. Hence, we adopt a universal threshold in our analysis.

We consider the following diagonal form of the metric around the peaks \cite{Musco:2018rwt}
\be
\mathrm{d} s^2=-A^2(r, t) \mathrm{d} t^2+B^2(r, t) \mathrm{d} r^2+R^2(r, t) \mathrm{d} \Omega^2~, 
\label{eq:metric-MS}
\ee
where $t$ is the cosmic time.
The areal radius $R$ is given by
\be
R(r, t) = a(t) r e^{\mathcal{R}(r)}.
\label{def:arealR}
\ee
Compaction function can be generally defined as the ratio of the Schwarzschild radius of the over-dense region to the areal radius as 
\be
\mathcal{C}(r, t) \equiv 2G\frac{\left[M(r, t)-M_b(r, t)\right]}{R(r, t)}\,,
\label{def:comfun}
\ee
where $G$ is the gravitational constant and $M_b$ is the mass contained in the radius $R$ in the homogeneous background.
Equivalently, the compaction function can be expressed in terms of the coarse-grained density contrast
\be
\mathcal{C}(r, t)=\frac{2G}{R(r, t)} \rho_b(t) \int_0^{R(r, t)} \mathrm{d} R\left[4 \pi R(r, t)^2\right] \delta(r, t)~,
\label{eq:coarse-grained}
\ee
where $\rho_b=3 M_{\mathrm{Pl}}^2 H^2=3 H^2 /8 \pi G$ is the background energy density with $H$ the Hubble parameter.
On comoving slices, the density contrast is related to the curvature perturbation by \cite{Harada:2015yda}
\be
\delta(r, t) \equiv \frac{\delta \rho}{\rho}=-\frac{8}{9}\left(\frac{1}{a H}\right)^2 \mathrm{e}^{-5 \mathcal{R}(r) / 2} \nabla^2 \mathrm{e}^{\mathcal{R} (r)/ 2}~.
\label{eq:contrast}
\ee
By substituting Eq.~\eqref{eq:contrast} into 
Eq.~\eqref{eq:coarse-grained}, one can obtain the relation between compaction function and the comoving curvature perturbation 
\be
\hat{\mathcal{C}}(r)=\frac{2}{3}\left[1-\left(1+r \hat{\mathcal{R}}'(r)\right)^2\right]~,
\label{eq:CFprofile}
\ee
where a prime stands for a derivative with respect to $r$. Notably, the time dependence cancels out, making $\mathcal{C}$ solely a function of $r$. This is a direct consequence of the conservation of $\mathcal{R}$ on super-horizon scales. 

Depending on whether the areal radius $R(r)$ given by Eq.~\eqref{def:arealR} is monotonic, the perturbation can be classified as Type I and Type II~\cite{Kopp:2010sh}. 
Type II perturbations are characterized by a non-monotonic $R(r)$, which implies that
\be
\left. 1+r\hat{\mathcal{R}}'\left(r,\mu,K,R_s\right)\right|_{\mathring{r}}=0
\label{eq:typeIIPT}
\ee
has one or more roots at specific values of $\mathring{r}$. For given $K$ and $R_s$, the minimal $\mu$ satisfying Eq.~\eqref{eq:typeIIPT} defines the boundary between Type I and Type II fluctuations, denoted as $\mu_{\mathrm{II}}=\mu_{\mathrm{II}}\left(K,R_s\right)$. Recent numerical studies indicate that Type II fluctuations do not always lead to the so-called Type B PBHs \cite{Inui:2024fgk,Shimada:2024eec}, and the resulting PBH mass depends significantly on the perturbation profile \cite{Uehara:2024yyp}, making it challenging to estimate their masses. For collapse from Type II fluctuations or relevant with Type B PBHs, one needs to go beyond Misner-Sharp equations. Baumgarte-Shapiro-Shibata-Nakamura (BSSN) formalism could be adopted and more details about GR simulation could be found in \cite{Aurrekoetxea:2024mdy}. 

As Type II fluctuations are extremely rare, in this paper we focus exclusively on Type A PBHs formed by Type I fluctuations, for which the compaction function always has a single maximum at $r_m$, determined by
\be
\left.\frac{\mathrm{d}\hat{\mathcal{C}}(r)}{\mathrm{d}r}\right|_{r_m}=0.
\label{def:rm}
\ee
The corresponding areal radius $R(r_m)$ gives the proper ``size'' of the over-dense region \cite{Kitajima:2021fpq}. For Type I fluctuations, Eq.~\eqref{def:rm} is equivalent to
\be
\left. \left(\hat{\mathcal{R}}'(r)+r\hat{\mathcal{R}}''(r) \right) \right|_{r_m} =0.
\label{eq:rm}
\ee
The threshold for PBH formation, independent of the specific profile, can be determined by the volume-averaged compaction function~\cite{Escriva:2019phb}
\be
\overline{\mathcal{C}}_{m}=  \dfrac{\int_0^{R\left(r_{\mathrm{m}}\right)} \mathrm{d} R(r)\left[4 \pi R^2(r)\right] \mathcal{C}(r)}{\frac{4 \pi}{3} R^3\left(r_{\mathrm{m}}\right)}~.
\label{eq:ACF}
\ee
A PBH will form if, on super-horizon scales, the volume-averaged compaction function satisfies
\be
\overline{\mathcal{C}}_{m} \geq \overline{\mathcal{C}}_{\mathrm{th}}= \dfrac{2}{5},
\label{eq:threshold}
\ee
from which we can get the threshold of the peak height $\mu_{\mathrm{th}}$ by solving
\be
\overline{\mathcal{C}}_{m}(\mu_{\mathrm{th}}) = \overline{\mathcal{C}}_{\mathrm{th}}.
\label{def:muth}
\ee
The range of peak heights that can produce Type A PBHs from Type I fluctuations is given by $\mu_{\mathrm{th}}<\mu<\mu_{\mathrm{II}}$.
We note that it was argued that the PBH formation threshold may depend on the choice of the window function or the profile of the perturbation~\cite{Young:2019osy}. 
Here, however, we avoid such profile dependence by adopting the criterion Eq.~\eqref{eq:threshold}, which is shown to be independent of the profile~\cite{Escriva:2019phb}.

\subsection{Black Hole Mass and Number Density}

As we mentioned, a peak with exactly the threshold height $\mu_\mathrm{th}$ will collapse into a PBH with zero mass. If the peak height $\mu$ exceeds the threshold $\mu_\mathrm{th}$, the corresponding PBH mass follows a universal power-law behavior, given by the critical collapse \cite{Choptuik:1992jv,Evans:1994pj,Koike:1995jm,Niemeyer:1997mt,Hawke:2002rf,Musco:2008hv}
\be
M_{\mathrm{PBH}}\left(\mu, K,R_s\right)= M_H\left(R_s\right) \mathcal{K} \Big(\mu-\mu_{\mathrm{th}}\left(K,R_s\right)\Big)^\gamma,
\label{eq:MPBHPT}
\ee
where $\mathcal{K}$ is the dimensionless mass ratio, and $\gamma$ is the critical index. In this paper, we will neglect the parameter dependence of $\mathcal{K}$ and $\gamma$, but consider the approximate reference values $\mathcal{K}= 6$ and $\gamma=0.36$~\cite{Escriva:2019nsa}.
$M_H(R_s)$ represents the horizon mass corresponding to a given comoving smoothing scale $R_s$, which scales as $R_s^2$. This can be related to the horizon mass associated with any comoving wavenumber $k$ as 
\be
\frac{M_H}{M_{k}}=\left( k R_s \right)^{2} ,
\label{eq:MH}
\ee
where $k$ is a reference wavenumber. It can be normalized \textit{e.g.} \cite{Kitajima:2021fpq}
\begin{equation}
M_{k}(k) \simeq 10^{20}\left(\frac{g_*}{106.75}\right)^{-1 / 6}\left(\frac{k}{1.56 \times 10^{13} \mathrm{Mpc}^{-1}}\right)^{-2} \mathrm{~g}~,
\label{def:Mkstar}
\end{equation}
with $g_*$ denoting  effective number of relativistic degrees of freedom\footnote{Enhanced PBH formation can be associated with rapid changes in the early Universe's effective relativistic degrees of freedom and equation of state, such as in scenarios involving novel high-temperature QCD first-order phase transitions~\cite{Lu:2022yuc}.}. 
For example, we can select the central wavenumber $k_*$ of the power spectrum, as defined in Eq.~\eqref{def:PR}, as the reference.

The resulting number density of PBHs of mass $M$ in a comoving volume can be related to the peak number density from Eq.~\eqref{eq:npeakk3} by applying the mass expression in Eq.~\eqref{eq:MPBHPT} as
\be 
\mathscr{N}_{\mathrm{PBH}}\left(M\right)  
\equiv  \int \frac{\mathrm{d}R_s}{R_s} \int \mathrm{d} K \int_{\mu_\mathrm{th}}^{\mu_\mathrm{II}} \mathrm{d} \mu  ~\delta_{\mathrm{D}}\left(\ln \frac{M}{M\left(\mu, K, R_s \right)}\right) \mathscr{N}_{\text{pk }}(\mu,K,R_s) \Theta\left(R_s-\Xi r_m\right), 
\label{def:nPBH}
\ee
where $\delta_{\mathrm{D}}$ is the Dirac delta function,
$\Theta$ is the Heaviside step function,
$\mathscr{N}_{\mathrm{PBH}}(M)$ remains conserved if subsequent evolutionary processes, such as merging, accretion, evaporation, etc., can be neglected. Consequently, the present-day PBH mass function, defined as the number density per logarithmic mass interval, is
\begin{equation}
f_{\mathrm{PBH}}\left(M\right) = \int \frac{\mathrm{d} R_s}{R_s} \int \mathrm{d} K \int \mathrm{d} \mu ~ \delta_{\mathrm{D}}\left(\ln \frac{M}{M\left(\mu, K, R_s \right)}\right) \frac{M\left(\mu, K, R_s \right)\mathscr{N}_{\text {pk}}\Theta\left(R_s-\Xi r_m\right)} {\rho_\mathrm{DM}}.
\label{eq:fPBHPT}
\end{equation}
The total PBH abundance is then given by
\be
f_{\mathrm{PBH}}^{\mathrm{tot}}\equiv \int \mathrm{d}\ln M ~ f_{\mathrm{PBH}}(M).
\label{def:ftot}
\ee

The additional step function $\Theta\left(R_s-\Xi r_m\right)$ in Eqs. \eqref{def:nPBH} and \eqref{eq:fPBHPT} can be understood as follows\footnote{Introducing instead a Delta function-type filter would be highly restrictive and sensitively rely on exact relation between $r_m$ and $R_s$.}. It is introduced to ensure that the maximum of the compaction function, $r_m$, lies approximately within the smoothing scale $R_s$. This condition guarantees that small PBHs with $R_s\lesssim r_m$ do not contribute to the PBH abundance, as expected. 
%\color{blue}{}This is directly related to potential over-counting of nested perturbation contributions that plague some of the previous analyses, and related to the so-called ``cloud-in-cloud'' problem.\color{black}{}
From another perspective, if the size of the over-dense region  $r_m$  is much larger than the horizon scale  $R_H\sim R_s$, the matter within the horizon cannot ``sense'' the curvature, 
and instead behaves as a rescaled background density. The parameter $\Xi \sim \mathcal{O}(1)$ reflects our ignorance of the exact relation between the overdense region's size and the horizon scale, which may depend on the specific shape or other properties of the profile. In this work, we adopt a simple and physically motivated prescription to set this parameter in Sec.~\ref{ss:JusXi}, leaving more detailed numerical studies for future work.

\subsection{Criteria for Selecting $\Xi$}
\label{ss:JusXi}

The aforementioned parameter $\Xi$ that relates $r_m/R_s$
is uncertain due to ambiguity in the exact timing of the horizon re-entry of the over-dense region. 
For a monochromatic power spectrum, there is only a single characteristic scale $k_*$, and the corresponding smoothing scale $R_s$ should be given by 
\be
\frac{M_H}{M_{k_*}}\simeq\left( k_* R_s \right)^{2}=1 , \quad R_s=1/k_*.
\label{eq:fixRs-mono}
\ee 
Substituting Eq.~\eqref{eq:fixRs-mono} into Eq.~\eqref{eq:fPBHPT}, we obtain the PBH mass function for the monochromatic power spectrum case
\be
f_{\mathrm{PBH}}\left(M(\mu)\right)
=\left| \frac{\mathrm{d}\ln M}{\mathrm{d} \mu}\right|^{-1}  \frac{M\left(\mu\right)  \mathscr{N}_{\text {pk }}(\mu)  }{\rho_{\mathrm{DM}}}.
\label{eq:fid-fPBH-mono}
\ee

\begin{figure}
\begin{center}
\includegraphics[width=0.65\textwidth]{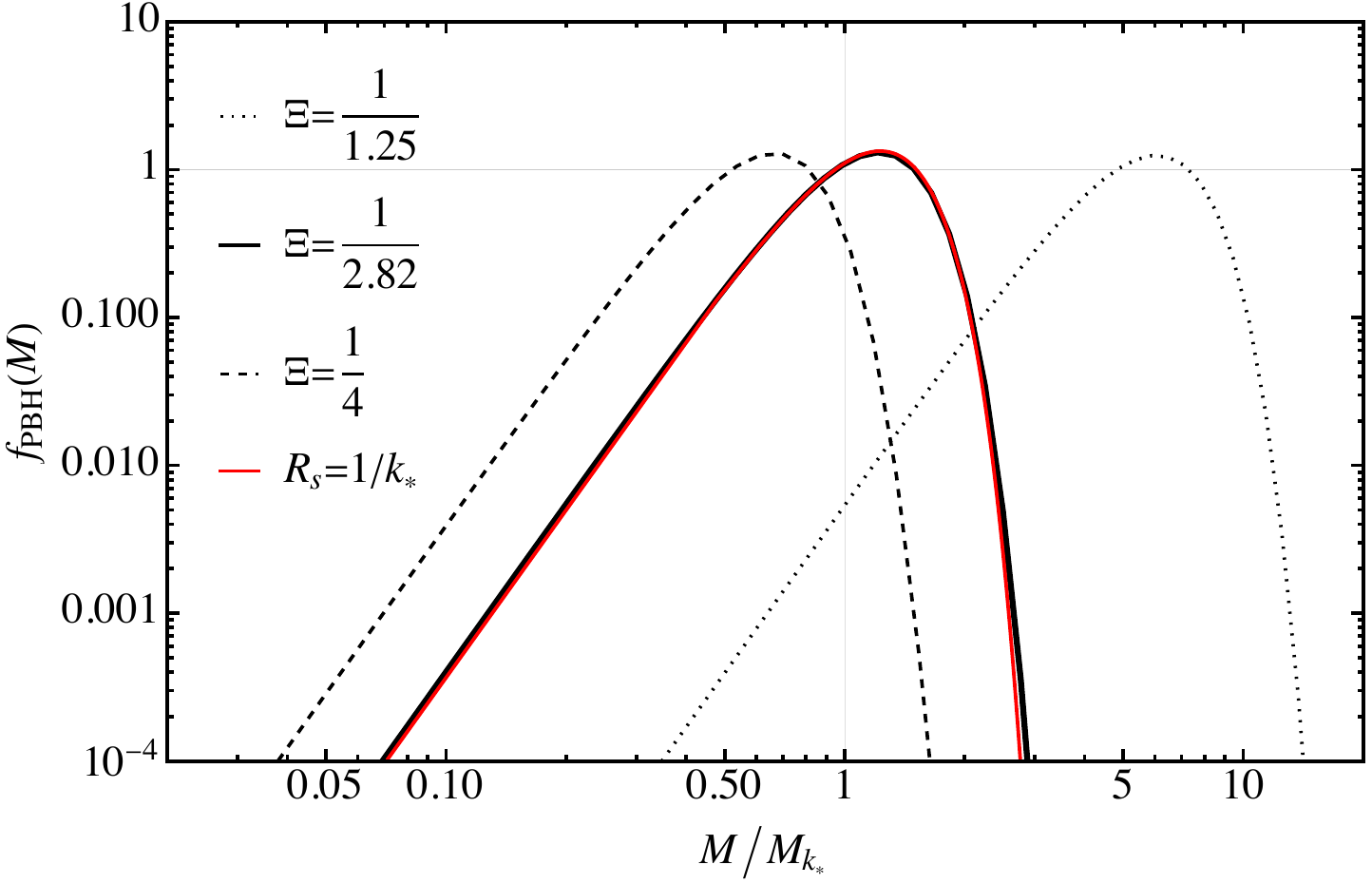}
\caption{PBH mass function for a monochromatic power spectrum obtained from the fiducial expression Eq.~\eqref{eq:fid-fPBH-mono} (red curve), and from the complete expression Eq.~\eqref{eq:fPBHPT} with different $\Xi$'s (dashed, dotted and solid black curves), considering a Gaussian window function of Eq.~\eqref{eq:winG}. Adjusting $\Xi$ such that central mass from the complete formula in Eq.~\eqref{eq:fPBHPT} is the same as the one from the fiducial expression Eq.~\eqref{eq:fid-fPBH-mono}, we have $\Xi_{\mathrm{G}}=1/2.82$. Note that we normalize all the curves by $f_{\mathrm{PBH}}^{\mathrm{tot}}=1$, which have different variances: $\mathcal{A}_{\mathcal{R}}=1.47\times 10^{-2}$ for the red curve and $\mathcal{A}_{\mathcal{R}}=1.55\times 10^{-2}$ for the black curve.}
\label{fig:diffXi}
\end{center}
\end{figure}

A larger $\Xi$ results in a more massive central mass $M_c$ of the PBH mass function $f_{\mathrm{PBH}}(M)$.
To determine $\Xi$ for different types of window functions, we require the general mass function in Eq.~\eqref{eq:fPBHPT} to yield the same central mass as the fiducial expression in Eq.~\eqref{eq:fid-fPBH-mono}. As shown in Fig.~\ref{fig:diffXi}, for the Gaussian window function, given in momentum space as
\be
\widetilde{W}_{\mathrm{G}}(k,R_s)=\exp\left[ -\frac{(k R_s)^2}{2}\right]~,
\label{eq:winG}
\ee
we find $\Xi_{\mathrm{G}}=1/2.82$. 
In Fig.~\ref{fig:diffXi} we also observe that central mass $M_c$ is shifting with respect to considered values of $\Xi$. 
The value of 
$\Xi$ may vary for other window function choices, which we discuss in Sec.~\ref{ss:diffWin}.

\section{Application to Log-normal Power Spectrum}\label{s:Applications and Results}

In many theories, the power spectrum is well-approximated by a log-normal function near its maximum (see \textit{e.g.}~\cite{Pi:2020otn}).
Here, we apply our framework to a log-normal power spectrum given by Eq.~\eqref{def:PR}, considering a Gaussian window function of Eq.~\eqref{eq:winG}, as an illustrative example demonstrating the practical implementation of our method. Other forms of power spectra and window functions can be readily analyzed similarly.

\subsection{General Scenario}\label{ss:Full}

\begin{figure}[t]
    \centering
    \includegraphics[width=0.65\linewidth]{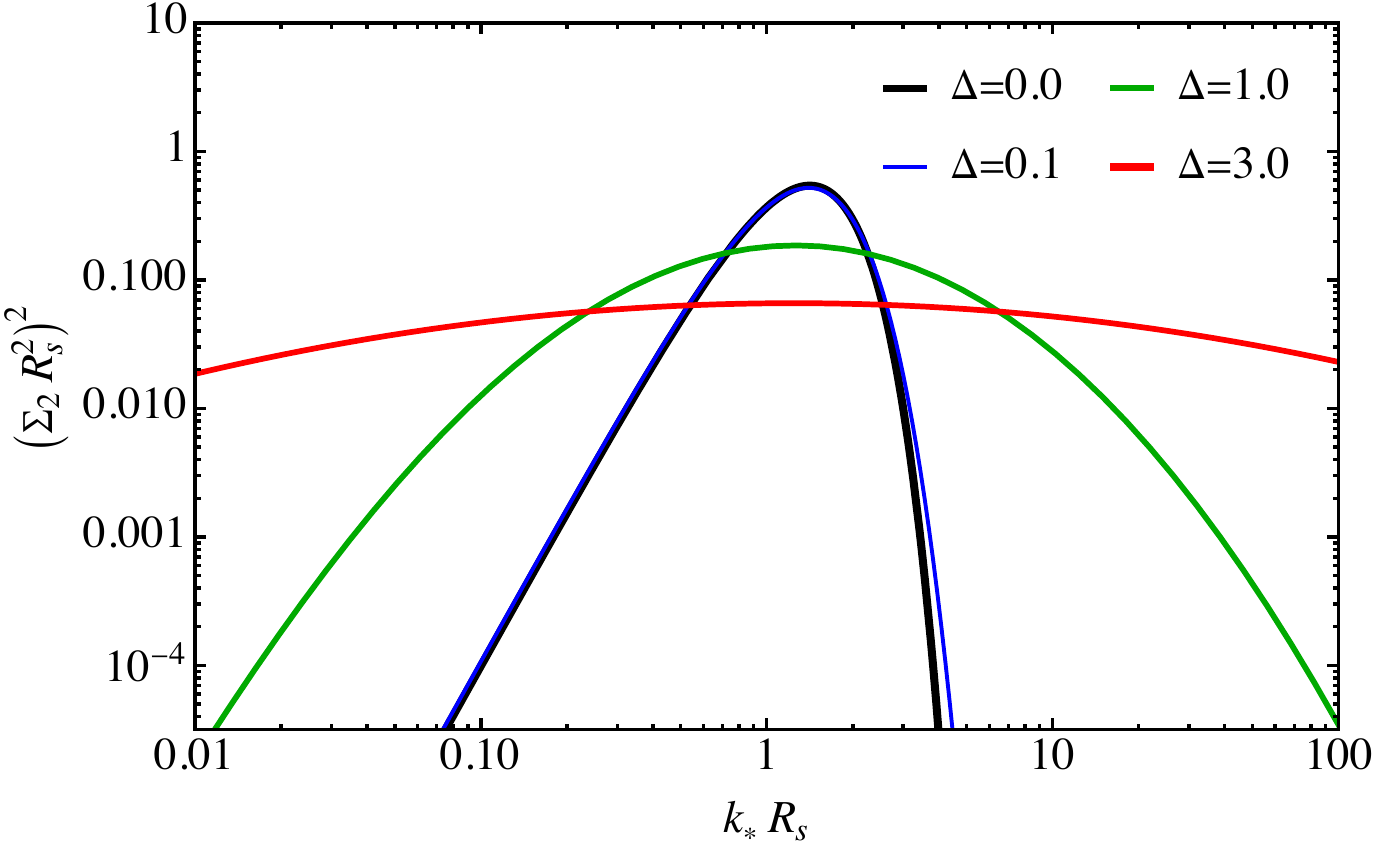}
    \caption{Illustration of $(\Sigma_2 R_s^2)^2$ as a function of $R_s$ for different choices of log-normal power spectrum widths $\Delta$, from Eq.~\eqref{eq:sigmasqu-logN}.}
    \label{fig:Sigma2squ}
\end{figure}

We first considering general log-normal power spectrum case.
Substituting Eq.~\eqref{def:PR} and Eq.~\eqref{eq:winG} into Eq.~\eqref{eq:kWenter}, we obtain 
\be
\label{eq:sigmasqu-logN}
\Sigma_n^2   \equiv \frac{\sigma_n^2 }{\mathcal{A}_{\mathcal{R}}} =\frac{R_s^{-2n}}{\sqrt{2 \pi \Delta^2}} \int  \mathrm{d} \kappa~ \kappa^{2 n-1} \exp \left(-\frac{\left(\ln \kappa-\ln \kappa_*\right)^2}{2 \Delta^2}-\kappa^2 \right),
\ee
and
\be
\gamma_3   =\frac{\Sigma_3^2}{\Sigma_{2} \Sigma_{4}},
\label{def:gamma3}
\ee
where 
\be
\kappa\equiv k R_s, \quad \kappa_*\equiv k_* R_s
\label{eq:kappa}
\ee
are dimensionless wave-numbers normalized by $R_s$. In the peaks theory, $\Sigma_n^2$ as well as $\sigma_n^2$ play a role analogous to the variance of the density contrast in the Press-Schechter formalism, with the key difference that $\Sigma_n^2$ and $\sigma_n^2$ have dimensions of $[M]^{2n}$ by definition. On the other hand, the variance of the density contrast is dimensionless, as shown in App.~\ref{s:PS} (see Eq.~\eqref{def:sigmadelta}). 
In Fig.~\ref{fig:Sigma2squ} we illustrate $(\Sigma_2 R_s^2)^2$ as a function of $ k_* R_s$. This quantity broadens for a larger width $\Delta$ of the log-normal spectrum, while its amplitude decreases as the {variance} remains conserved.

The number density of peaks, from Eq.~\eqref{eq:npeakk3}, is found to be
\be
\begin{aligned}
\mathscr{N}_{\text {pk }}\left(\mu,K,R_s\right)  
=&~2\left(\frac{1}{6 \pi}\right)^{3 / 2} \frac{\Sigma_2^2}{\Sigma_1^4} \frac{\Sigma_4^3}{\Sigma_3^3} \frac{\mu}{2 \pi \mathcal{A}_{\mathcal{R}}} \frac{K}{\sqrt{1-\gamma_3^2}} f\left(\frac{\Sigma_2}{\Sigma_1^2} \frac{\mu}{\sqrt{\mathcal{A}_{\mathcal{R}}}} K^2\right) \\
&~\times 
\exp \left[-\frac{\mu^2}{2 \mathcal{A}_{\mathcal{R}}}\left(\frac{\Sigma_2}{\Sigma_1^2}\right)^2\left(1+\frac{\left(K^2-\gamma_3\right)^2}{1-\gamma_3^2}\right)\right].
\end{aligned}
\label{eq:logN-npeak}
\ee
The peak number density in Eq.~\eqref{eq:logN-npeak} primarily depends on $\mu$, $K$, $R_s$ through the dominant exponential factor, which clearly shows that $K$ is centered approximately at  $\sqrt{\gamma_3(R_s)}$. 

Considering the following dimensionless radius
\be
x\equiv k r, \quad x_*\equiv k_* r , \quad x_W\equiv R_s^{-1} r, 
\label{eq:x}
\ee
we can recast the two-point correlation functions of Eq.~\eqref{eq:kWenter} as
\be
\psi_n(r,R_s)=\frac{1}{\sqrt{2 \pi \Delta^2}} \frac{k_*^{2 n}}{\Sigma_n^2} \int \frac{\mathrm{d} x}{x}\left(\frac{x}{x_*}\right)^{2 n} \frac{\sin (x)}{x} \exp \left[-\frac{\left(\ln x-\ln x_*\right)^2}{2 \Delta^2}-\left(\frac{x}{x_W}\right)^2\right].
\label{eq:logNpsi1}
\ee
Substituting Eq.~\eqref{eq:sigmasqu-logN} and Eq.~\eqref{eq:logNpsi1} into Eq.~\eqref{def:q1q2} and Eq.~\eqref{eq:profilefullbare}, we obtain the smoothed peak profile 
\be
\hat{\mathcal{R}}(r,\mu,K,R_s) = \mu  \sum_{n=1}^2 q_n(K,R_s) \psi_n\left(r, R_s\right).
\label{eq:profilefull}
\ee
Since $\psi_n \to 1$ when $r\to 0$, the height of $\hat{\mathcal{R}}$ is given by
\be
\hat{\mathcal{R}}(r\to 0) = \mu   \sum_{n=1}^2 q_n(K,R_s).
\ee

\begin{figure}[t]
    \centering
    \includegraphics[width=0.65\linewidth]{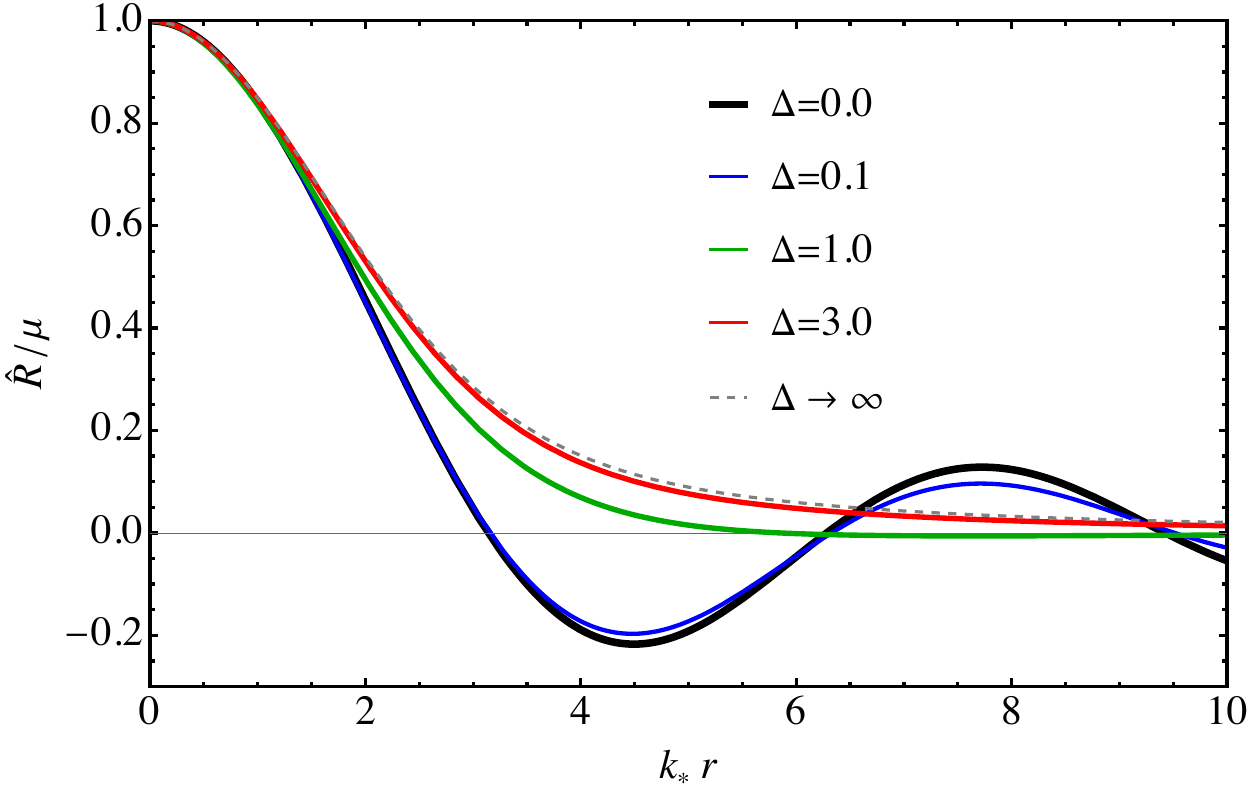}    \caption{Normalized peak profile $\hat{\mathcal{R}}/\mu$ of Eq.~\eqref{eq:profilefull} for different choices of log-normal spectrum widths $\Delta$, considering $K=\sqrt{\gamma_3}$ and $R_s^{-1}=k_*$. Window-function-dominated approximation (WFD) of Eq.~\eqref{eq:app-w-dom} is displayed for reference (gray dashed line).}
    \label{fig:profilediffDelta1}
\end{figure}

\begin{figure}[t]
    \includegraphics[width=0.49\linewidth]{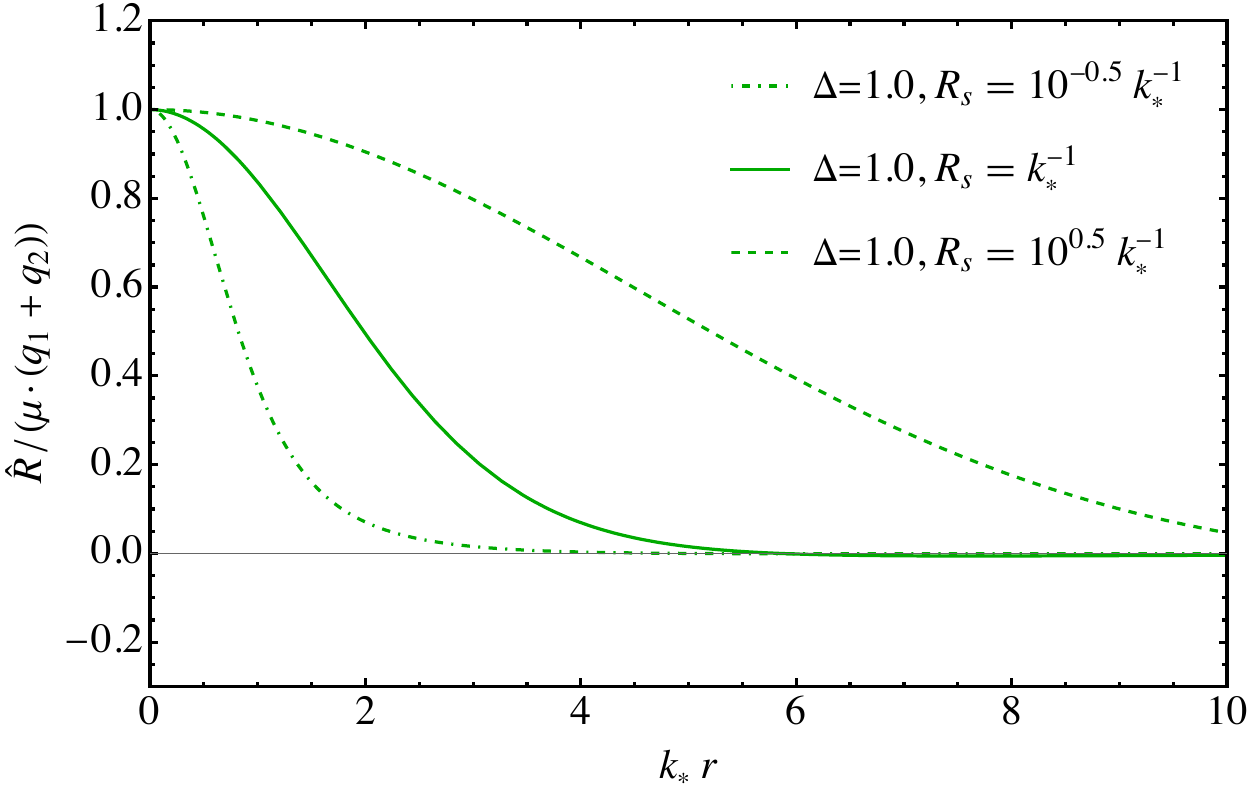} 
    \includegraphics[width=0.49\linewidth]{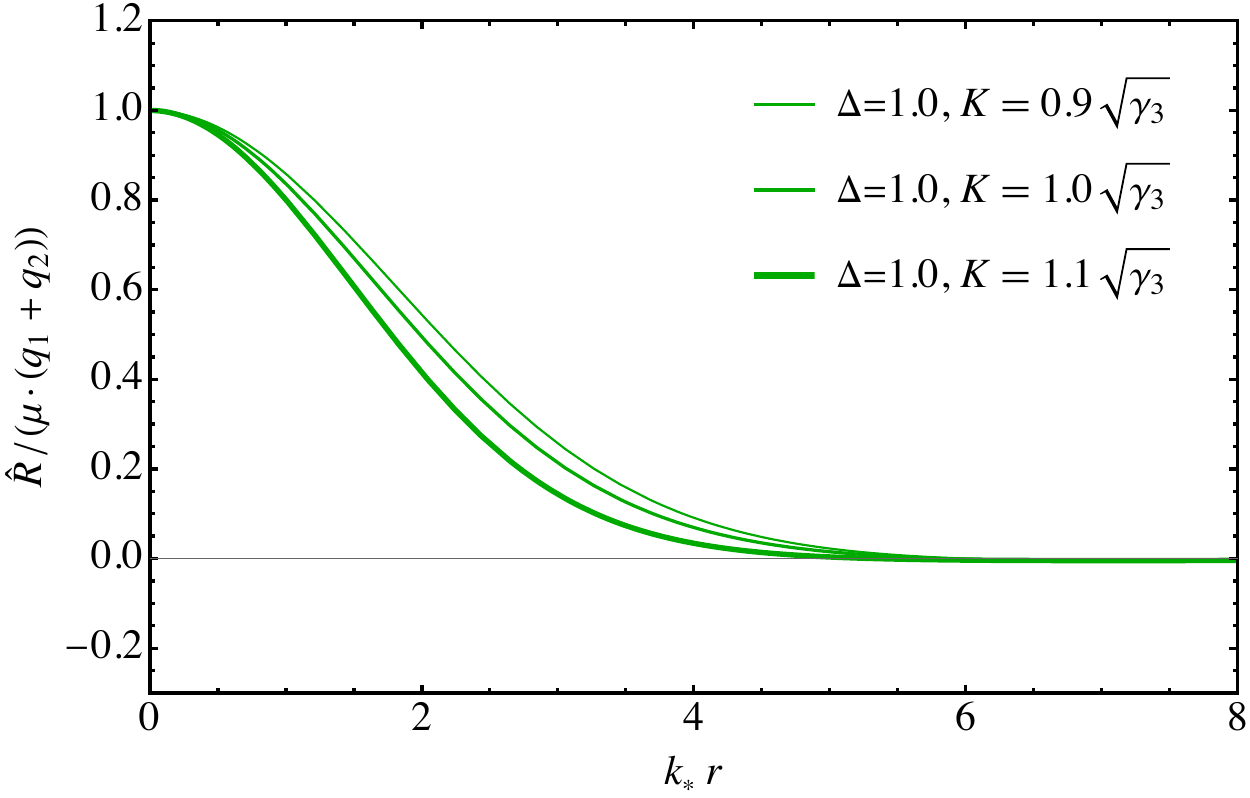}    
    \caption{[Left] Normalized peak profile $\hat{\mathcal{R}}/\mu(q_1+q_2)$ of Eq.~\eqref{eq:profilefull} for different choices of $R_s$, considering $K=\sqrt{\gamma_3}$. [Right] Normalized peak profile $\hat{\mathcal{R}}/\mu(q_1+q_2)$ of Eq.~\eqref{eq:profilefull} for different choices of $K$, considering $R_s^{-1}=k_*$.}
    \label{fig:profilediffDelta2}
\end{figure}

In Fig.~\ref{fig:profilediffDelta1} we display normalized peak profile $\hat{\mathcal{R}}/\mu$ of Eq.~\eqref{eq:profilefull} for different choices of log-normal spectrum widths $\Delta$. We observe that a broader power spectrum results in a wider profile. In Fig.~\ref{fig:profilediffDelta2} we display normalized peak profile $\hat{\mathcal{R}}/\mu(q_1+q_2)$ for different choices of $R_s$ as well as $K$.
Considering smoothing on larger scales $R_s$ is seen to result in wider profile. Further, as suggested by the definition of $K$, changes in $K$ result in profiles of different widths.

To focus our study on Type I fluctuations, it is necessary to determine the boundary between Type I and Type II fluctuations, denoted as $\mu_{\mathrm{II}}$. 
This represents the minimum value of $\mu$ corresponding to a non-monotonic areal radius. By rearranging Eq.~\eqref{eq:typeIIPT}, we can obtain $\mu_{\mathrm{II}}$ by solving
\be
\mu_{\mathrm{II}}(K,R_s)
=\mathrm{Min} \left\{ -\left(\sum_{n=1}^2 q_n(K,R_s)   r\frac{\mathrm{d}\psi_n\left(r, R_s\right)}{\mathrm{d}r} \right)^{-1} \right\}.
\label{eq:muII-logN-Gau}
\ee
Here, $\mathrm{Min}\{...\}$ represents the minimum value when $r$ varies. Additionally, the radius of the first maximum of the compaction function $r_m$ can be readily determined using Eq.~\eqref{eq:rm} as
\begin{equation}
0
= \sum_{n=1}^2 q_n(K,R_s)   \left.\left( \frac{\mathrm{d}\psi_n\left(r, R_s\right)}{\mathrm{d}r}+r\frac{\mathrm{d}^2\psi_n\left(r, R_s\right)}{\mathrm{d}r^2} \right)\right|_{r_m}.
\label{eq:rm-logN-Gau}
\end{equation}
We observe the dependence $r_m=r_m(K,R_s)$, since the upper limit of the integral over $r$ in the averaged compaction function of Eq.~\eqref{eq:ACF} is independent of $\mu$. 

For the smoothed compaction function of Eq.~\eqref{eq:CFprofile} and the areal radius Eq.~\eqref{def:arealR}, we have
\begin{align}\label{eq:Com-logN-Gau}
\hat{\mathcal{C}}(r,\mu,K,R_s)&= \frac{2}{3}\left[1-\left(1+ \mu  \sum_{n=1}^2 q_n(K,R_s)   r\frac{\mathrm{d}\psi_n\left(r, R_s\right)}{\mathrm{d}r} \right)^2\right], \\
R&= a r  \exp\left( \mu   \sum_{n=1}^2 q_n(K,R_s) \psi_n\left(r, R_s\right)\right)~.
\label{eq:RGau}
\end{align}
This allows us to determine the averaged compaction function of Eq.~\eqref{eq:ACF}. Then, from Eq.~\eqref{def:muth} we can obtain the threshold value in terms of peak height $\mu_{\mathrm{th}}(K,R_s)$. Both $\mu_{\mathrm{II}}$ and $\mu_{\mathrm{th}}$ increase when the power spectrum has a larger $\Delta$, a smaller smoothing scale $R_s$, or a narrower profile for larger $K$.

Considering the above determined parameters in mass formula Eq.~\eqref{eq:MPBHPT}, we can derive the PBH mass as a function of $\mu$, $K$, and $R_s$. Combining this with the comoving number density from Eq.~\eqref{eq:logN-npeak} and substituting the results into Eqs.~\eqref{eq:fPBHPT} and \eqref{def:ftot}, we can determine the PBH mass function for various values of log-normal spectrum width $\Delta$, along with the total PBH abundance.

\subsection{Monochromatic Spectrum Limit}\label{ss:Mono}

To facilitate comparison with other works, we present the general formulas in the limit of vanishing spectrum width $\Delta\rightarrow 0$ when a log-normal spectrum resembles narrow monochromatic power spectrum. This not only simplifies the expressions significantly, but also provides a clearer physical interpretation.
Further, the monochromatic spectrum serves as a key reference because it allows to define the unique parameter $\Xi$ in the step function of $\Theta\left(R_s - \Xi r_m\right)$, which is used to exclude undesired contributions from small PBHs.

\begin{figure}[t]
\begin{center}
\includegraphics[width=0.65\textwidth]{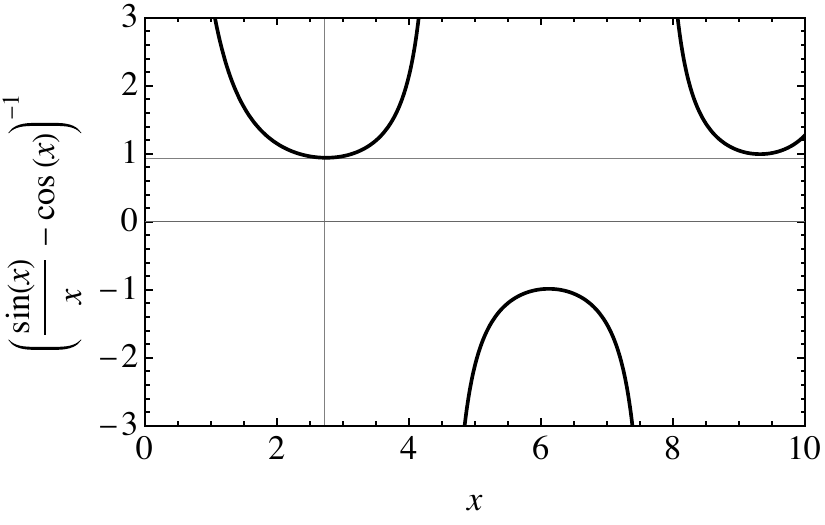}
\caption{Illustration of boundary determination between Type I and Type II fluctuations, with $\mu_{\mathrm{II}} \simeq 0.941$ (horizontal grid-line) for the case of monochromatic spectrum from Eq.~\eqref{eq:muII}.}
\label{fig:Mono-Gau-mu2II}
\end{center}
\end{figure}

We consider the monochromatic power spectrum  given by
\be
\mathcal{P}_{\mathcal{R}}(k)=\mathcal{A}_{\mathcal{R}}  \delta_{\mathrm{D}}\left(\ln k-\ln k_*\right).
\label{eq:monoPS}
\ee
We again focus on the Gaussian window function of Eq.~\eqref{eq:winG}. Following our analysis of log-normal power spectrum and substituting Eq.~\eqref{eq:monoPS} in Eq.~\eqref{eq:kWenter}, we obtain
\be
\Sigma_n^2 =k_*^{2n} e^{-\kappa_*^2} ,\quad \gamma_n  =1, \quad K=\sqrt{\gamma_3}=1,
\label{eq:sigmagamma-mono}
\ee
where $\kappa_*$ is defined in Eq.~\eqref{eq:kappa}. From Eq.~\eqref{eq:sigmagamma-mono} we have $q_2=0$ and the peak profile is described by $\psi_1$ only, yielding
\be
\hat{\mathcal{R}}(r)=\mu \psi_1(r) = \mu\frac{\sin (x_*)}{x_*},
\label{eq:psi1-mono}
\ee
where $x_*\equiv k_*r$ is defined in Eq.~\eqref{eq:x}. Resulting Eq.~\eqref{eq:psi1-mono} implies that the profile is universal and the smoothing scale $R_s$ is not relevant. The upper bound of Eq.~\eqref{eq:typeIIPT}, or that of Eq.~\eqref{eq:muII-logN-Gau}, for Type A PBHs reduces to just a constant value 
\be
\mu_{\mathrm{II}}=\mathrm{Min}\left[  \left( \frac{\sin x_*}{x_*}-\cos x_* \right)^{-1}  \right] \simeq 0.941.
\label{eq:muII}
\ee
In Fig.~\ref{fig:Mono-Gau-mu2II}, we illustrate the functional behavior and the determination of the boundary between the Type I and Type II fluctuation regions. The condition for the value to be positive arises from the requirement that the peak height must be positive. Additionally, the condition of first solution ensures that scenarios where the peak height changes across singularities are not considered.

\begin{figure}[t]
\begin{center}
\includegraphics[width=0.65\textwidth]{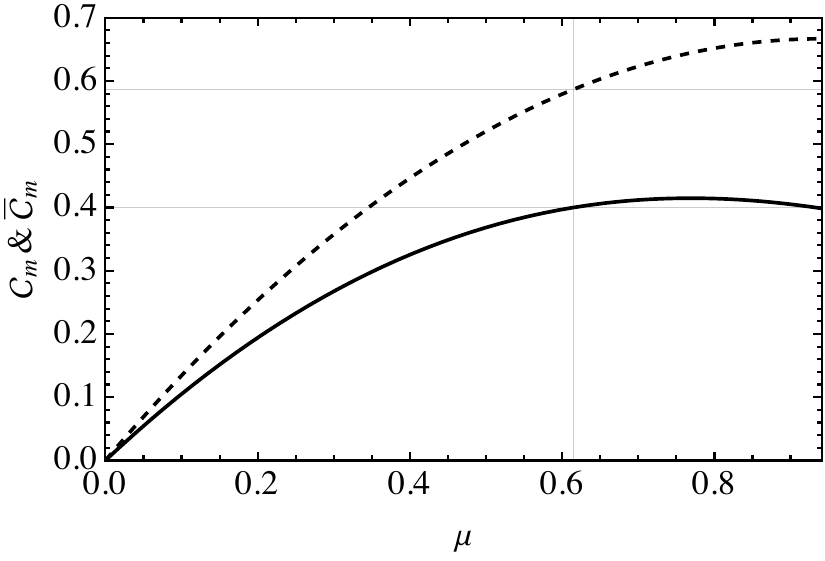}
\caption{The averaged compaction function (solid line), $\overline{\mathcal{C}}_m$, and the maximum compaction function (dashed line), $\mathcal{C}_m$, for a monochromatic power spectrum. The threshold value is defined by $\overline{\mathcal{C}}_m = 0.4$, with the corresponding $\mathcal{C}_m = 0.587$, which matches the threshold condition in the PSC method based on the compaction function.}
\label{fig:Mono-Gau-ACF}
\end{center}
\end{figure}

From Eq.~\eqref{eq:rm} or Eq.~\eqref{eq:rm-logN-Gau}, and considering $x_m\equiv k_* r_m$, we find that the radius of the compaction function maximum is also determined~(see also Ref.~\cite{Kitajima:2021fpq})
\be
 \frac{\sin x_m}{x_m^2}- \frac{\cos x_m}{x_m} -\sin x_m  = 0 \quad\Longrightarrow\quad x_m\simeq 2.74.
\label{eq:xmmono}
\ee
Substituting Eq.~\eqref{eq:psi1-mono} into Eq.~\eqref{eq:Com-logN-Gau} and Eq.~\eqref{eq:RGau}, we find
\begin{equation}
\begin{aligned}
\hat{\mathcal{C}}(r)
=&~ \frac{2}{3}\left[1-\left(1- \mu  \left( \frac{\sin x_*}{x_*} - \cos x_* \right) \right)^2\right] ,\\
\frac{R}{a}=&~ \frac{x_*}{k_*}   \exp \left(\mu \frac{\sin (x_*)}{x_*}  \right).
\end{aligned}
\end{equation}
Then, we can calculate the averaged compaction function $\overline{\mathcal{C}}_m$ inside a sphere of areal radius $R(r_m)$ according to Eq.~\eqref{eq:ACF}. Requiring that
\be
\overline{\mathcal{C}}_m\left(\mu_{\mathrm{th}}\right) = \dfrac{2}{5},
\ee
we find $\mu_{\mathrm{th}} \simeq 0.615$ and $\mathcal{C}_m(\mu_\mathrm{th})\simeq 0.587$. In Fig.~\ref{fig:Mono-Gau-ACF} we display the averaged compaction function $\overline{\mathcal{C}}_m$ and the maximum compaction function  $\mathcal{C}_m$ for monochromatic power spectrum as a function of $\mu$.

Substituting Eq.~\eqref{eq:sigmagamma-mono} into Eq.~\eqref{eq:logN-npeak}, 
we obtain the comoving number density of peaks 
\be
\mathscr{N}_{\text{pk}}\left(\mu,R_s\right) = \left(\frac{1}{6 \pi}\right)^{\frac32} e^{\frac{1}{2}\kappa_*^2} f\left( \frac{\mu}{\sqrt{\mathcal{A}_{\mathcal{R}}}} e^{\frac{1}{2}\kappa_*^2} \right)  \frac{ k_*^3 }{\sqrt{2 \pi \mathcal{A}_{\mathcal{R}}}}   \exp \left[-\frac{\mu^2}{2 \mathcal{A}_{\mathcal{R}} } e^{\kappa_*^2} \right],
\label{eq:npeak-mono}
\ee
where function $f(x)$ is defined in Eq.~\eqref{eq:f(x)}.
Unlike the universal profile in Eq.~\eqref{eq:psi1-mono}, the comoving number density depends on $R_s$, leading to a result that differs from the non-window-function approach, even for a monochromatic power spectrum that is discussed in Ref.~\cite{Kitajima:2021fpq}. In Ref.~\cite{Kitajima:2021fpq}, the relation between the peak number density and the PBH number density is simply linked via the Jacobian factor $\left| \mathrm{d}\ln M(\mu) / \mathrm{d}\mu \right|^{-1}$. 
Tracing the $R_s$ dependence in Eq.~\eqref{eq:npeak-mono}, we find that
\be\label{Npklimit}
 \mathscr{N}_{\mathrm{pk}} \sim \exp \left( 2\kappa_*^2-e^{\kappa_*^2} \right),
\ee
where we use the approximation from Ref.~\cite{Bardeen:1985tr}
\be\label{eq:f(x)->x^3}
f(x) \to x^3 \text{ when }x \to \infty.
\ee
When $R_s\to 0$, the comoving number density of peaks in Eq.~\eqref{Npklimit} converges to a constant, and the mass function goes to zero as $R_s^2$. 
If the PBH mass does not approach zero, the PBH abundance diverges in the UV limit, as indicated by Eq.~\eqref{eq:fPBHPT}. The maximum of the number density is estimated by taking maximum of the term $\left( 2\kappa_*^2-e^{\kappa_*^2}\right) $, \textit{i.e.} 
$\kappa_*=\sqrt{\ln 2}\simeq 0.833$.
On the other hand, the mass decreases proportionally to $R_s^{2}$, as is shown in Eq.~\eqref{eq:MH}. Therefore, the abundance can be approximated as
\be
f_{\mathrm{PBH}}\sim M_{\mathrm{PBH}}  \mathscr{N}_{\mathrm{pk}} \sim \kappa_*^{2} \exp \left( 2\kappa_*^2-e^{\kappa_*^2} \right).
\label{eq:peakk*}
\ee
The maximum PBH abundance is reached approximately around $\kappa_*\sim \mathcal{O}(1)$, with
$\kappa_*\simeq 1.037$. With the $R_s^2$ UV tail and exponential suppressed IR tail ($\sim\exp(-e^{k_*^2R_s^2})$), we can see easily that the total PBH abundance converges. 

A robust method for calculating PBH abundance must satisfy certain physical conditions. First, the central mass $M_c$ in the mass spectrum $f_{\mathrm{PBH}}$ should correspond closely to the central frequency of the perturbation $k_*$, with $M_c/M_{k_*}\sim \mathcal{O}(1)$. This can be verified for our approach by analyzing $\kappa_*$ as discussed earlier. Second, the total PBH abundance given by Eq.~\eqref{eq:fPBHPT} must converge. Some earlier approaches define the PBH mass based on the size of the overdense region, characterized by $r_m$ or $k_*$, such that $M_\mathrm{PBH}\sim r_m^{2}\sim k_*^{-2}$~\cite{Young:2019yug,Young:2019osy,Young:2020xmk,Kitajima:2021fpq,Ferrante:2022mui,Iovino:2024uxp}. However, this constant-mass assumption, characterized by $r_m$, leads to a divergent total abundance when the entire $R_s$ interval is considered. In contrast, our method demonstrates that the PBH mass is well determined by the smoothing scale, $M_{\mathrm{PBH}}\propto R_s^2$, rather than $M_{\mathrm{PBH}}\propto r_m^2$. As the smoothing scale becomes smaller, the PBH mass approaches zero, ensuring a convergent total abundance in the UV limit.

\subsection{High Peak Limit}\label{ss:HPL}

The full expression for the peak number density is complex, but a practical simplification is considering the ``high peak limit'' (HPL). This approximation is valid when the peak height $\mu$ is large. Under this approximation, the dependence on $K$ can be neglected by adopting its most probable value  $K \simeq \sqrt{\gamma_3}$. And the other formulas like the number density and profile are also much simplified. Therefore, it is worth working in the HPL, of which the simplified formulas can help us to understand the physical interpretations. We will show later that the result obtained from the HPL approximation agrees sufficiently well with the full calculation.

The HPL approximation suggests that the peak width $K$ can be approximately fixed by evaluating Eq.~\eqref{eq:npeakk3} considering the following expression for $P_1^{(3)}$,
\be
 P_1^{(3)}\left(\frac{\sigma_2}{\sigma_1^2} \mu, \frac{\sigma_2}{\sigma_1^2} \mu K^2\right) \simeq \frac{1}{\sqrt{2 \pi}} \frac{1}{\sqrt{\gamma_3}} \frac{\sigma_1^2}{\sigma_2}\frac{1}{2 \mu}  \exp \left[-\frac{\mu^2}{2}\left(\frac{\sigma_2}{\sigma_1^2}\right)^2\right] \delta_{\mathrm{D}} \left(K-\sqrt{\gamma_3}\right). 
 \label{eq:P13HPL}
\ee
Here we mention that in the monochromatic spectrum limit, we have
\begin{equation}
\frac{\sqrt{2}}{\mu} \frac{\sigma_1^2}{\sigma_2} \sqrt{1-\gamma_3^2} \to 0.
\label{eq:to0}
\end{equation}
In the case of an extended spectrum, the window function plays the decisive role in determining the moments of the power spectrum. In this case, as we will see in the next subsection, we have $\gamma_3\to \sqrt{2/3}$. 

By setting $K=\sqrt{\gamma_3}$, 
Eq.~\eqref{def:q1q2} simplifies to $
q_1=1$ and $q_2=0$
and the next-to-leading order contribution $\nabla^2 \psi_1$ vanishes. By substituting this into Eq.~\eqref{eq:profilefullbare}, the HPL profile is obtained directly as
\be
\hat{\mathcal{R}}_{\mathrm{HPL}}(r)=\mu   \psi_1(r).
\label{eq:zetaGHPL}
\ee

To derive the HPL peak number density, we substitute Eq.~\eqref{eq:P13HPL} into Eq.~\eqref{eq:npeakk3}, resulting in a function that depends solely on $\mu$
\be
\begin{aligned}
\mathscr{N}_{\mathrm{pk,HPL}}(\mu)\equiv &\int \mathscr{N}_{\mathrm{pk}}(\mu,K) \mathrm{d}K \\
=&~\left(\frac{1}{6 \pi}\right)^{3 / 2} \frac{\sigma_2}{\sigma_1^2} \frac{\sigma_4^3}{\sigma_3^3}  f\left(\frac{\sigma_2}{\sigma_1^2} \mu \gamma_3 \right) \frac{1}{\sqrt{2 \pi}} \exp \left[-\frac{\mu^2}{2}\left(\frac{\sigma_2}{\sigma_1^2}\right)^2\right].
\label{eq:npeakHPL}
\end{aligned}
\ee
This approach simplifies our calculations while maintaining accuracy, as PBH formation primarily occurs for high peaks. 
With $\sigma_n^2$ and $\psi_n$ still given by Eq.~\eqref{eq:sigmasqu-logN} and Eq.~\eqref{eq:logNpsi1}, the peak profile and number density in the HPL are provided by Eq.~\eqref{eq:zetaGHPL} and Eq.~\eqref{eq:npeakHPL}, respectively. Due to the absence of $\psi_2$ in the profile, the expression for $\mu_{\mathrm{II}}$ and $r_m$ can be directly obtained by substituting $q_1$ and $q_2$ into Eq.~\eqref{eq:muII-logN-Gau} and Eq.~\eqref{eq:rm-logN-Gau}.
As we demonstrate below, the HPL approximation provides valuable insights into the physical interpretation of key quantities.

\subsection{Profile Restrictions and Smoothing Scale}\label{ss:cut-off}

Building on the technical details outlined above, we discuss the dependence of our method on smoothing scale $R_s$. Different smoothing scales correspond to compaction functions with distinct shapes, each having a maximum point that does not necessarily coincide with the smoothing scale. Here, for concreteness, we focus on a log-normal power spectrum and a Gaussian window function. Our analysis can be readily extended to other types of power spectra and window functions.  

\begin{figure}[t]
    \centering
\includegraphics[width=0.65\linewidth]{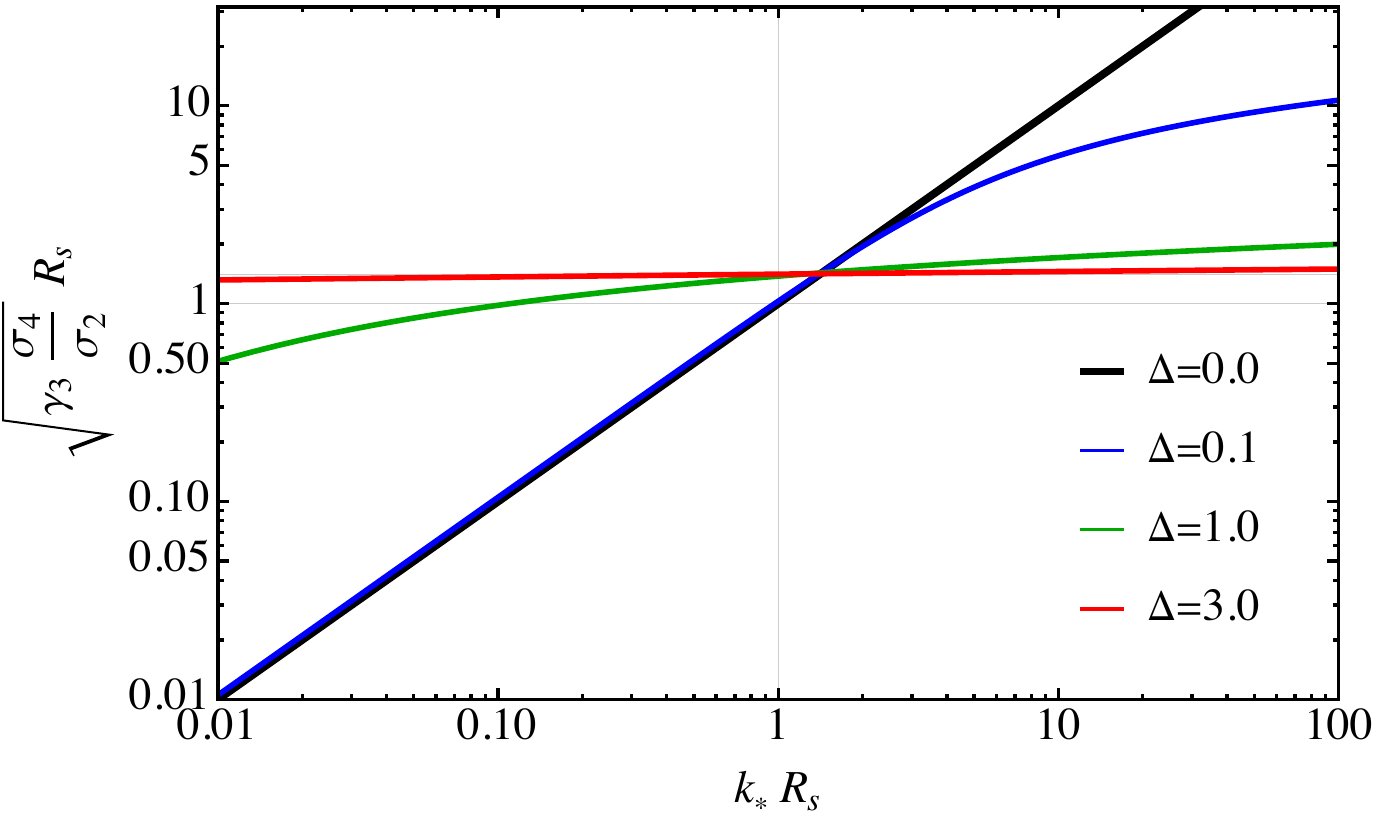}
    \caption{Dependence of the peak profile mean width, $\sqrt{\gamma_3 \sigma_4/\sigma_2}$, on the smoothing scale, as given by Eq.~\eqref{def:muK}. For a monochromatic power spectrum  $\sqrt{\gamma_3 \sigma_4/\sigma_2}=k_*$. The horizontal grid-lines are $1$ and $\sqrt{2}$. For really large $\Delta$, we use Eq.~\eqref{eq:app-w-dom-sig} to derive the limit value of $\sqrt{\gamma_3 \sigma_4/\sigma_2} ~R_s \to \sqrt{2}$. }
    \label{fig:Kc}
\end{figure}

An additional constraint is applied to the profile for a given smoothing scale to ensure that the innermost maximum of the compaction function lies within the sphere of radius $R_s$. 
First, in Fig.~\ref{fig:Kc} we display how the peak profile mean width $K_3 = \sqrt{\gamma_3 \sigma_4/\sigma_2}$ depends on the smoothing scale, as given by Eq.~\eqref{def:muK}. 
The profile mean width decreases as the smoothing scale shrinks. This effect is more pronounced for narrow power spectra, while broad power spectra are less sensitive. In the large-scale limit, for a fixed $R_s$, a broader power spectrum results in a wider profile.
We now focus on HPL approximation in Sec.~\ref{ss:HPL} for simplicity. The position of the innermost maximum of the compaction function, $r_m$, is approximately the size of the over-dense region, which is given by
\begin{equation}
\left.\left( \frac{\mathrm{d}\psi_1\left(r, R_s\right)}{\mathrm{d}r}+r\frac{\mathrm{d}^2\psi_1\left(r, R_s\right)}{\mathrm{d}r^2} \right)\right|_{r_m} = 0.
\label{eq:rm-logN-Gau-HPL}
\end{equation}

\begin{figure}
    \centering
    \includegraphics[width=0.65\linewidth]{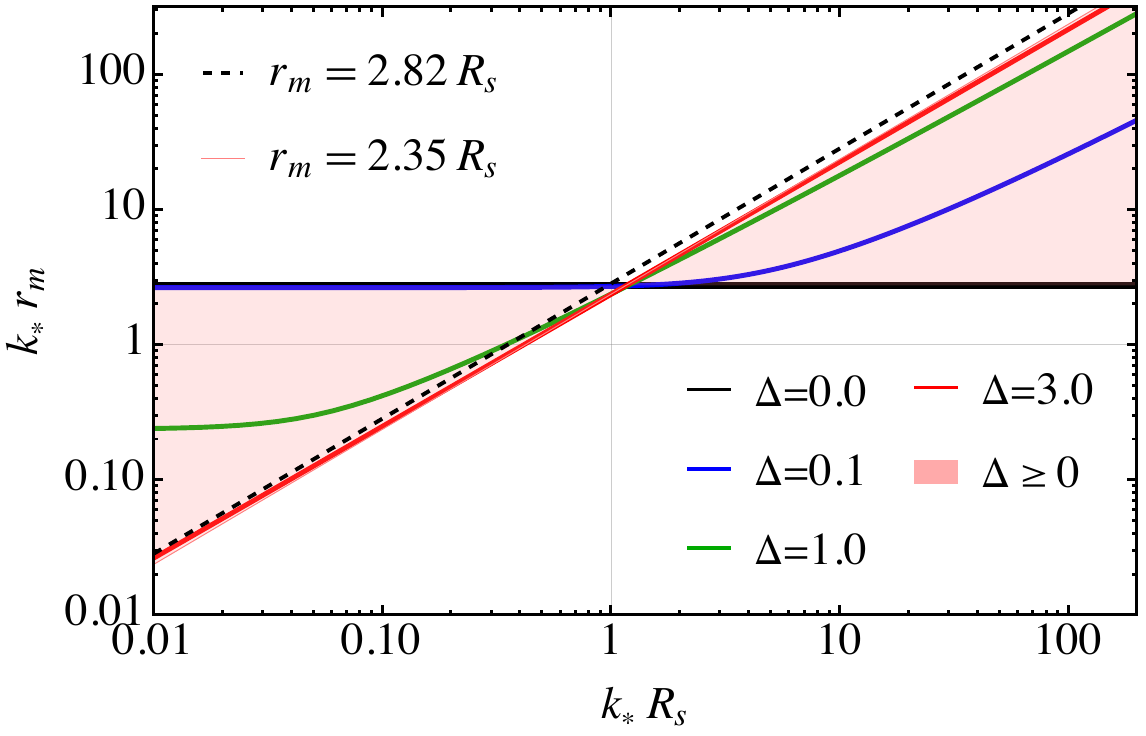}
    \caption{Dependence of compaction function maximum radius $r_m$ on smoothing scale. The pink shaded region is where all the curves with $\Delta\geq0$ can cover. The $\Delta=0$ (monochromatic) case is given by $k_*r_m\simeq2.74$ and the large width limit is given by $k_*r_m \simeq 2.35 k_*R_s$. 
    The dashed line represents the cutoff condition used in this work, $R_s= \Xi r_m$, where we adopt $\Xi_{\mathrm{G}}=1/2.82$ for the Gaussian window function (see Sec.~\ref{ss:JusXi}). }
    \label{fig:rmkW}
\end{figure}

In Fig.~\ref{fig:rmkW} we display the dependence of the maximum radius $r_m$ on smoothing scale, $r_m=r_m(\Delta,R_s)$. In the general formulation of our method, $r_m$ also depends on the peak width $K$ (see Eq.~\eqref{eq:rm-logN-Gau}). However, in the HPL approximation, this dependence is omitted for simplicity.
We observe in Fig.~\ref{fig:rmkW} the presence of a plateau when $R_s$ is small. For a narrow peak, this plateau emerges immediately when $k_*R_s \lesssim 1$, while for larger widths $\Delta$ it appears at smaller scales.
However, the existence of such a plateau is unphysical, as it suggests that for a smoothing scale $R_s$ much smaller than $1/k_*$, the innermost maximum $r_m \gg R_s \sim R_H$ remains constant. Obviously, PBHs can not form when $R_H$ is significantly smaller than $r_m$. To address this issue, we introduce a step function  $\Theta( R_s-\Xi r_m)$. This enforces the condition that PBHs only form when the innermost maximum of the compaction function $r_m$  lies approximately within the smoothing scale $R_s$. The $\mathcal{O}(1)$ coefficient of $\Xi$ depends on the details of the collapse, which can only be accurately determined through numerical relativity investigations. In this analysis, we adopt the principle that the central mass is primarily determined by $R_s\sim 1/k_*$, in particular for the monochromatic case. See Sec.~\ref{ss:JusXi} for details.

In order to understand the different behaviors of $r_m(R_s)$ for $k_* R_s >1$ and $k_* R_s <1$, we note that the key factor in the definitions of $\Sigma_n$ in Eq.~\eqref{eq:sigmasqu-logN} and $\psi_n$ in Eq.~\eqref{eq:logNpsi1} is the following exponential function component that we mark as $f_e$,
\begin{equation}
\ln f_e \equiv    - \frac{\left(\ln k-\ln k_*\right)^2}{2 \Delta^2}-(k R_s)^2  .
\label{eq:terms}
\end{equation}
Depending on whether the first or the second term dominates in Eq.~\eqref{eq:terms}, the exponential function $f_e$ exhibits different behaviors as a function of $k$. If the second term (from Window function) in Eq.~\eqref{eq:terms} is negligible, $\sigma_n$ and $\psi_n(r)$ become homogeneous expressions of $k/k_*$. Conversely, if the first term (from the power spectrum) is negligible, $\sigma_n$ and $\psi_n(r)$ become homogeneous expressions of $k R_s$. In both cases, $r_m$ can be explicitly solved, as discussed below in Eq.~\eqref{eq:rminf} and the text in the vicinity. To gain insight, we consider three limiting cases below.
\begin{enumerate}[label=(\arabic*)] 
    \item The ``window-function dominated" (WFD) case in HPL, which describes the large-$R_s$ behavior seen in Fig.~\ref{fig:rmkW};
    \item The $k_*$-dominated case, which captures the small-$R_s$ limit; 
    \item The WFD case in full expression, which is similar to (1) except that the non-trivial $K$-dependence is retained.
\end{enumerate}

In the case (1), the WFD in HPL, the first term in Eq.~\eqref{eq:terms} is negligible. Hence we obtain
\be
\label{eq:app-w-dom-sig}
\sigma_n^2 
\simeq \sigma_{n,\mathrm{WFD}}^2 = \frac{1}{\sqrt{2 \pi }}\frac{\mathcal{A}_{\mathcal{R}}}{\Delta} \frac{R_s^{-2n}}{2}\Gamma(n).
\ee
This gives $\gamma_3 =\sigma_3^2/(\sigma_2\sigma_4)=\sqrt{2/3}$,
and \eqref{eq:logNpsi1} gives
\be
\psi_{n}(r)\simeq \psi_{n,\mathrm{WFD}}(r)
= {}_1 F_1\left(n,\frac{3}{2},-\frac{(r/R_s)^2}{4}\right),
\label{eq:app-w-dom}
\ee
where ${}_1 F_1(a,b,z)$ is the confluent hypergeometric function. 
The WFD limit is more prominent when $\Delta\gtrsim1$, as can be seen from Fig.~\ref{fig:rmkW}. Apparently, this is because when the power spectrum becomes flatter and flatter, the $k$-dependence is mainly provided by the window function. 
Substituting Eq.~\eqref{eq:app-w-dom} into Eq.~\eqref{eq:rm-logN-Gau-HPL}, we obtain the equation for $r_m$ in the WFD limit in HPL
\begin{equation}
1+\frac{(r_m / R_s)^2}{2} -\left(\frac{(r_m / R_s)^3}{4}+\frac{1}{r_m / R_s}\right)\sqrt{\pi}  \mathrm{erfi}\left( \frac{r_m / R_s}{2} \right)\exp\left[ -\frac{(r_m / R_s)^2}{4} \right] = 0,
\label{eq:WFDHPLrm}
\end{equation}
where $\mathrm{erfi}(x)$ is the imaginary error function. The solution of Eq.~\eqref{eq:WFDHPLrm} is found to be 
$r_m\simeq 2.35 R_s$.
Hence, for the case of WFD limit in HPL, $r_m$ is proportional to $R_s$. As is shown in Fig.~\ref{fig:rmkW}, the deviation between the precise result and $r_m \simeq 2.35 R_s$ of WFD in HPL reflects the contribution from the power spectrum where the first term in Eq.~\eqref{eq:terms} is not ignored. The precise result indicates that a broader power spectrum produces a larger innermost maximum at large smoothing scales but results in an even smaller innermost maximum compared to a narrow power spectrum at small smoothing scales.

\begin{figure}
    \centering
    \includegraphics[width=0.65\linewidth]{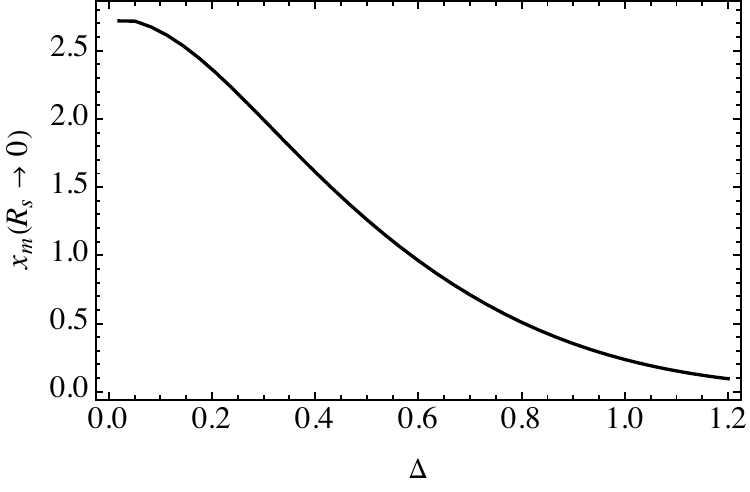}
    \caption{Relation between 
    maximum compaction function radius $r_m$ in the limit of $R_s \rightarrow 0$ as a function of spectrum width $\Delta$, in terms of $x_m\equiv k_* r_m$ determined by Eq.~\eqref{eq:rminf}. }
    \label{fig:lnxmkW}
\end{figure}

In case (2), where the power spectrum dominates in Eq.~\eqref{eq:terms}, the smoothing scale $R_s$ no longer plays a role.
In this limit, Eq.~\eqref{eq:sigmasqu-logN} simplifies to
\begin{equation}
\sigma_n^2 
\simeq \mathcal{A}_{\mathcal{R}}  k_*^{2n}e^{2n^2\Delta^2},
\end{equation}
with $\gamma_3\to 1$, and the general method reduces to HPL. Here, $\psi_1(r)$ cannot be expressed analytically. From Eq.~\eqref{eq:rm-logN-Gau-HPL} we can obtain $r_m$, by considering $x_m = k_* r_m$ and solving
\begin{equation}
 \int \frac{\mathrm{d} x}{x} \left(\frac{x}{x_m}\right)^{3} \left[ \frac{\sin x}{x^2} -\frac{\cos x}{x} -\sin x \right] \exp \left(-\frac{\left(\ln x-\ln x_m\right)^2}{2 \Delta^2} \right) = 0 ,
\label{eq:rminf}
\end{equation}
which only depends on $\Delta$. This corresponds to the constant $r_m$ observed for different values of spectrum widths $\Delta$ in Fig.~\ref{fig:rmkW} when $R_s$ is small. These plateaus indicate that the window function is no longer effective, resulting in a constant $r_m$ that represents the minimum
attainable value after the window function ceases to be relevant.
Covering these plateaus in the final result essentially involves redundant accumulation, since in the actual physical processes $r_m$ associated with these plateaus corresponds to the smallest over-dense size dictated by the input power spectrum. In Fig.~\ref{fig:lnxmkW} we further illustrate the relation between
$r_m(R_s \to 0)$ and $\Delta$. This demonstrates that the window function loses its effectiveness at a smaller smoothing scale for a broader power spectrum.

Better understanding of the form of $r_m(R_s)$ can be inferred  from Eq.~\eqref{eq:kWenter}. The blue-tilted part of  $k^{2n}\mathcal{P}_{\mathcal{R}}(k)$ is dominated by the smoothing scale, while the red-tilted part is approximately governed by the central frequency $k_*$. While the specific shape of the power spectrum and the form of the window function can alter the exact expression for $r_m(R_s)$, they do not change the general trends described above.

\begin{figure}
    \centering
    \includegraphics[width=0.65\linewidth]{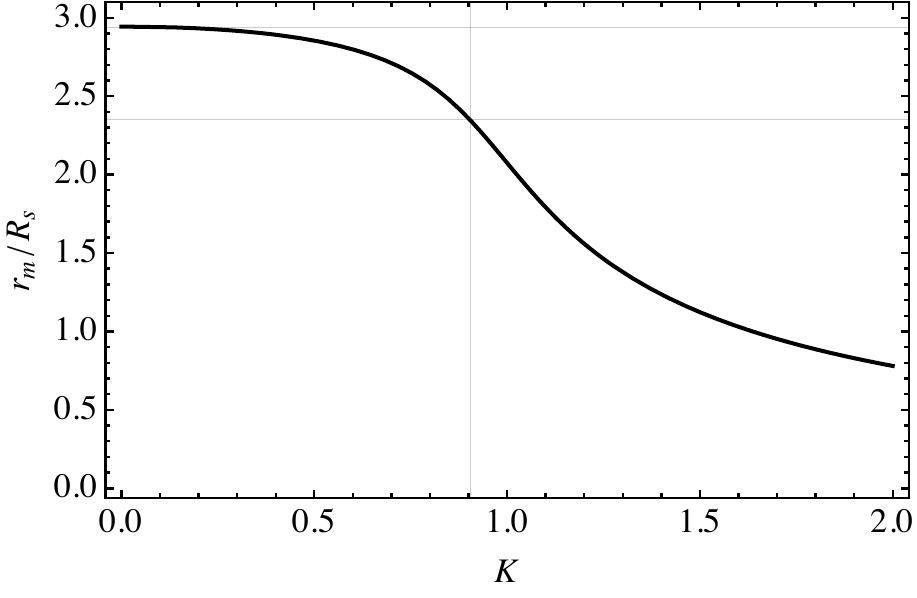}
    \caption{Dependence of $r_m/R_s$ on $K$ in WFD limit considering next-to-leading order terms in profile, obtained by solving Eq.~\eqref{eq:solveXiK}. The vertical line denotes $K=(2/3)^{1/4}$, $r_m / R_s\simeq 2.35$ being solution of WFD in HPL and the horizontal line is the solution when $K=0$, $r_m / R_s\simeq 2.94$.}
    \label{fig:Xi}
\end{figure}

In case (3), the dependence on the peak width $K$ can be retained in WFD limit by considering the next-to-leading order terms in the profile from Eq.~\eqref{eq:profilefull}. As displayed in Fig.~\ref{fig:Xi}, $r_m / R_s$ is no longer constant, as it is in the HPL case, but instead becomes dependent on $K$. In WFD case, the two coefficients $q_{n=1,2}$ are independent of $R_s$ due to $\gamma_3 \rightarrow \sqrt{2/3}$, giving
\begin{equation}
 q_1=3\left(1-\sqrt{\frac{2}{3}}   K^2\right) , \quad q_2=\sqrt{\frac{3}{2}}   K^2-1.
\end{equation}
Considering Eq.~\eqref{eq:rm-logN-Gau}, $r_m$ can be found from
\be
\begin{aligned}
& q_1\left(K\right) \left(-\frac{(r_m / R_s)^2+2}{2}\right)+q_2\left(K\right)\left(\frac{(r_m / R_s)^4-8 (r_m / R_s)^2-4}{8}\right) \\
&+ \left\{ q_1\left(K\right) \frac{(r_m / R_s)^4+4}{4 (r_m / R_s)} +q_2\left(K\right)\left(\frac{-(r_m / R_s)^6+10 (r_m / R_s)^4-4 (r_m / R_s)^2+8}{16 (r_m / R_s)} \right)\right\} \\
&+ \sqrt{\pi}   \operatorname{erfi}\left(\frac{r_m / R_s}{2}\right) \exp \left[-\frac{(r_m / R_s)^2}{4}\right] = 0
\end{aligned}
\label{eq:solveXiK}
\ee
In Fig.~\ref{fig:Xi} we display numerical solution of Eq.~\eqref{eq:solveXiK}.

We highlighted the impact of the smoothing scale on the entire set of peaks theory statistical quantities, particularly when the window function fails to appropriately modulate the full frequency range. 
The limitation of the window function in effectively smoothing out small scales reflects the failure to capture the central frequency $k_*$. This can be utilized to simulate the process of approximating a flat spectrum by a extremely broad log-normal power spectrum. Naturally, our method can also be directly readily applied to a flat spectrum.

\subsection{Choice of Window Function}\label{ss:diffWin}

We now discuss the influence of the choice of different window functions, focusing on the monochromatic power spectrum for simplicity. While the Gaussian window function given by Eq.~\eqref{eq:winG} has been the primary focused thus far, variety of alternatives exists \cite{Ando:2018qdb, Young:2019osy}. We consider two additional characteristic cases, the real-space top-hat window function 
\be
\widetilde{W}_{\mathrm{rTH}}\left( k,R_s\right)=3 \frac{\sin \left(k R_s\right)-(k R_s) \cos (k R_s)}{(k R_s)^3},
\label{eq:winrTH}
\ee
and $k$-space top-hat window function 
\be
\widetilde{W}_{\mathrm{kTH}}\left( k,R_s\right)=\Theta\left(R_s^{-1}-k\right).
\label{eq:winkTH}
\ee
Here, $\Theta(x)$ is Heaviside step-function, with $\Theta(x=0)=1$. 
Correspondingly, we have
\be
\Sigma_n^2 \equiv \frac{\sigma_n^2}{\mathcal{A}_{\mathcal{R}}}=k_*^{2 n}~\widetilde{W}\left( k_*,R_s\right)^2, \quad \gamma_n=1,
\label{eq:SigmadiffWin}
\ee
and
\be
\mathscr{N}_{\text{pk}}= \left(\frac{1}{6 \pi}\right)^{\frac32} \widetilde{W}^{-1}\left( k_*,R_s\right) f\left(  \frac{\mu}{\sqrt{\mathcal{A}_{\mathcal{R}}}} \widetilde{W}^{-1}\left( k_*,R_s\right) \right)  \frac{ k_*^3 }{\sqrt{2 \pi \mathcal{A}_{\mathcal{R}}}}   \exp \left[-\frac{\mu^2}{2 \mathcal{A}_{\mathcal{R}} } \widetilde{W}^{-2}\left( k_*,R_s\right) \right]~.
\label{eq:npeak-diffWin}
\ee
Note that Eq.~\eqref{eq:SigmadiffWin} and Eq.~\eqref{eq:npeak-diffWin} are valid for all types of window functions, and can be evaluated by substituting Eq.~\eqref{eq:winG}, Eq.~\eqref{eq:winrTH} or Eq.~\eqref{eq:winkTH} as $\widetilde{W}$. The peak profile is the same as that of Eq.~\eqref{eq:psi1-mono}.   
Following procedure of Sec.~\ref{ss:JusXi} we can identify filtering conditions on $\Xi$ for each window function as 
\be
\Xi_{\mathrm{G}}=1/2.82, \quad \Xi_{\mathrm{rTH}}=1/2.90, \quad \Xi_{\mathrm{kTH}}=1/2.87.
\label{eq:3Xi}
\ee
With these results, we can estimate the PBH abundance using different window function types. We show and discuss this further in Sec.~\ref{s:results}.

Importantly, as smoothing scale approaches zero ($R_s \rightarrow 0$), we find that Eq.~\eqref{eq:SigmadiffWin} and Eq.~\eqref{eq:npeak-diffWin} result in
\be
\Sigma_n^2 \equiv \frac{\sigma_n^2}{\mathcal{A}_{\mathcal{R}}}=k_*^{2 n}, \quad \gamma_n=1,
\ee
and
\be
\mathscr{N}_{\text{pk}}= \left(\frac{1}{6 \pi}\right)^{\frac32} f\left(  \frac{\mu}{\sqrt{\mathcal{A}_{\mathcal{R}}}} \right)  \frac{ k_*^3 }{\sqrt{2 \pi \mathcal{A}_{\mathcal{R}}}}   \exp \left[-\frac{\mu^2}{2 \mathcal{A}_{\mathcal{R}} } \right]~.
\ee
These results are identical to those obtained when considering a bare power spectrum. This further supports the idea that PBH formation methods ignoring the window function are effectively equivalent to taking the limit $R_s \rightarrow 0$. However, such extremely small PBHs are not expected to form from perturbations with a wavelength of $1/k_*$. 

\begin{figure}[t]
\begin{center}
\includegraphics[width=0.65\textwidth]{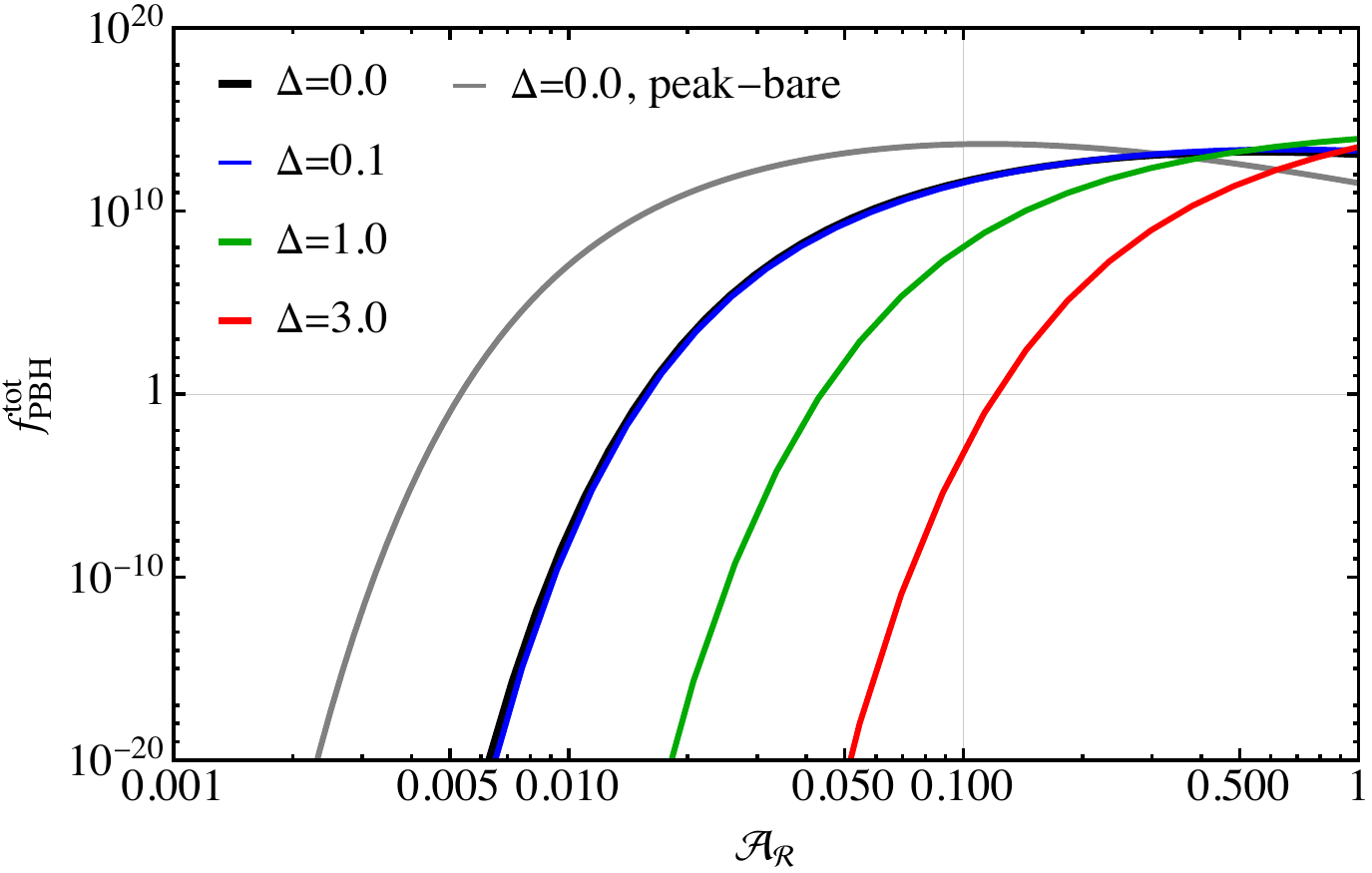}
\caption{Total PBH abundance calculated by the peaks theory smoothed by
a Gaussian window function and filtered by a filter function (solid lines, ``peak-filtered'') for different choices of the log-normal power spectrum widths $\Delta$. Result following method of Ref.~\cite{Kitajima:2021fpq} without smoothing or filtering is overlaid for reference (gray line, ``peak-bare'').}
\label{fig:ftotallPTFull}
\end{center}
\end{figure}

\section{Results and Discussion}\label{s:results}

We now present the PBH mass function and its total abundance calculated using our method developed in this work, along with results obtained considering HPL and for different choices of window functions. We highlight several notable features of the results and provide a comparison with those derived from previous methods in the literature.

\subsection{PBH Abundance}

In Tab.~\ref{tab:PTFullchart}  we summarize  the {variance} $\mathcal{A}_{\mathcal{R}}$ required to produce PBH as all the dark matter for power spectra with different widths, along with their corresponding central mass $M_{c}$.
Similar to other methods, broader power spectra result in lower PBH abundance for a given {variance} $\mathcal{A_R}$, as the power spectrum amplitude around the central frequency $k_*$ decreases with increasing width. 

\begin{figure}[t]
\begin{center}
\includegraphics[width=0.65\textwidth]{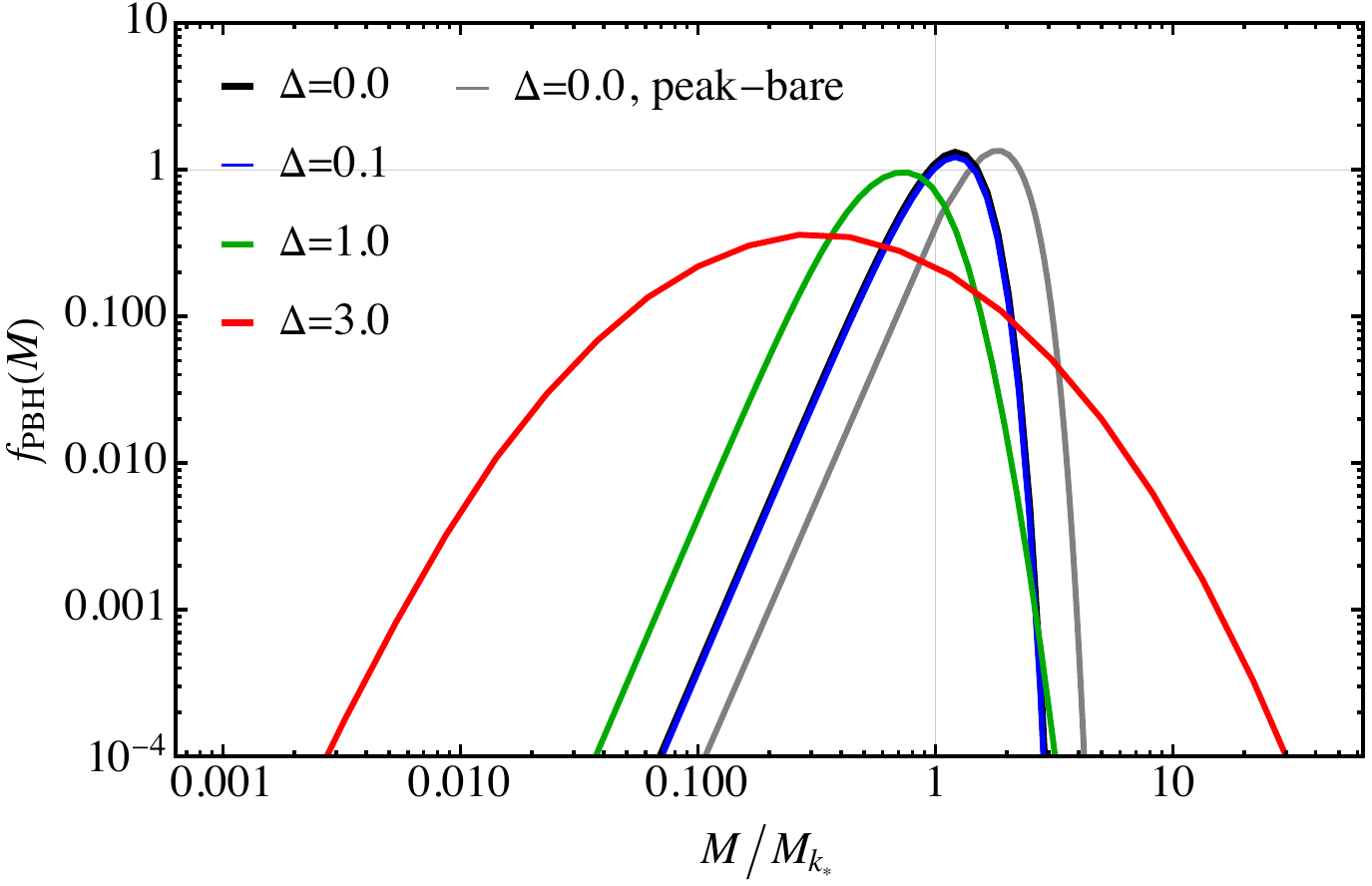}
\caption{PBH mass function calculated by the peaks theory smoothed by
a Gaussian window function and filtered by a filter function (solid lines, ``peak-filtered'') for different choices of the log-normal power spectrum widths $\Delta$, normalized such that PBHs constitute all DM with $M_{k_*}= 10^{20}$~g. Result following method of Ref.~\cite{Kitajima:2021fpq} without smoothing or filtering is overlaid for reference (gray line, ``peak-bare'').  
}
\label{fig:fPBHMnew}
\end{center}
\end{figure}

\begin{figure}[t]
\begin{center}
\includegraphics[width=0.49\textwidth]{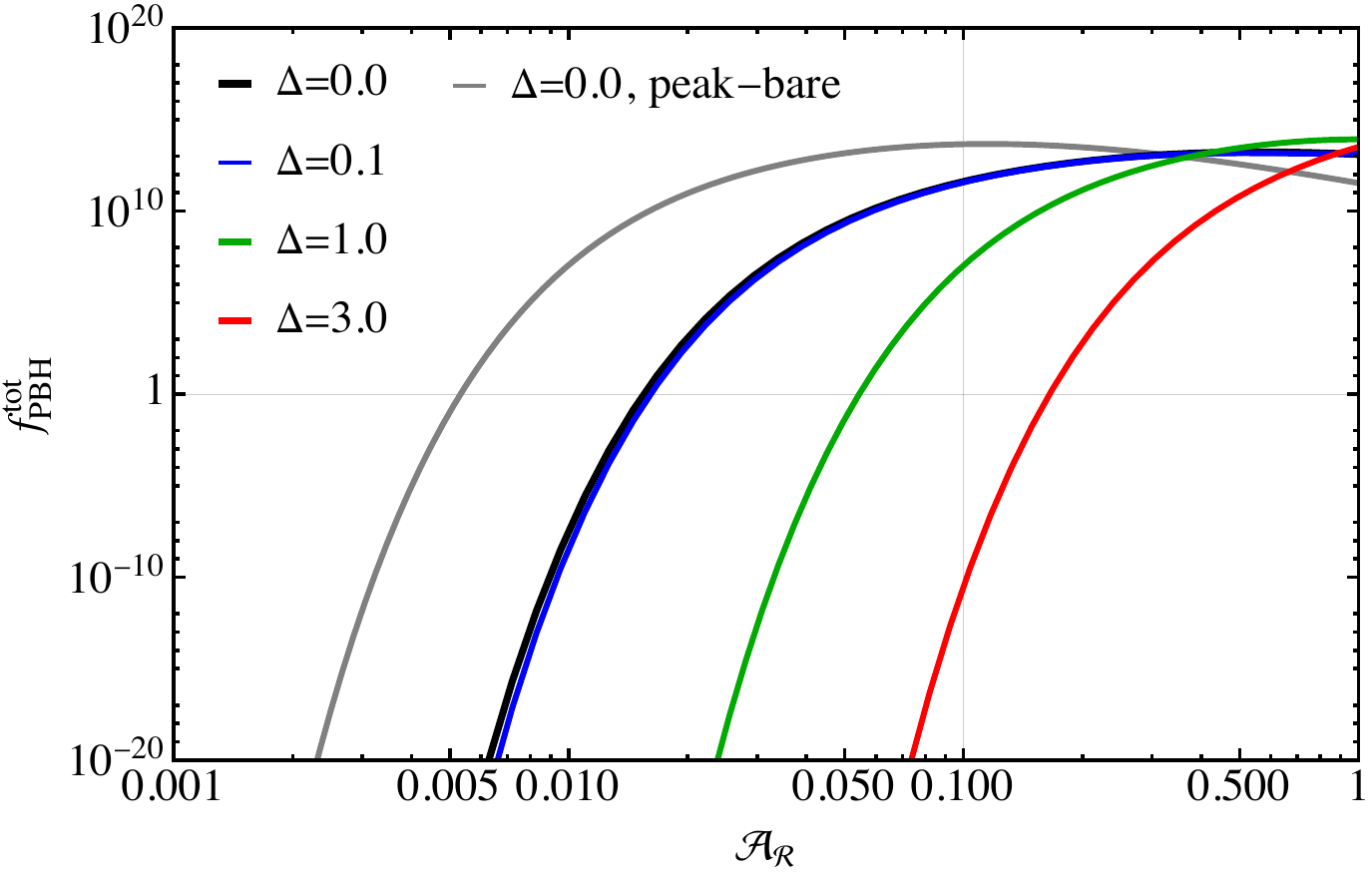} 
\includegraphics[width=0.49\textwidth]{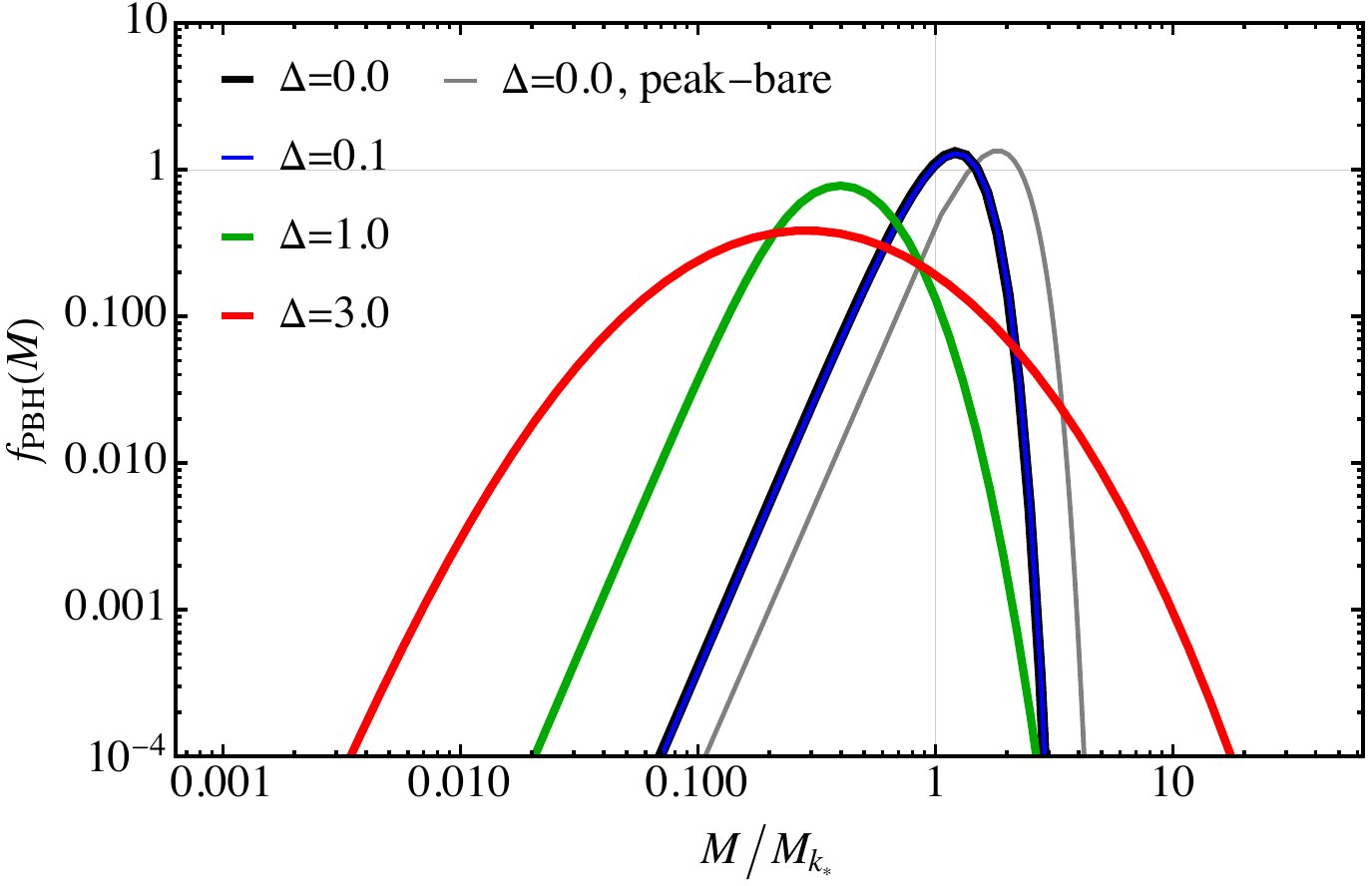}
\caption{[Left] Analogous to Fig.~\ref{fig:ftotallPTFull}, but considering HPL.
[Right] Analogous to Fig.~\ref{fig:fPBHMnew}, but considering HPL.}
\label{fig:ftotallPTHPL}
\end{center}
\end{figure}

In Fig.~\ref{fig:ftotallPTFull}, we present the total PBH abundance calculated using the peaks theory method with Gaussian window function, as developed in this study. For comparison, we also include the results from \cite{Kitajima:2021fpq}, where neither smoothing nor filtering was applied, and a lower mass parameter $\mathcal{K} = 1$ was used, in contrast to $\mathcal{K} = 6$ adopted in our work. The mass functions shown in Fig.\ref{fig:fPBHMnew} exhibit some universal features. All follow the same scaling law, $f_{\mathrm{PBH}}\propto M^{1+1/\gamma}\sim M^{3.8}$, arising from critical collapse in IR, and displaying exponential decay dictated by statistical effects in UV. Additionally, the mass functions become broader for broader power spectra. A mass shift is also observed in both the Press-Schechter formalism, see Fig.\ref{f:fpbhwide}, and in the peaks theory method we develop smoothed with a window function and including a filter function (denoted as ``peak-filtered''), as displayed in Figs.~\ref{fig:fPBHMnew} and \ref{fig:ftotallPTHPL}.

\begin{table}[t]
\begin{center}
\label{tab:PTFullchart}
\begin{equation*}
\begin{array}{lcccccccc}
\hline\hline
\Delta & \text{color} & \mathcal{A}_{\mathcal{R}} & M_{c}/M_{k_*} & \mathcal{A}_{\mathcal{R}} \text{(HPL)} & M_{c}/M_{k_*} \text{(HPL)} & e^{-\Delta/2} & \\
\hline 
0~(\text{bare})& \text{Gray} & 5.33\times 10^{-3}  & 1.80 & 5.33\times 10^{-3} & 1.80 & 1.00 &  & \\
0~(\text{filtered-Gau}) & \text{Black} & 1.55\times 10^{-2}  & 1.24 & 1.55\times 10^{-2} & 1.24 & 1.00 & \\
0~(\text{filtered-rTH}) & \text{Cyan} & 6.98\times 10^{-3}  & 1.24 & 6.98\times 10^{-3}  & 1.24 & 1.00 & \\
0~(\text{filtered-kTH}) & \text{Magenta} & 5.93\times 10^{-3}  & 1.24 & 5.93\times 10^{-3} & 1.24 & 1.00 & \\
0.1~(\text{filtered-Gau}) & \text{Blue} & 1.56\times 10^{-2}  & 1.21  & 1.62\times 10^{-2}  & 1.21 & 0.95 & \\
0.4~(\text{filtered-Gau})& \text{Purple} & 2.24\times 10^{-2}   & 0.98 &  2.46\times 10^{-2}   & 0.91 & 0.82 & \\
1.0~(\text{filtered-Gau})& \text{Green} & 4.35\times 10^{-2}  & 0.76 &  5.44\times 10^{-2}  & 0.40 & 0.61 & \\
2.0~(\text{filtered-Gau})& \text{-} & 8.14\times 10^{-2}  & 0.45 &  1.11\times 10^{-1}  & 0.39 &  0.37 & \\
3.0~(\text{filtered-Gau})& \text{Red} & 1.21\times 10^{-1}  & 0.27 & 1.65\times 10^{-1}  & 0.26 & 0.22 & \\
\hline\hline
\end{array}
\end{equation*}
 \caption{\label{tab:PTFullchart}
The amplitude $\mathcal{A_R}$, central mass $M_c$, and the estimated central mass (with a normalization from monochromatic case) as displayed in Fig.~\ref{fig:ftotallPTFull}, Fig.~\ref{fig:fPBHMnew} and \ref{fig:ftotallPTHPL}, normalized such that PBH abundance is unity and constitutes all DM, with $M_{k_*}= 10^{20}$~g in Eq.~\eqref{def:Mkstar}. 
}
\end{center} 
\end{table}

Broad power spectra result in smaller central masses where the peak of the mass functions is located. See the fourth row of Tab.~\ref{tab:PTFullchart} for details. This observation is consistent with Ref.~\cite{Yoo:2020dkz} and can be explained as follows: A log-normal spectrum with width~$\Delta$, centered at $k_*$, can be viewed as a distribution of perturbations spanning from $k_\mathrm{min}\sim k_*e^{-\Delta}$ to $k_\mathrm{max}\sim k_*e^{\Delta}$, with an approximately uniform amplitude. By dividing this range into $n$ narrow intervals such that each bin is small enough to apply Eq.~\eqref{eq:peakk*} and Eq.~\eqref{eq:npeak-mono} to estimate PBH abundance, we find that the maximum in the $j$-th interval corresponds to $R_s^{-1}\sim k_*\exp((-1+j/n)\Delta)$, where $j=1,~2, \dots ,2n$. While the amplitude $\mathcal{A}_\mathcal{R}/(\sqrt{2\pi}\Delta)$ remains approximately constant across intervals, the extra factor $k_*^3\exp(3(-1+j/n)\Delta)$ factor in $\mathscr{N}_\mathrm{pk}$ of Eq.~\eqref{eq:npeak-mono} ensures that the global maximum PBH abundance always occurs at the largest wave number $ke^{\Delta}$, corresponding to the smallest mass $M_*e^{-\Delta/2}$. This reasoning is appropriate for a top-hat power spectrum \cite{Saito:2009jt,DeLuca:2020ioi,DeLuca:2020agl,Sugiyama:2020roc,Fumagalli:2024kxe}, and works well for a log-normal power spectrum with width $\Delta$, as demonstrated in Tab.~\ref{tab:PTFullchart}.

\subsection{PBH Abundance in High Peak Limit}

In Fig.~\ref{fig:ftotallPTHPL}, we present the total PBH abundance and mass function calculated using HPL, both with a Gaussian window function. For reference, the results from Ref.~\cite{Kitajima:2021fpq}, which do not include smoothing or filtering, are also shown. The small differences between the general full method calculation and HPL approximation are summarized in the third and fifth rows of Tab.~\ref{tab:PTFullchart}. As discussed in Sec.~\ref{ss:HPL}, HPL is useful for simplifying calculations, particularly for narrow power spectra, but may not provide reliable results for broad power spectra.

\begin{figure}[t]
    \centering
    \includegraphics[width=0.65\linewidth]{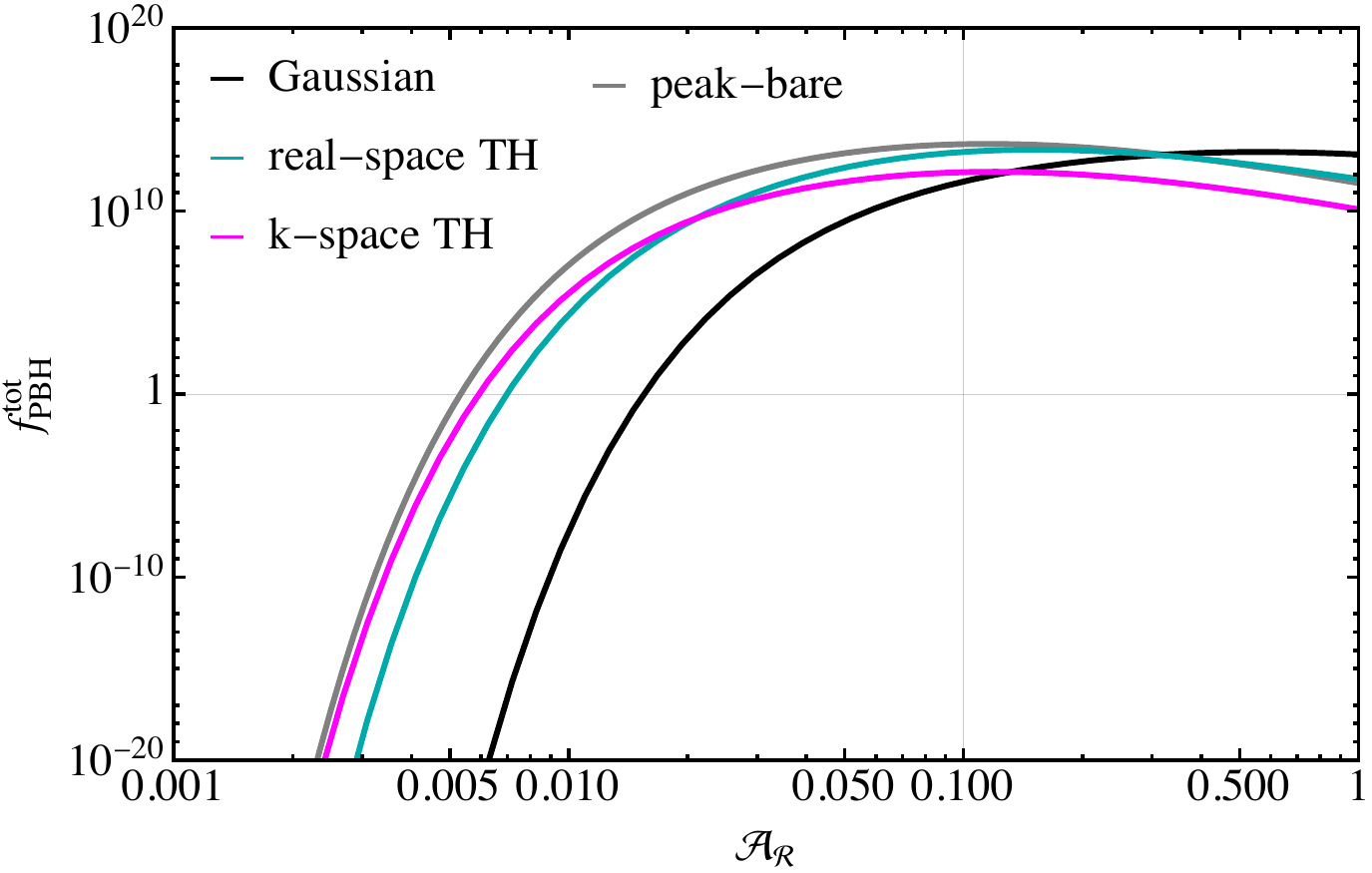}
    \caption{Total PBH abundance calculated by the peaks theory smoothed by a window function and filtered by a filter function (solid lines, ``peak-filtered'') method using different types of window function, all the lines are for the monochromatic power spectrum. Result following method of Ref.~\cite{Kitajima:2021fpq} without smoothing or filtering is overlaid for reference (gray line, ``peak-bare'').}
    \label{fig:diffWinftot}
\end{figure}

\subsection{Window Function Dependence}

\begin{figure}[t]
\begin{center}
\includegraphics[width=0.49\textwidth]{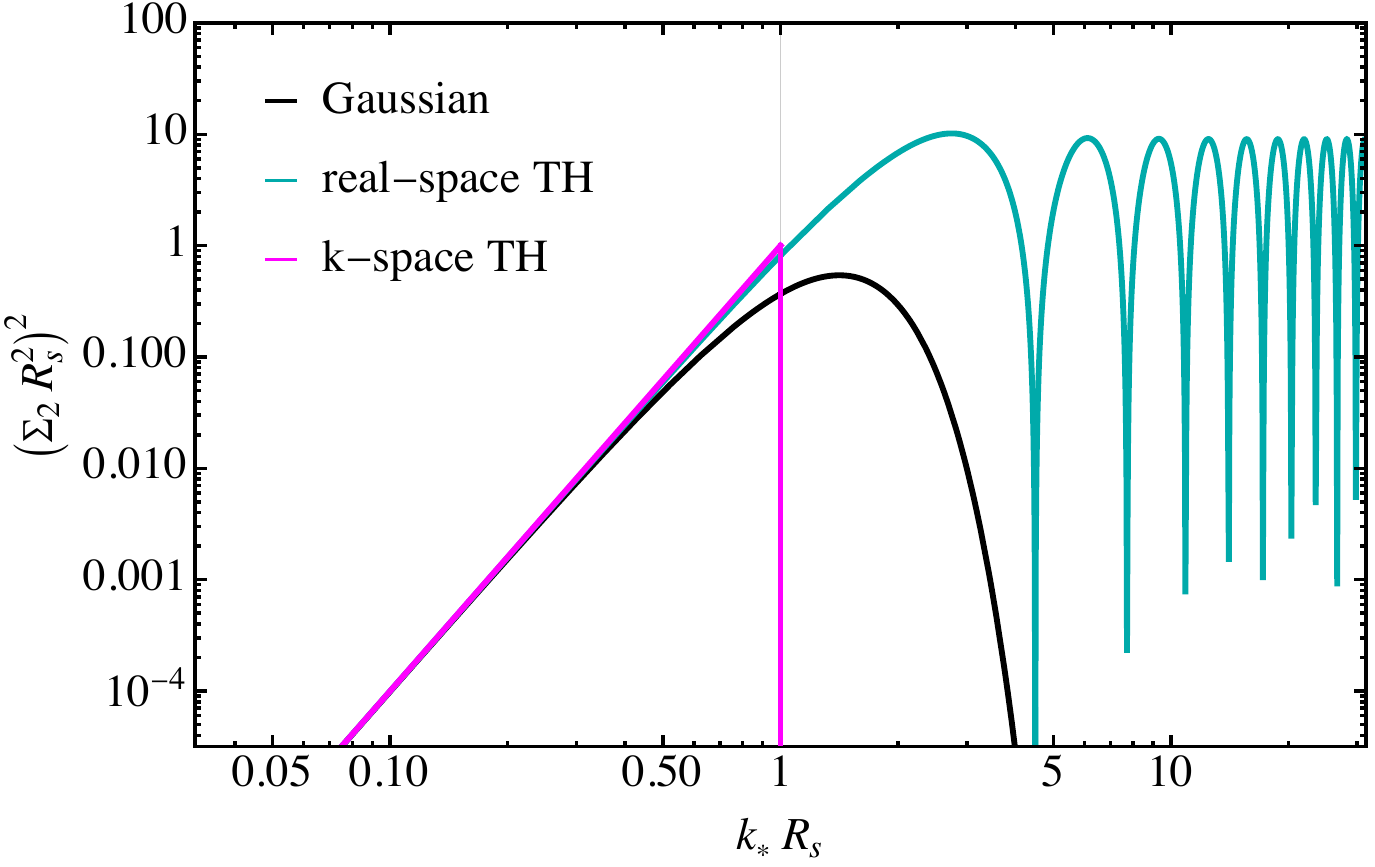}
\includegraphics[width=0.49\textwidth]{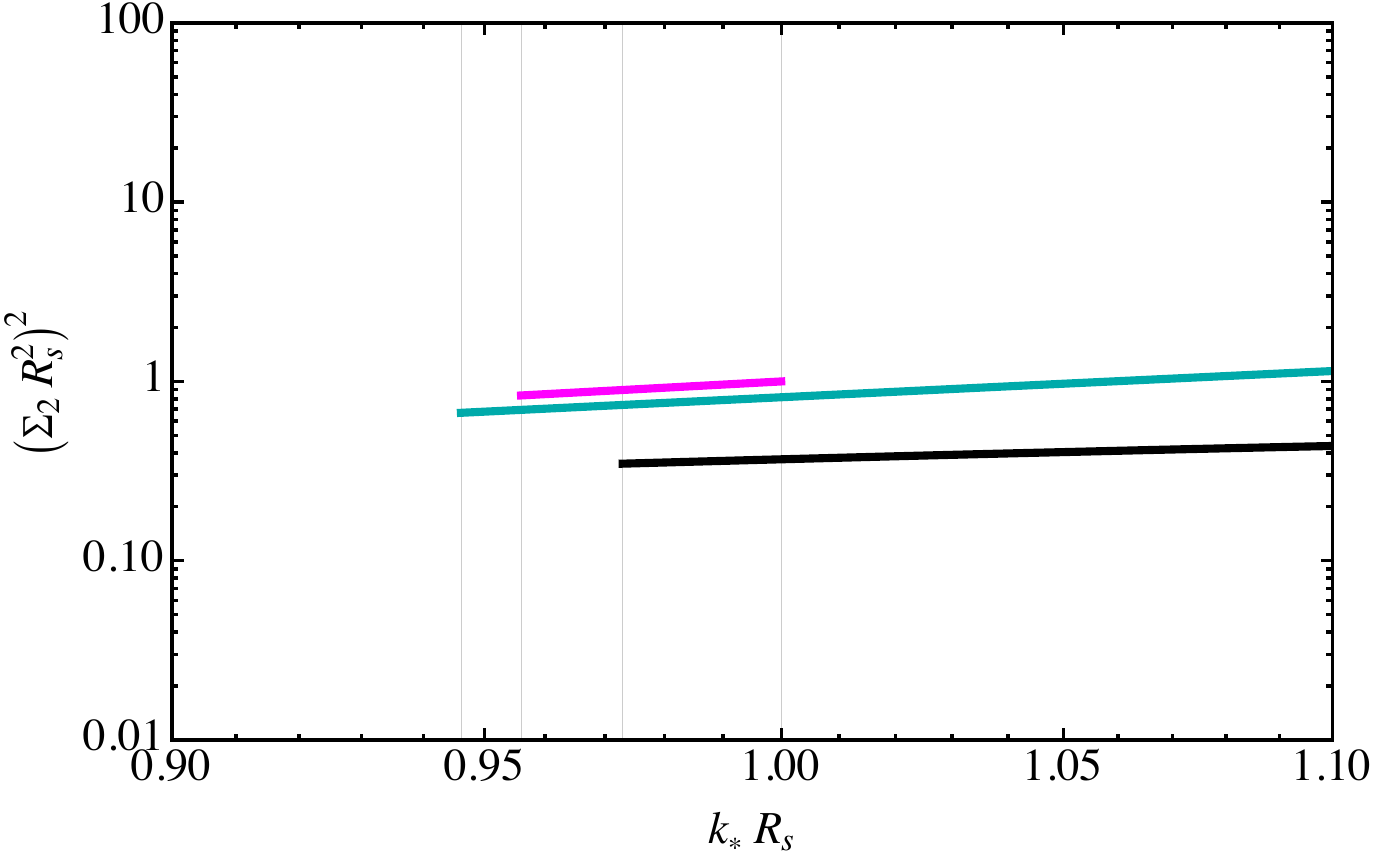}
\caption{[Left] Illustration of $(\Sigma_2 R_s^2)^2$ for monochromatic power spectrum considering different types of window functions, see Eq.~\eqref{eq:sigmasqu-logN}. [Right] Zoom in illustration of $(\Sigma_2 R_s^2)^2$ for monochromatic power spectrum considering different types of window functions around $k_* R_s=1$. }
\label{fig:diffWin-Sigma2}
\end{center}
\end{figure}

In Fig.~\ref{fig:diffWinftot}, we compare the total PBH abundance estimated for a monochromatic power spectrum using our method with different window functions: real-space top-hat, $k$-space top-hat, and Gaussian. The total abundance produced by the $k$-space top-hat window function is larger than that obtained using the real-space top-hat window function, while the Gaussian window function yields the most suppressed result. The {variance} required to achieve $f_{\mathrm{PBH}}^{\mathrm{tot}}=1$ is $\mathcal{A}_\mathcal{R}=6.98\times 10^{-3}$, for the real-space top-hat window function and $\mathcal{A}_\mathcal{R}=5.93\times 10^{-3}$ for the $k$-space top-hat window function. These findings align with the trends reported in Ref.~\cite{Yoo:2020dkz}, confirming that $k$-space top-hat window function allows for more efficient PBH formation than the Gaussian window function. In Fig.~\ref{fig:diffWin-Sigma2} we display results for a monochromatic power spectrum. Here, the differences in PBH production caused by the choice of window function can be explained by the differences in the variance $(\Sigma_2 R_s^2)^2$ around $R_s \sim 1/k_*$, which approximates the variance of the density contrast $\sigma_\delta^2$. As shown in Fig.~\ref{fig:diffWin-Sigma2}, a larger variance generally leads to a larger PBH abundance.

\begin{figure}[t]
    \centering
    \includegraphics[width=0.65\linewidth]{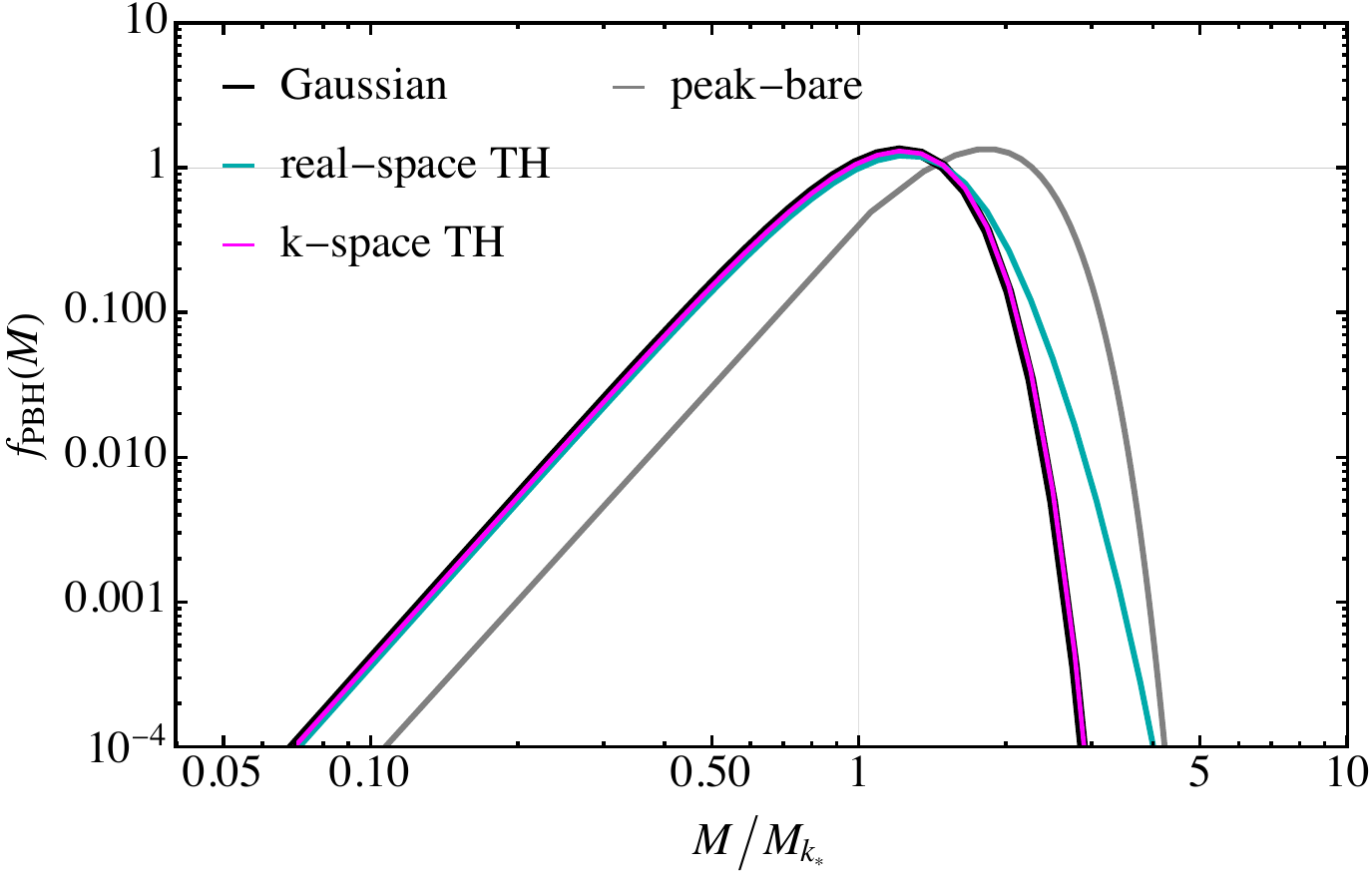}
    \caption{PBH mass function calculated by the peaks theory smoothed by a window function and filtered by a filter function (solid lines, ``peak-filtered'') method using different types of window function, all considering the monochromatic power spectrum, normalized such that PBHs are all DM with $M_{k_*}= 10^{20}$~g of Eq.~\eqref{def:Mkstar}. Result following method of Ref.~\cite{Kitajima:2021fpq} without smoothing or filtering is overlaid for reference (gray line, ``peak-bare''). }
    \label{fig:diffWinfM}
\end{figure}

In Fig.~\ref{fig:diffWinfM}, we show the PBH mass function calculated using the peaks theory smoothed by a window function and filtered by a filter function method with different types of window functions for the monochromatic power spectrum. All choices produce the same central mass $M_c / M_{k_*} = 1.24$, which follows from the scaling $R_s \sim 1/k_*$, as shown in Eq.~\eqref{eq:fixRs-mono}. For calculations without smoothing, the central mass shifts to the higher mass side. The mass function obtained with a $k$-space top-hat window function is very similar to that with a Gaussian window function, while the real-space top-hat window function allows for the production of more massive PBHs, as seen in Fig.~\ref{fig:diffWinfM}.

\subsection{Comparison with Earlier Methods}

We now compare the results of PBH mass function and abundance in different methods: the Press-Schechter formalism based on the density contrast (PS$\delta$), the Press-Schechter-type formalism using the compaction function (PSC), the peaks theory without a window function or filtering function (peak-bare) following Ref.~\cite{Kitajima:2021fpq}, as well as the peaks theory smoothed by a window function and filtered by a filter function (peak-filtered) studied in this paper, and its high peak limit (peak-HPL).

\begin{figure}[t]
    \centering
    \includegraphics[width=0.65\linewidth]{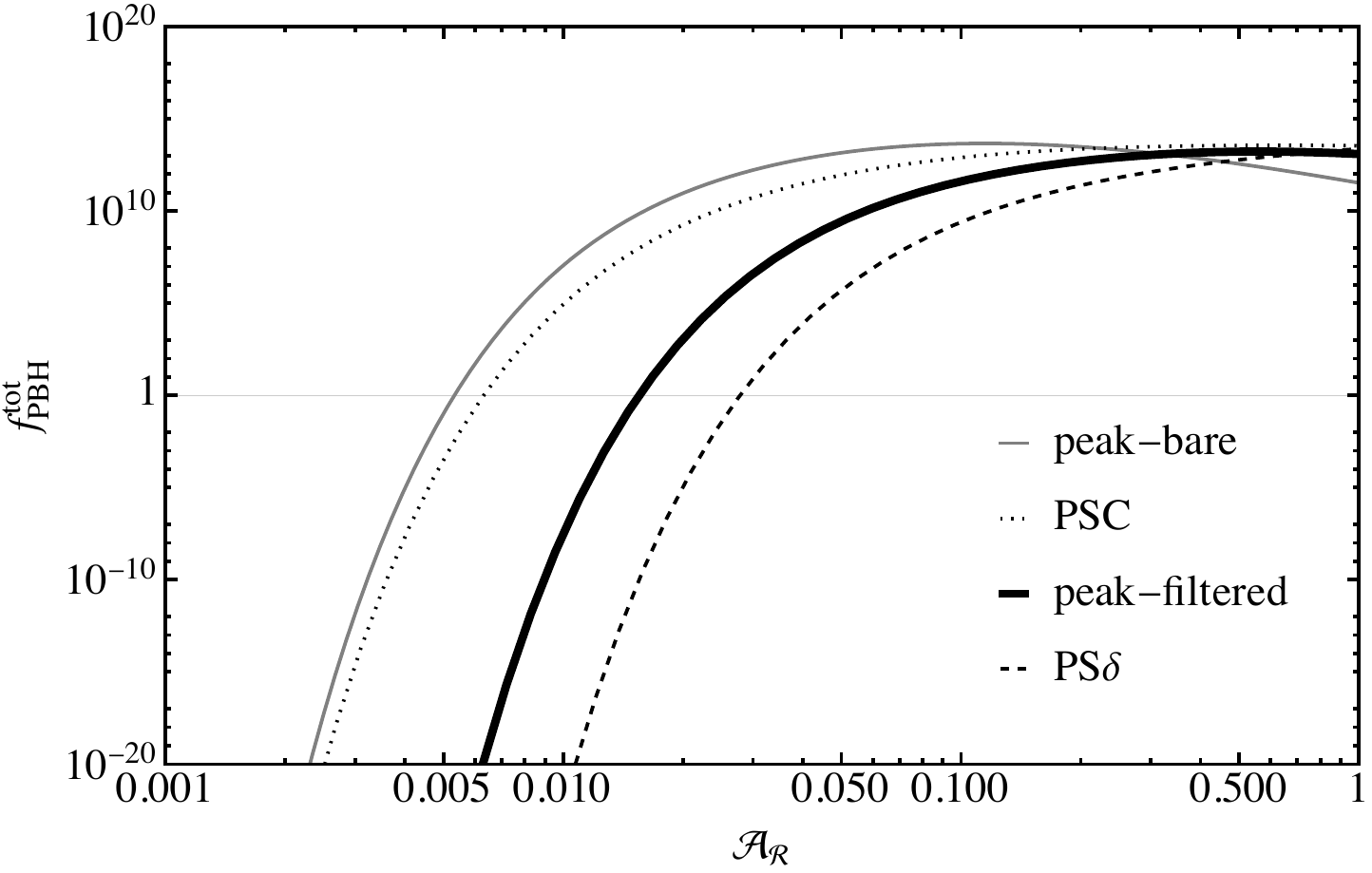}
    \caption{Comparison of total PBH abundance calculated by different methods for a monochromatic power spectrum with $\Delta=0$. Result following method of Ref.~\cite{Kitajima:2021fpq} without smoothing or filtering is overlaid for reference (gray line, ``peak-bare''). For ``peak-filter" and ``PS$\delta$", we are using a Gaussian window function. For ``peak-bare" and ``PSC", there is no window function.}
    \label{fig:diffmethods-D00}
\end{figure}

\begin{figure}[t]
\begin{center}
\includegraphics[width=0.5\textwidth]{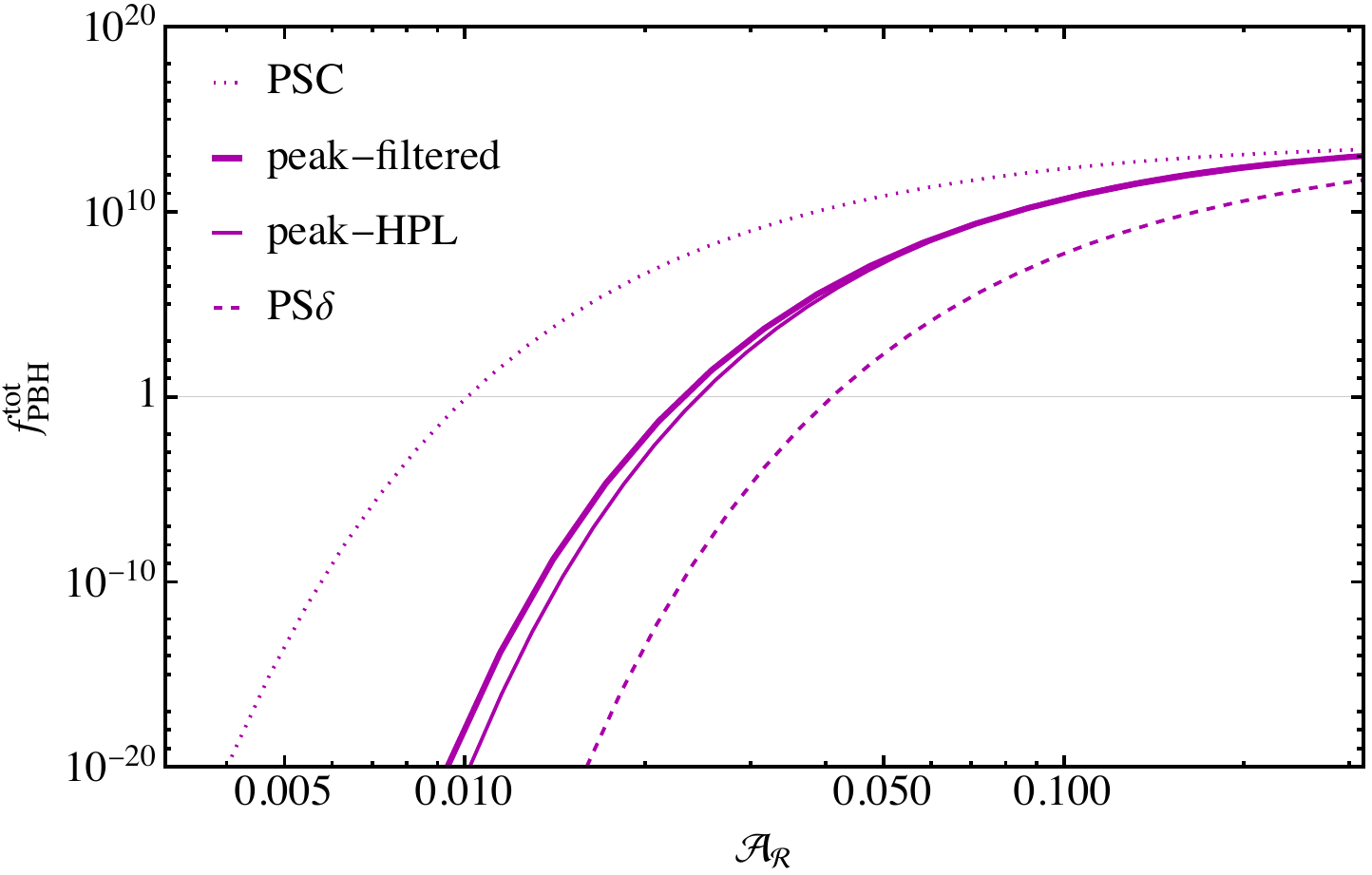}\includegraphics[width=0.5\textwidth]{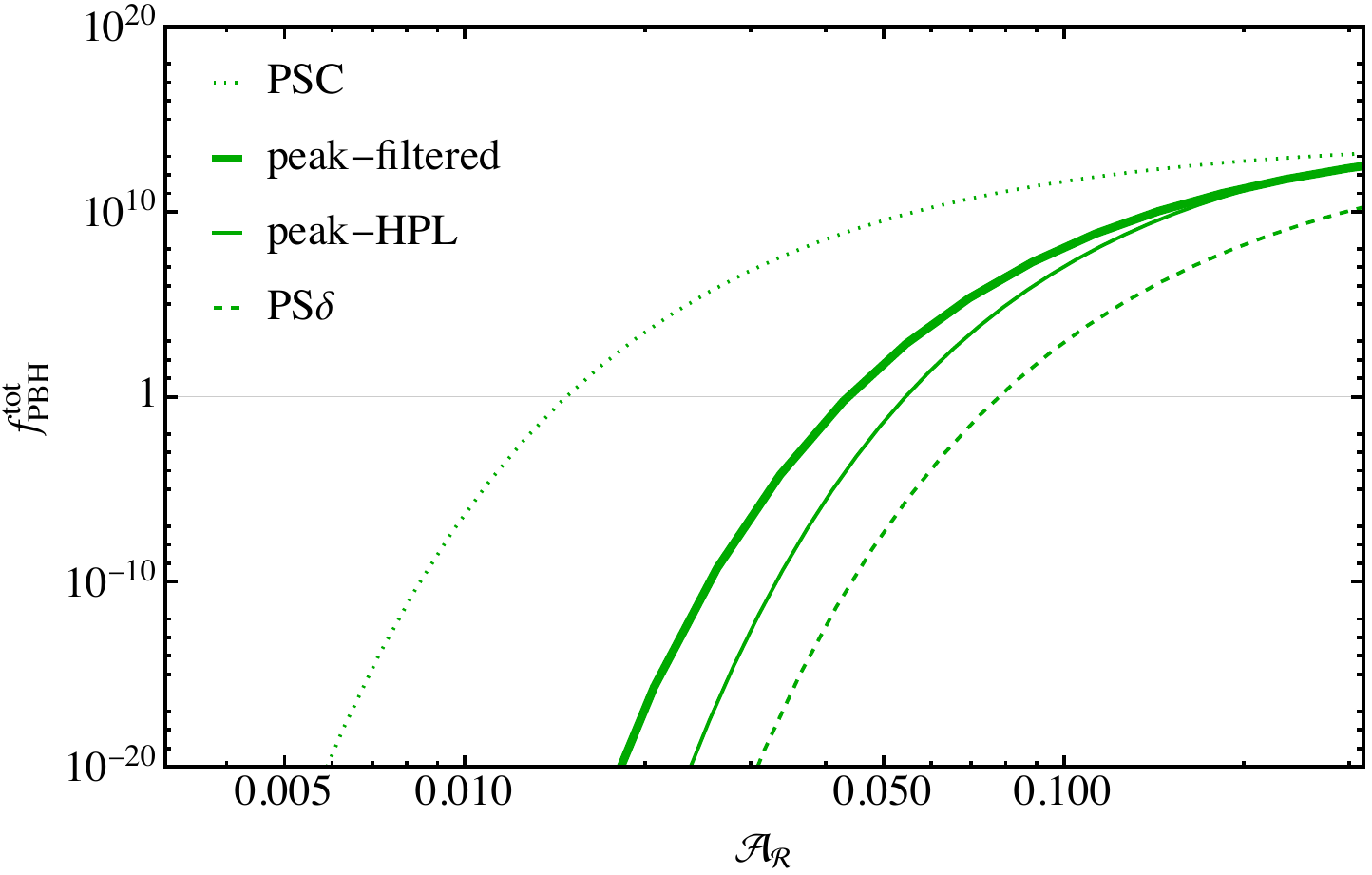}
\caption{Comparison of total PBH abundance calculated by different methods, for power spectrum widths $\Delta=0.4$ [Left] and $\Delta=1$ [Right]. Full calculation of the peaks theory smoothed by
a Gaussian window function and filtered by a filter function (solid lines, ``peak-filtered'') is always more productive than the high peak limit, and the difference is more pronounced in the case of a broad spectrum. }
\label{fig:diffmethods-D041}
\end{center}
\end{figure}

In Fig.~\ref{fig:diffmethods-D00}, we focus exclusively on the monochromatic case using a Gaussian window function. It is evident that the Press-Schechter-type formalism based on the compaction function (PSC) predicts a similar PBH abundance to that of the peaks theory without a window function (peak-bare). This is not surprising, as the compaction function threshold is the dominant factor in such calculations, and the PSC method adopts this directly from the peaks theory. However, there is a significant discrepancy between the peaks theory without a window function (peak-bare) and the peaks theory smoothed by a window function and filtered by a filter function (peak-filtered). In the latter, all unwanted small PBHs with $R_s \ll r_m$ are excluded. %\color{blue}{}Actually, compared with the window function, the filter function plays a more important role in the final result.\color{black}{}
Thus, we conclude that the peaks theory without a window function (peak-bare) substantially overestimates the PBH abundance. 
The peaks theory smoothed by a window function and filtered by a filter function (peak-filtered) predicts moderate PBH abundance lying between the predictions of the peaks theory without a window function (peak-bare) and the Press-Schechter formalism based on density contrast (PS$\delta$).

In Fig.~\ref{fig:diffmethods-D00}, we see that $f_\mathrm{PBH}^\mathrm{tot}(\mathcal{A_R})$ from peak theory decreases slightly when $\mathcal{A_R}$ is large ($\sim\mathcal{O}(0.1)$), while the Press-Schechter results (both PSC and PS$\delta$) approaches a constant but never drops. (See also Fig.~\ref{fig:diffWinftot} for $f_\mathrm{PBH}^\mathrm{tot}(\mathcal{A_R})$ from peaks theory with different window functions.) The main reason is that in peaks theory, besides the Gaussian function, there is a $\mu^3$ prefactor in the formula of number density. (See \eqref{eq:npeak-mono} and \eqref{eq:f(x)->x^3}.) Because of this prefactor, the number density (thus the abundance) has a peak, which maximizes the PBH abundance when it locates in the strip of PBH formation, \textit{i.e.} $\mu_\mathrm{th}<\mu_\mathrm{max}<\mu_\mathrm{II}$. When $\mu_\mathrm{max}$ lies beyond $\mu_\mathrm{II}$, it mainly generate Type B PBHs which are not counted in our formula, while the type A PBHs become less. On the other hand, in the Press-Schechter formalism, there is only the Gaussian function in the number density, so PBH abundance always increases when $\mu$ gets larger and larger, but saturates when $\mu\gg\mu_\mathrm{II}$. 

\begin{figure}[t]
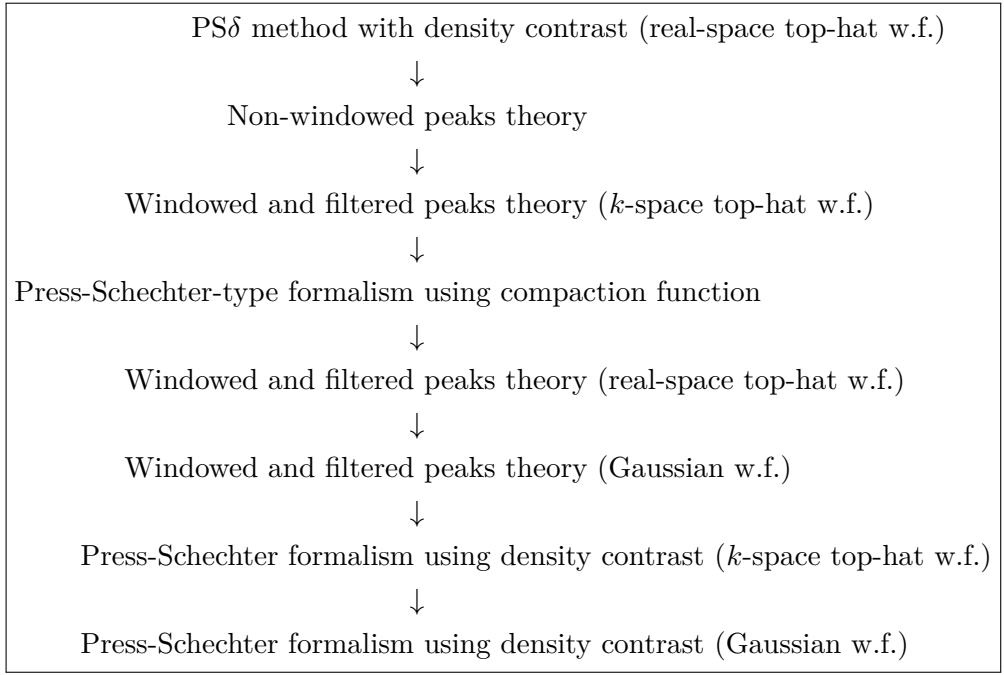

\begin{equation*}
\boxed{
\begin{aligned}
 \text{PS$\delta$ method with}~& \text{density contrast (real-space top-hat w.f.)} \\
 \downarrow&\\
 \text{Non-windowed}~& \text{peaks theory} \\
  \downarrow&\\
 \text{Windowed and filtered}~& \text{peaks theory ($k$-space top-hat w.f.)} \\
  \downarrow&\\
 \text{Press-Schechter-type formalism}~& \text{using compaction function}\\
  \downarrow&\\
 \text{Windowed and filtered}~& \text{peaks theory (real-space top-hat w.f.)} \\
  \downarrow&\\
 \text{Windowed and filtered}~& \text{peaks theory (Gaussian w.f.)} \\
  \downarrow&\\
 \text{Press-Schechter formalism}~& \text{using density contrast ($k$-space top-hat w.f.)} \\
  \downarrow&\\
 \text{Press-Schechter formalism}~& \text{using density contrast (Gaussian w.f.)} 
\end{aligned}
}
\end{equation*} 
\caption{\label{fig:summarize} Summary of PBH abundance calculated by different methods, considering monochromatic power spectrum normalized at $f_{\rm PHB}^{\rm tot} = 1$. Arrows indicate decreased PBH formation efficiency and abundance.} 
\end{figure}

In Fig.~\ref{fig:diffmethods-D041}, we made the same comparison by considering log-normal power spectra with $\Delta=0.4$ and $\Delta=1$. Press-Schechter-type formalism using the compaction function always predicts a much higher abundance, mainly because no window function is considered. The results from the high peak limit is slightly lower but basically the same with the full calculation, of which the difference becomes larger for a broader power spectrum. As in Fig.~\ref{fig:diffmethods-D00}, all the ``peak-filtered" and ``PS$\delta$" results shown in Fig.~\ref{fig:diffmethods-D041} use a Gaussian window function.

In Fig.~\ref{fig:summarize} we summarize comparison of PBH abundance calculations from different methods, showcasing decreased PBH formation efficiency and abundance. Here, we rank the PBH production efficiencies, with lower {variance} $\mathcal{A}_{\mathcal{R}}$ indicating higher efficiency, across all methods discussed in this work, for a monochromatic power spectrum normalized at $f_{\mathrm{PBH}}^{\mathrm{tot}} = 1$. Notably, when a window function is applied, the PBH abundances estimated using the peaks theory are consistently lower than those obtained without a window function (peak-bare).

\section{Conclusions}\label{s:conclusion}

We presented a novel refined method based on peaks theory to calculate the abundance of PBHs that can form from an enhanced power spectrum of the curvature perturbation in the early Universe. Our approach emphasizes the impact of power spectrum dispersion and incorporates a proper treatment of window and filter functions that can significantly improve accuracy.

We demonstrated that the PBH mass function should be determined by the smoothing scale introduced through the window function, rather than by the size of the over-dense region. The latter approach inadvertently includes contributions from extremely small PBHs that are unphysical and should not be considered. We demonstrated how this issue can be addressed by appropriately introducing a filter function that ensures the central mass in the PBH mass function aligns with the central frequency of the power spectrum. This correction avoids over-counting PBHs associated with extremely small smoothing scales, leading to a suppressed PBH abundance compared to previous methods that lacked a window function. While the exact results depend on the choice of the window function, our method provides a robust framework for more reliable estimates. The uncertainty associated with the choice of window function reflects our limited understanding of the nonlinear transfer function.  
Aside from this aspect, the other components of our framework are more precisely formulated than in previous analytical approaches. Our analysis indicates that previous studies based on other approaches might have significantly overestimated PBH abundances, potentially by as much as  $\mathcal{O}(10)$ orders of magnitude. The tension between peaks theory and the Press–Schechter formalism is alleviated by incorporating a smoothing procedure into the peak theory framework. This is analogous to what is done in the Press–Schechter approach, but was previously omitted in conventional peak theory treatments.  
In addition, our use of a filter function removes nonphysical contributions from super-horizon-scale overdensities, further reconciling the two methods.

Our method opens promising avenues for applications and motivates a re-examination of observables related to PBH abundance, such as their connection to induced gravitational waves (GWs) that can arise from second-order. Notably, the recently reported stochastic GW background signals, aside from possible astrophysical or alternative origins, have also been proposed to be associated with induced GWs generated at second order from cosmological perturbations. However, the relatively large amplitude required for these waves raises concerns about PBH overproduction. A broader power spectrum could mitigate this issue, as PBH abundance decreases with larger $\Delta$ for a given $\mathcal{A_R}$. By applying the advanced peaks theory with window and filter functions developed in this work, PBH production efficiency can be further reduced, offering a potential resolution to the PBH overproduction problem for Gaussian curvature perturbations in a radiation-dominated universe. This highlights the interplay between our method and PBH observables.

Another potential direction involves further extending the analysis, such as to non-Gaussian scenarios. However, applying the peaks theory method to non-Gaussian cases is not straightforward. While some attempts have been made in this direction\footnote{See Ref.~\cite{Kitajima:2021fpq,Escriva:2022pnz,Yoo:2019pma}.}, previous works primarily focused on a monochromatic spectrum of curvature perturbations, making those methods unsuitable for spectra with finite widths. 
The main challenge lies in the complicated correspondence between Fourier space and real space in the presence of non-Gaussianities, \textit{e.g.} how to reformulate a ``peaks theory of non-Gaussian random fields''. Simply replacing the Gaussian profile of $\mathcal{R}(r)$ by the non-Gaussian local relation seems not appropriate when the power spectrum is broad. A potentially fruitful starting point could be to address non-Gaussianities perturbatively. This remains an open avenue for future investigation.

%%%%%%%%%%%%%%%%%%%%%%%%%%%%%%%%%%%%%%%%%%%%%%
\section*{Acknowledgment}
\addcontentsline{toc}{section}{Acknowledgments}

We would like to thank J. Richard Bond, Chul-Moon Yoo, Albert Escrivà, Masahide Yamaguchi, Cristiano Germani, Tomohiro Harada, Naoya Kitajima, Kazunori Kohri, Takahiko Matsubara, Teruaki Suyama, Takahiro Tanaka, Sebastian Bahamonde, Cristian Joana and Sam Young for valuable discussions. 
JN.W. thanks the hospitality of Masahide Yamaguchi during her visit to IBS.
S.P. is supported in part by the National Key Research and Development Program of China Grant No. 2020YFC2201502, and by National Natural Science Foundation of China No. 12475066 and No. 12047503. 
This work was supported in part by JSPS KAKENHI grant Nos. 20H05853 (MS), 23K13109 (VT), and 24K00624 (MS).
This works was also supported by the World Premier International Research Center Initiative (WPI), MEXT, Japan.

%%%%%%%%%%%%%%%%%%%%%%%%%%%%%%%%%%%%%%%%%%%%%%

\appendix

\begin{figure}[t]
\begin{center}
Table of Additional Notation
\begin{equation*}
\setlength{\arraycolsep}{0pt}
\begin{array}{cccccccc}
\hline\hline
\text{Notation} & \text{Meaning} & \text{Definition} & \\
\hline 
\rho & \text{Energy density} & {\rm ~~Eq.~}\eqref{def:densitycontrast} & \\
\delta & \text{Density contrast} & {\rm ~~Eq.~}\eqref{def:densitycontrast}   & \\
\beta_{\mathrm{PBH}} & \text{PBH abundance at formation moment} & {\rm ~~Eq.~}\eqref{def:beta1}  & \\
\delta_{\mathrm{cr}} & \text{Critical value of $\delta$} & {\rm ~~Eq.~}\eqref{def:beta1}  & \\
\mathbb{P}_G (\delta) & \text{Probability distribution function of $\delta$} & {\rm ~~Eq.~}\eqref{def:beta1}  & \\
\mathcal{P}_\delta(k) & \text{Power spectrum of density contrast} & {\rm ~~Eq.~}\eqref{def:PowSdelta}  & \\
\sigma_\delta^2 & \text{Variance of $\mathbb{P}_G (\delta)$} & {\rm ~~Eq.~}\eqref{def:sigmadelta0}  & \\
\alpha & \text{Suppression factor in PBH mass} & {\rm ~~Eq.~}\eqref{M-R}  & \\
\mathcal{C}_\ell& \text{Linear part of compaction function} & {\rm ~~Eq.~}\eqref{def:Cell}  & \\
\mathcal{C}_{\ell,\mathrm{II}}& \text{Type II fluctuation boundary for $\mathcal{C}_{\ell}$} & {\rm ~~Eq.~}\eqref{eq:ClII}  & \\
\mathcal{C}_{\ell,\mathrm{th}} & \text{Threshold value for the linear part of compaction function} & {\rm ~~Eq.~}\eqref{eq:massEXPS} & \\
\mathbb{P}_G(\mathcal{C}_{\ell})& \text{PDF of the linear part of compaction function} & {\rm ~~Eq.~}\eqref{eq:PDFofCell}  & \\
\sigma_{\ell}^2  & \text{Variance of }\mathbb{P}_G(\mathcal{C}_{\ell}) & {\rm ~~Eq.~}\eqref{eq:PDFofCell}  & \\
\boldsymbol{r}_p & \text{Position of the maxima point} & {\rm ~~Eq.~}\eqref{eq:pointprocess}  & \\
n_{\mathrm{pk}}(\boldsymbol{r})& \text{Density field of maxima points} & {\rm ~~Eq.~}\eqref{eq:pointprocess}  & \\
\boldsymbol{\eta} & \text{First derivative of }F(\boldsymbol{r}), \quad \eta_i, i=1,2,3 \text{ are the components} & {\rm ~~Eq.~}\eqref{eq:expandofFgrad} & \\
\boldsymbol{\zeta} & \text{Second derivative of }F(\boldsymbol{r}),\quad \zeta_{ij}, i,j=1,2,3 \text{ are the components} & {\rm ~~Eq.~}\eqref{eq:expandofFgrad} & \\
\xi\left(\boldsymbol{r}_1, \boldsymbol{r}_2\right) & \text{Two-point correlation function of }F(\boldsymbol{r}) & {\rm ~~Eq.~}\eqref{def:xi}  & \\
\nu & \text{Variable defined with }F & {\rm ~~Eq.~}\eqref{eq:lambda123toxyz}  & \\
x,y,z & \text{Variables defined with }\zeta_{11},\zeta_{22},\zeta_{33} & {\rm ~~Eq.~}\eqref{eq:lambda123toxyz}  & \\
\gamma_1 & \left\langle \nu x \right\rangle & {\rm ~~Eq.~}\eqref{def:gamma} & \\
\boldsymbol{M} & \text{Covariance matrix of Gaussian variables in }F,\boldsymbol{\eta},\boldsymbol{\zeta} & {\rm ~~Eq.~}\eqref{def:M1M2} & \\
\boldsymbol{V}_1 & \left(\eta_1, \eta_2, \eta_3, \zeta_{23}, \zeta_{13}, \zeta_{12} \right) & {\rm ~~Eq.~}\eqref{eq:V2}  & \\
\boldsymbol{V}_2 & \left(\nu, x,y,z \right) & {\rm ~~Eq.~}\eqref{eq:V2}  & \\
\boldsymbol{V}_0 & \left( \boldsymbol{V}_1,\boldsymbol{V}_2 \right) & {\rm ~~Eq.~}\eqref{eq:V2}  & \\
\boldsymbol{M}_i, ~i=1,2 & \text{Covariance matrix of }\boldsymbol{V}_i, \quad i=1,2 & {\rm ~~Eq.~}\eqref{eq:M1M2}  & \\
Q & \frac{1}{2}\boldsymbol{V}_0^{\mathrm{T}} \boldsymbol{M}^{-1} \boldsymbol{V}_0 & {\rm ~~Eq.~}\eqref{eq:Q} & \\
x_\star & \gamma_1   \nu & {\rm ~~Eq.~}\eqref{eq:Q} & \\
\zeta_A, ~A=1,2...6 & \text{Redefinition of }\zeta_{ij}  & {\rm ~~Eq.~}\eqref{def:zetaA}  & \\
\boldsymbol{\lambda}& \text{Diagonalized }-\boldsymbol{\zeta} & {\rm ~~Eq.~}\eqref{def:hatS}  & \\
\lambda_A, ~A=1,2,3 & \text{Eigenvalues of }-\boldsymbol{\zeta}  & {\rm ~~Eq.~}\eqref{def:lambdaA}  & \\
\mathrm{d}\Omega_{S^3} & \text{Volume element of three-sphere} & {\rm ~~Eq.~}\eqref{def:dOmega3}  & \\
\boldsymbol{\chi}&\text{Constrain conditions for peaks} & {\rm ~~Eq.~}\eqref{eq:Npkvxyz} & \\
R_* & \sqrt{3} \sigma_1 / \sigma_2 & {\rm ~~Eq.~}\eqref{def:Rstar}  & \\
\hline\hline
\end{array}
\end{equation*}
\end{center}
\end{figure}

\section{PBH Formation in Earlier Methods}\label{s:PS}

\subsection{PS$\delta$ Method Using Density Contrast}\label{s:sPS}

PBHs can form from gravitational collapse of large density perturbations in the early Universe.
Since the PDF of density perturbations typically has a tail extending to very large values albeit whose probability is highly suppressed, PBHs can form when the density perturbation exceeds a critical density contrast, $\delta_{\mathrm{cr}}$. In the PS$\delta$ formalism, this threshold is typically analyzed in the comoving slicing, where the density perturbation, as defined in Eq.~\eqref{eq:contrast}, is approximated by its linear component 
\be
\delta \simeq \delta_\ell =-\frac{2}{3} \frac{3(1+w)}{5+3 w}\frac{1}{H^2 a^2}\nabla^2\calR
\label{def:densitycontrast}
\ee
in a Universe dominated by a single-component fluid, with $w$ being the equation of state\footnote{$w = 1/3$ for radiation-dominated era.}.

The PBH abundance at formation is calculated then using the Press-Schechter formalism, which integrates the PDF of the density contrast in the comoving gauge starting from a critical density $\delta_\text{cr}$. Several analytical thresholds have been proposed. For example, $\delta_\text{cr} \simeq w$ was initially proposed in Ref.~\cite{Carr:1975qj}, subsequently refined to $\delta_\text{cr} = \sin^2(\pi\sqrt{w}/(1+3w))$ in Ref.~\cite{Harada:2013epa}. The critical density contrast, considering formalism of Ref.~\cite{Harada:2013epa}, is
\be
\delta_\text{cr}=\frac{3(1+w)}{5+3w}\sin^2\left(\frac{\pi\sqrt w}{1+3w}\right)
\xRightarrow{w = 1/3}0.41.
\label{eq:HYKlimit}
\ee
However, this approximation is valid only when the pressure around the density peak's compaction function is negligible. 

As a simplified estimate, we will focus on a radiation-dominated Universe and adopt the threshold value put forth in Ref.~\cite{Harada:2013epa}. The PBH abundance is then given by 
\be\label{def:beta1}
\beta_{\mathrm{PBH}}=\int^\infty_{\delta_\text{cr}}\mathbb{P}_G (\delta)\dif\delta.
\ee
Here, the Gaussian PDF of density contrast is given by
\be\label{def:P-delta}
\mathbb{P}_G (\delta)=\frac1{\sqrt{2\pi\sigma_\delta^2}}\exp\left(-\frac{\delta^2}{2\sigma_\delta^2}\right),
\ee
where $\sigma_\delta$ is the smoothed variance defined considering a window function $\widetilde{W}(k,R_s)$ and smoothing scale $R_s$
\begin{align}\nn
\sigma^2_\delta(R_s)\equiv\langle\delta^2(\boldsymbol{r},R_s)\rangle=\int\frac{\dif k}{k}\mathcal{P}_\delta(k)\widetilde{W}^2(k,R_s)~.
\label{def:sigmadelta0}
\end{align}
Note that the smoothed variance is always smaller than the ``bare'' variance $\sigma_{\delta, {\rm bare}}$, which is defined as the integral of the power spectrum
\be
\sigma^2_\delta(R_s) < \sigma^2_{\delta, {\rm bare}} = \int\frac{\dif k}{k}\mathcal{P}_\delta(k)=\langle\delta^2(\boldsymbol{r})\rangle.
\ee

The power spectrum of $\delta$ is connected to the power spectrum of the curvature perturbation $\calR$ by Eq.~\eqref{def:densitycontrast}, which gives
\be
\calP_\delta(k)=\frac{16}{81}\left(\frac{k}{Ha}\right)^4\calP_\calR(k).
\label{def:PowSdelta}
\ee
Therefore, the variance of density contrast is
\begin{align}\nn
\sigma_\delta^2(R_s)&=\frac{16}{81}\int\frac{\dif k}{k}\left(\frac k{Ha}\right)^4\calP_\calR(k)\widetilde{W}^2(k,R_s)\notag\\
&=\frac{16}{81}\int\frac{\dif k}{k}\left(kR_s\right)^4\calP_\calR(k)\widetilde{W}^2(k,R_s),
\label{def:sigmadelta}
\end{align}
where the second step, we set the smoothing scale of the window function to correspond to the comoving horizon, that is $R_s = \mathcal{H}^{-1} = 1/(aH)$. Note that the comoving horizon  increases over time
\be\label{R-a}
R_s = \frac{1}{aH} \propto a.
\ee
Consequently, the ``smoothing kernel" $(kR_s)^4 \widetilde{W}_{\mathrm{G}}^2(k, R_s)$ is maximized around $k = \sqrt{2}/R_s$ for the Gaussian window function $\widetilde{W}_{\mathrm{G}}^2$ defined in Eq.~\eqref{eq:winG}. This peak sweeps through momentum space as the Universe expands, that is as $\sqrt{2}/R_s$ decreases.

\begin{figure}[t]
\begin{center}
\includegraphics[width=0.49\textwidth]{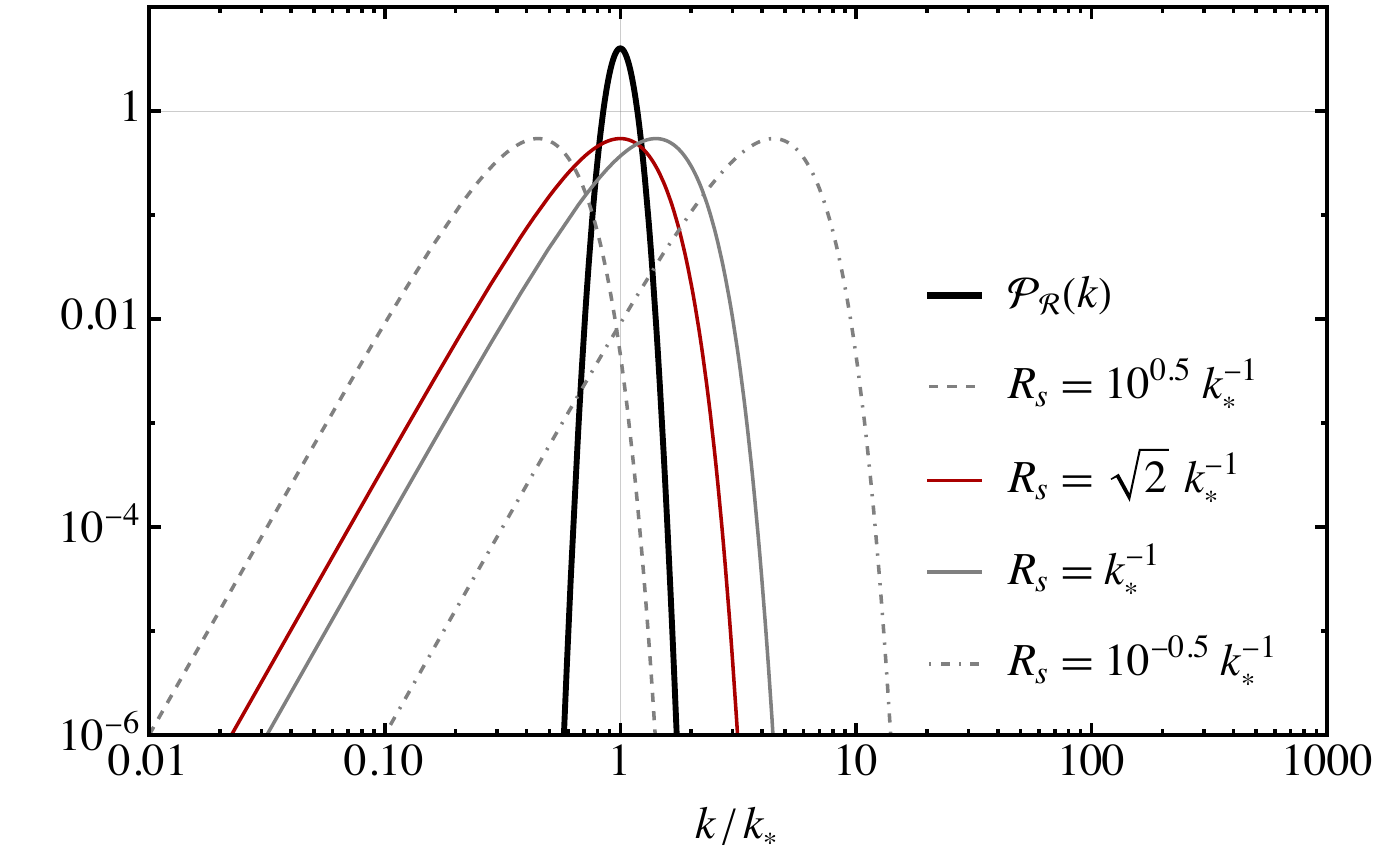}
\includegraphics[width=0.49\textwidth]{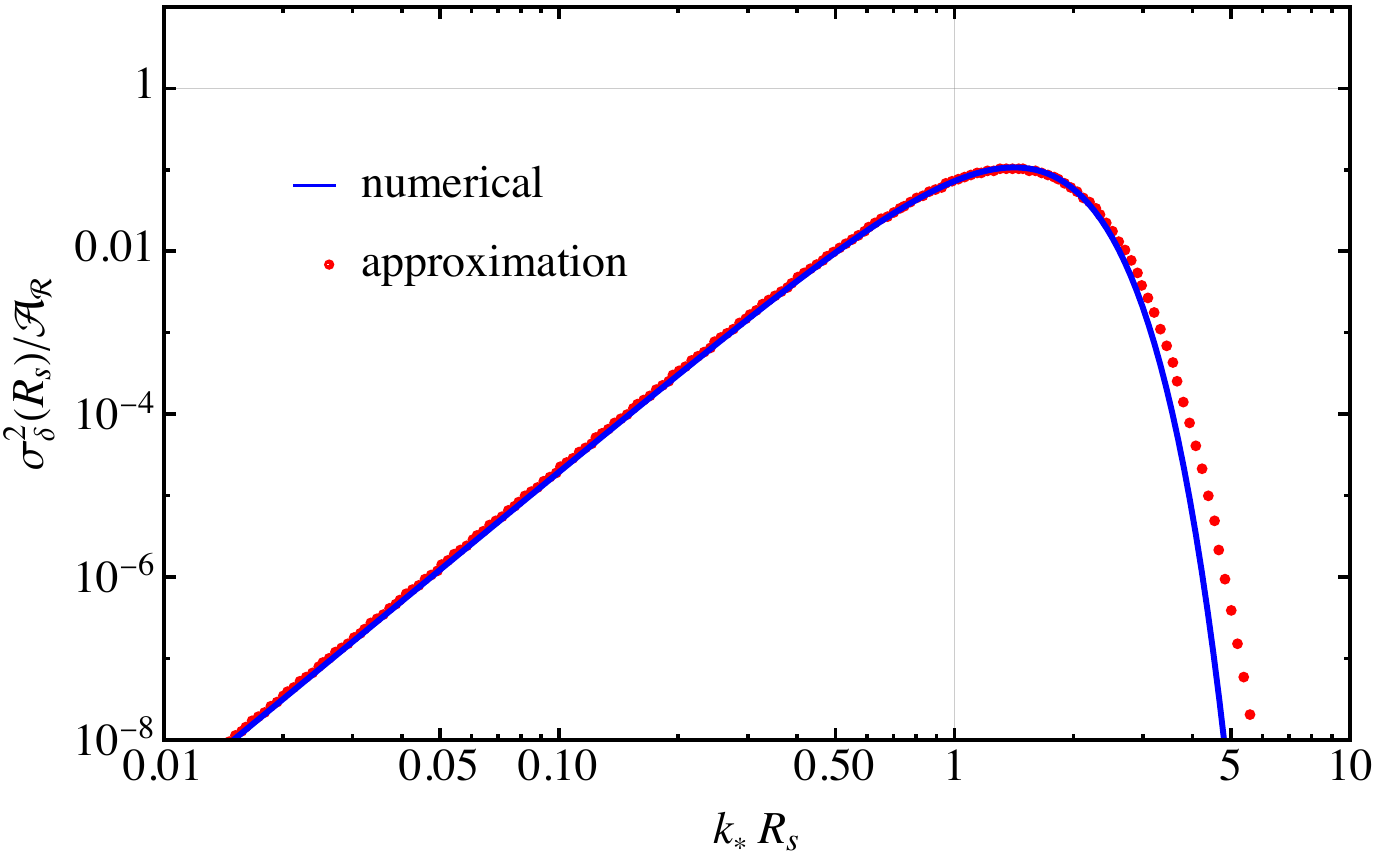}
\caption{[Left] The smoothing kernel $(kR_s)^4\widetilde{W}_{\mathrm{G}}^2(k,R_s)$ of variance $\sigma_\delta^2(R_s)$ considering different smoothing scales $R_s$ and a Gaussian window function $\widetilde{W}_{\mathrm{G}}$ given by Eq.~\eqref{eq:winG}. When $R_s$ corresponds to the comoving horizon that increases with time, the kernel curve sweeps the entire momentum space from right to left. The $\sigma_\delta^2(R_s)$ reaches its maximum when the peak of the kernel at $\sqrt2/R_s$  coincides with the peak of the power spectrum $\mathcal{P}_{\mathcal{R}}(k)$ at $k_*$, which we display here as a log-normal power spectrum with a width of $\Delta = 0.1$, given by Eq.~\eqref{def:PR}. [Right] The variance $\sigma_\delta^2(R_s)$ as a function of $R_s$. The numerical integral of Eq.~\eqref{sigmadelta1} (red dot curve) is overlaid with  the narrow-spectrum approximation Eq.~\eqref{sigmadelta2} (blue curve).}
\label{f:kernel}
\end{center}
\end{figure}

If the integrand in Eq.~\eqref{def:sigmadelta}, corresponding to the power spectrum of the curvature perturbation in the comoving slice, has a narrow peak at $k_*$, the variance $\sigma_\delta^2(R_s)$ reaches its maximum when the two peaks at $k \simeq \sqrt{2}/R_s$ and $k \simeq k_*$ coincide, that is when $k_* R_s = \sqrt{2}$. This overlap between the two peaks provides the dominant contribution to the integral in Eq.~\eqref{def:sigmadelta}, corresponding to the scale of the most significant PBH formation. In Fig.~\ref{f:kernel} we display the behavior of the smoothing kernel in momentum space for different smoothing scales $R_s$ as well as variance $\sigma_{\delta}^2$ as a function of $R_s$. Using the same log-normal power spectrum described in Sec.~\ref{s:Applications and Results} and Eq.~\eqref{def:PR}, along with the window function of Eq.~\eqref{eq:winG}, we obtain
\be\label{sigmadelta1}
\sigma_\delta^2(R_s)=\frac{16}{81}\frac{\AR}{\sqrt{2\pi}\Delta}\int\frac{\dif k}{k}\left(kR_s\right)^4\exp\left(-\frac{\ln^2(k/k_*)}{2\Delta^2}-(kR_s)^2\right).
\ee
For a general case, this integral can be evaluated numerically or approximated analytically using the stationary phase method.

The mass of PBHs is approximately the total mass enclosed within the comoving horizon, with a suppression factor that can be estimated to be $\alpha \simeq 0.2$ from a simple analytical calculation~\cite{Carr:1975qj}. Thus, the mass of a PBH formed from the collapse of all the matter within the comoving horizon is given by
\be\label{M-R}
M_\text{PBH}=\alpha M_H, \quad M_H=\rho_b \frac{4\pi}{3}\left(\frac{H^{-1}}{2}\right)^3
=\frac{1}{16G H}~,
\ee
where we used $\rho_b = 3 H^2/8 \pi G$.
Approximately, the PBH mass scales as $M_\text{PBH} \sim \mpl^2 / H \propto t \propto a^2$. By this proportionality, the horizon scale can be related to the horizon mass as
\be\label{R-MPBH}
k_*R_s=\frac{H_*a_*}{Ha}=\frac{a}{a_*}=\left(\frac{M_H}{M_{k_*}}\right)^{1/2}.
\ee
Here, $M_{k_*}$ represents the horizon mass at $k_*R_s = 1$, as given in Eq.~\eqref{eq:fixRs-mono}, which is half of the central mass corresponding to $k_*R_s = \sqrt{2}$.
 
Thus, the $R_s$-dependence in Eq.~\eqref{sigmadelta1} and related equations can be replaced with the dependence on $M_\text{PBH}$. For simplicity, we will omit the ``PBH" subscript from $M_\text{PBH}$ in the following discussion. By substituting Eq.~\eqref{R-MPBH} into Eq.~\eqref{sigmadelta1}, Eq.~\eqref{def:P-delta}, and Eq.~\eqref{def:beta1}, we obtain the following expression for the PBH abundance at formation 
\be\label{beta(M)1}
\beta_{\mathrm{PBH}}(M)=\frac{\alpha}{\sqrt{2\pi\sigma_\delta^2(M)}}\int_{\delta_\text{cr}}^\infty \dif\delta\exp\left(-\frac{\delta^2}{2\sigma_\delta^2(M)}\right) .
\ee
The dependence of $R_s$ in $\sigma_\delta^2(R_s)$ is converted to a dependence on $M$ using Eq.~\eqref{R-MPBH}. The integral above simplifies to a complementary error function
\be
\int_{\delta_\text{cr}}^\infty\dif\delta \exp\left(-\frac{\delta^2}{2\sigma_\delta^2(M)}\right) =\sqrt{\frac\pi2}\sigma_\delta(M)\text{erfc}\left(\frac{\delta_\text{cr}}{\sqrt2\sigma_\delta(M)}\right)\simeq\frac{\sigma_\delta^2(M)}{\delta_\text{cr}}\exp\left(-\frac{\delta_\text{cr}^2}{2\sigma_\delta^2(M)}\right),
\ee
hence
\be\label{eq:beta(M)2}
\beta_{\mathrm{PBH}}(M)\simeq\frac{\alpha}{\sqrt{2\pi}}\frac{\sigma_\delta(M)}{\delta_\text{cr}}\exp\left(-\frac{\delta_\text{cr}^2}{2\sigma_\delta^2(M)}\right)\equiv\frac{\alpha}{\sqrt{2\pi}\nu(M)}\exp\left(-\frac{\nu(M)^2}{2}\right).
\ee
In the last step of Eq.~\eqref{eq:beta(M)2}, we simplify the expression by defining $\nu(M) \equiv \delta_\text{cr} / \sigma_\delta(M)$, and we use $\delta_\text{cr} = 0.41$ as given in Eq.~\eqref{eq:HYKlimit} for the subsequent calculations.

\begin{figure}[t]
\begin{center}
\includegraphics[width=0.65\textwidth]{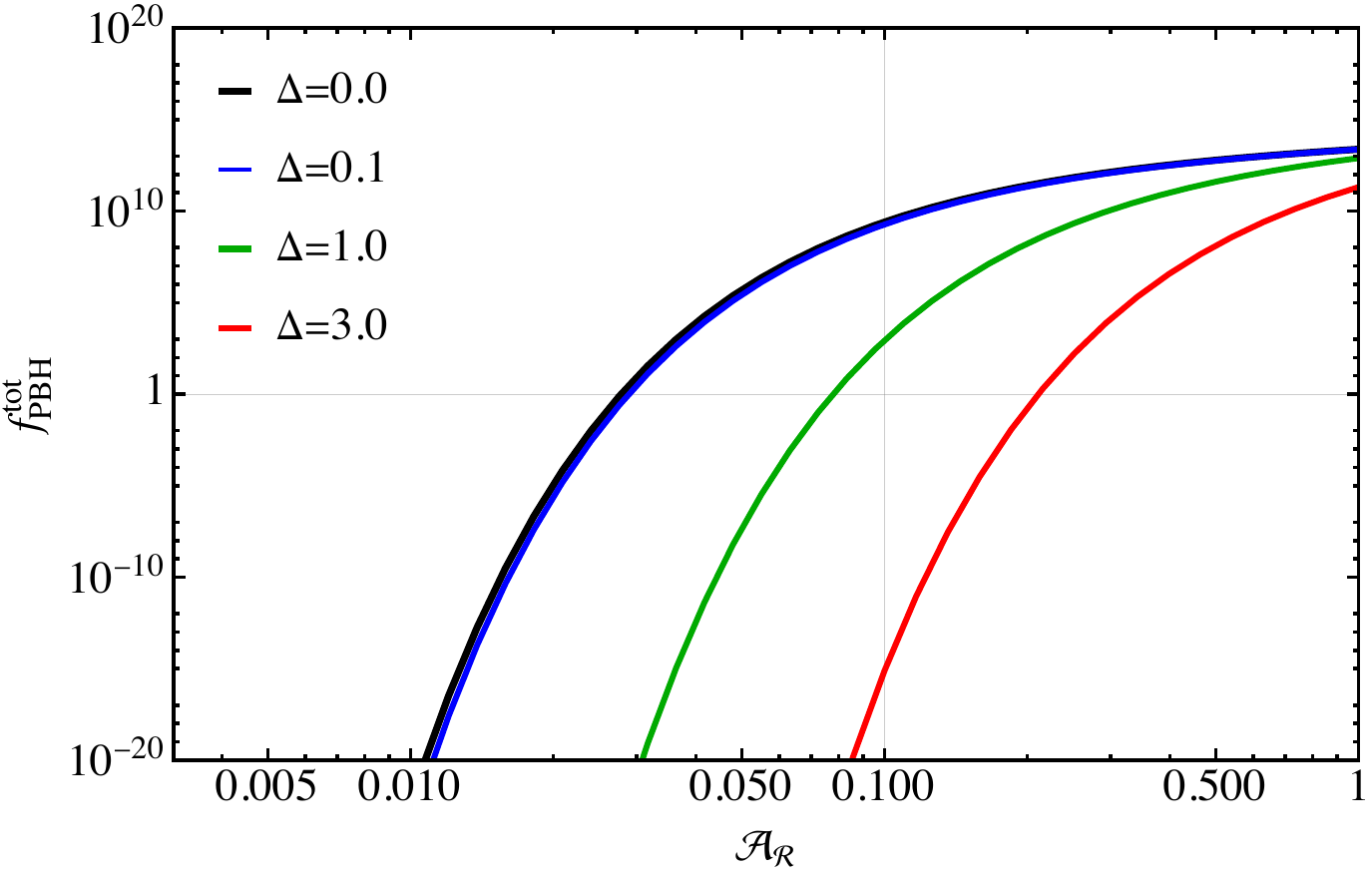}
\caption{The PBH abundance calculated using the PS$\delta$ method with density contrast and a Gaussian window function for different choices of width $\Delta$.}
\label{f:ftotPS}
\end{center}
\end{figure}

\subsubsection{Narrow spectrum case}

First, let us consider the narrow-spectrum case when the width of the log-normal curvature perturbation $\Delta$, is much smaller than 1, as in Fig.~\ref{f:kernel} where $\Delta = 0.1$. Since the width of the kernel in Eq.~\eqref{sigmadelta1} is typically 1, it can be safely factored out of the integral by setting $k \to k_*$, except for the neighborhood of log-normal peak 
\be
\begin{aligned}
\sigma_\delta^2(R_s)&\simeq
\frac{16}{81}\AR(k_*R_s)^4e^{-(k_*R_s)^2}
\frac{1}{\sqrt{2\pi}\Delta}\int_{-\infty}^{+\infty}\dif\ln k~\exp\left(-\frac{\ln^2(k/k_*)}{2\Delta^2}\right)\\
\label{sigmadelta2}
&=\frac{16}{81}\AR(k_*R_s)^4e^{-(k_*R_s)^2}.
\end{aligned}
\ee
We immediately observe that the window function factor becomes independent of the variance, as the power is always concentrated around the narrow peak, which depends solely on $R_s$. As mentioned earlier, although the form of Eq.~\eqref{sigmadelta2} resembles that of the kernel $\left(k R_s\right)^4 \widetilde{W}_{\mathrm{G}}^2\left(k , R_s\right)$ in Eq.~\eqref{def:sigmadelta}, the interpretation is  different. Using Eq.~\eqref{sigmadelta2} we derive the variance, the characteristic density contrast, as a function of the comoving horizon as shown in the right panel of Fig.~\ref{f:kernel}. 
As time progresses, the comoving Hubble horizon expands as a power law of the scale factor, as given by Eq.~\eqref{R-a}. Therefore, the right panel of Fig.~\ref{f:kernel} can also be interpreted as a function of time.

\begin{figure}[t]
\begin{center}
\includegraphics[width=0.65\textwidth]{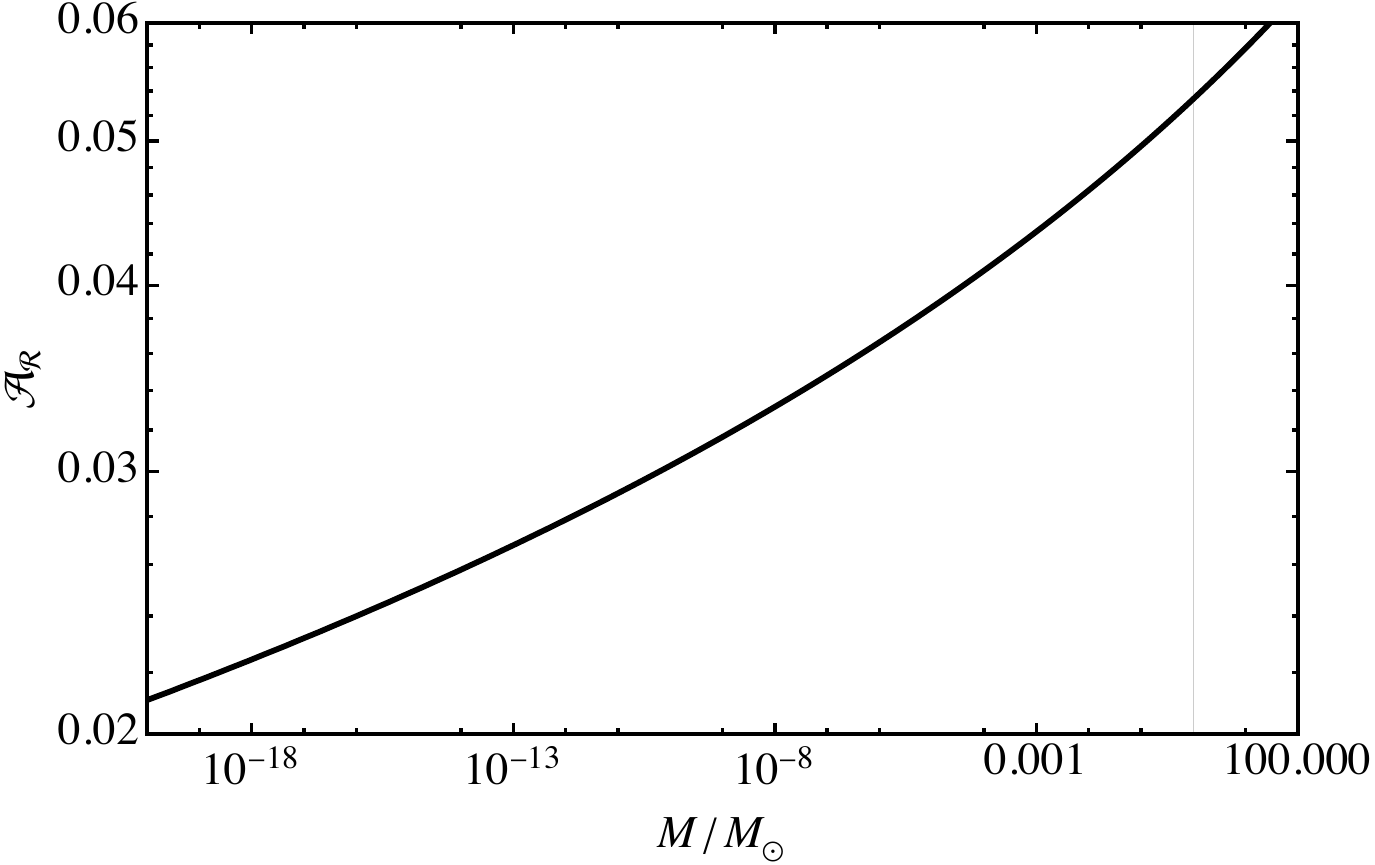}
\caption{The amplitude of the narrow power spectrum of the curvature perturbation on comoving slice required to produce PBHs as all DM with $f_{\text{PBH}} = 1$, as given by Eq.~\eqref{f-beta}. It depends weakly on $M_{\mathrm{PBH}}$.}
\label{f:A-M}
\end{center}
\end{figure}
 
At any given moment, PBHs form from the high-$\sigma$ tail of the PDF, where $\delta > \delta_\text{cr}$, as described by Eq.~\eqref{def:P-delta}. This tail is always Gaussian-suppressed. The moment with the least suppression corresponds to the maximum PBH formation abundance.
This occurs when the variance $\sigma_\delta^2(R_s)$ reaches its maximum at $R_s = \sqrt{2}/k_*$, as shown in the left panel of Fig.~\ref{f:kernel}. A spectrum of PBHs forms as $R_s$ evolves from $R_s < \sqrt{2}/k_*$ to $R_s > \sqrt{2}/k_*$. The PBH abundance peaks at $R_s = \sqrt{2}/k_*$ and rapidly decreases for $R_s \gtrsim \sqrt{2}/k_*$.

The PBH abundance, $\beta_{\mathrm{PBH}}(M)$, for a narrow spectrum exhibits a sharp peak at the minimum of $\nu(M)$, corresponding to the maximum of $\sigma_\delta(M)$. From Eq.~\eqref{sigmadelta2}, we know that this maximum occurs at $R_s = R_* \equiv \sqrt{2}/k_*$, which gives
\be
\sigma_\delta^2(R_*)\simeq\frac{16}{81} 4e^{-2}\AR,
\ee
then 
\be
\nu(R_*)=\frac{\delta_\mathrm{cr}}{\sigma_\delta(R_*)}=\frac{9e}{8\AR^{1/2}}\delta_\mathrm{cr}\xRightarrow{~\delta_{\rm cr} = 0.41~}\frac{1.254}{\AR^{1/2}}.
\ee
The PBH abundance depends sensitively on the amplitude of $\AR$ 
\be
\beta_*\equiv \beta_{\mathrm{PBH}}(R_*)=\frac{8\alpha\AR^{1/2}}{9\sqrt{2\pi}e\delta_\mathrm{cr}}\exp\left(-\frac{81e^2}{128\AR}\delta_\mathrm{cr}^2\right).
\ee 
This is related to the PBH mass function, the PBH abundance at the current epoch, as~\cite{Carr:2020gox} 
\be\label{f-beta}
f_\text{PBH}(M)=3.81\times10^{8}\alpha^{1/2}\left(\frac{g_{*i}}{106.75}\right)^{-1/4}\left(\frac{h}{0.67}\right)^{-2}\beta_*\left(\frac{M}{M_\odot}\right)^{-1/2},
\ee
where $g_{*i}$ represents the effective relativistic degrees of freedom, and $h$ is the normalized Hubble constant. The $M^{-1/2}$ dependence in Eq.~\eqref{f-beta} arises from Eq.~\eqref{R-MPBH}, accounting for the redshift from horizon reentry to matter-radiation equality.

In Fig.~\ref{f:ftotPS}
the function $f_{\mathrm{PBH}}^{\mathrm{tot}}(\AR)$ is displayed for the monochromatic case. From this figure, we observe that for $f_{\text{PBH}}^{\text{tot}} = 1$, the curvature perturbation power spectrum on comoving slices must be enhanced to a $\sim $few$\times 10^{-2}$. To examine this explicitly, the amplitude $\AR$ can be solved inversely using numerical methods, as shown in Tab.~\ref{tab:PSchart}. As illustrated in Fig.~\ref{f:A-M}, there is weak dependence on $M_{\mathrm{PBH}}$.

\begin{figure}[t]
\begin{center}
\includegraphics[width=0.65\textwidth]{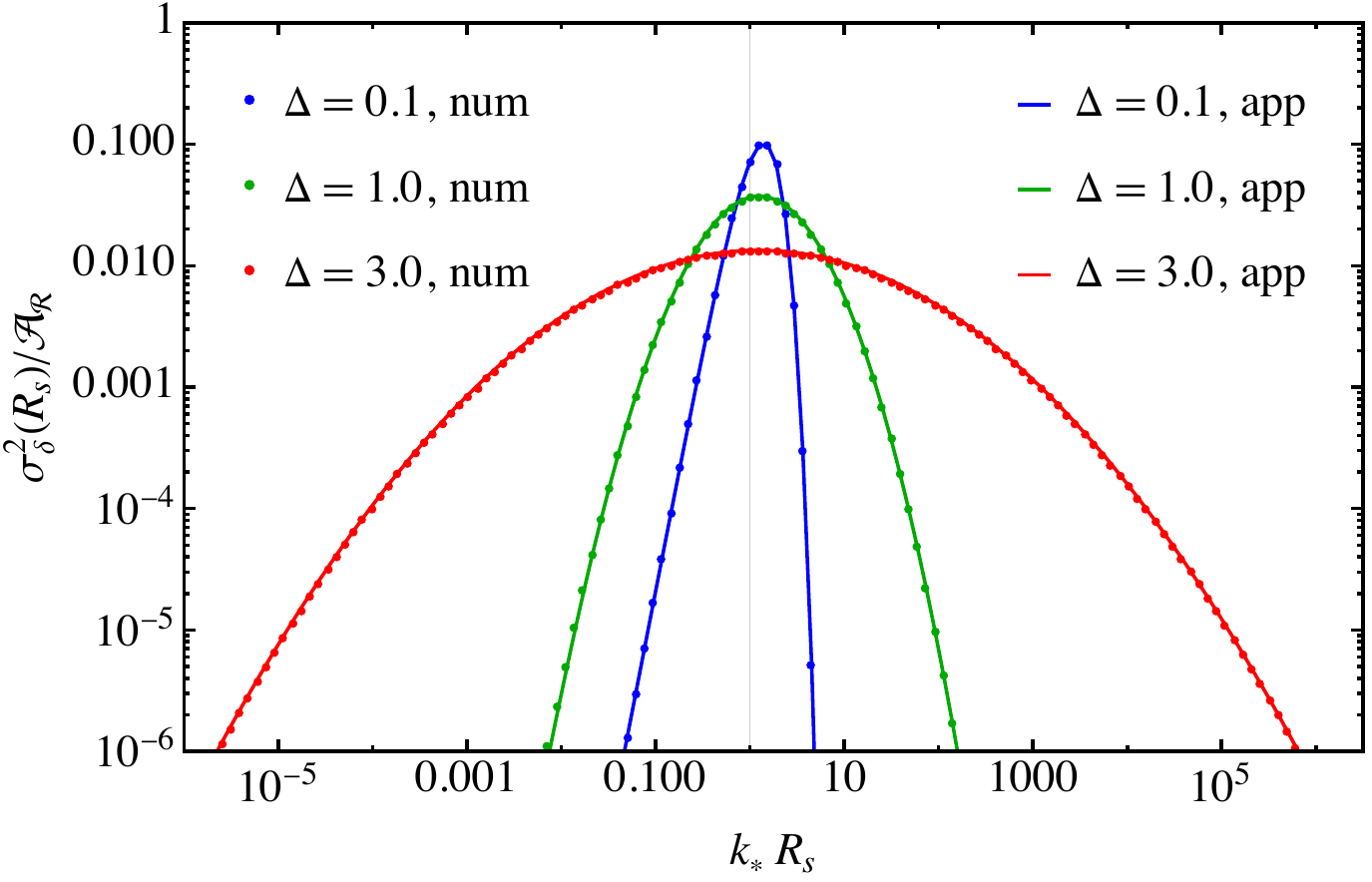}
\caption{The variance $\sigma_\delta^2(R_s)$ as a function of $R_s$, for different choices of log-normal curvature perturbation widths $\Delta$. The numerical integral of Eq.~\eqref{sigmadelta1} (dotted lines) and the stationary phase approximation of Eq.~\eqref{eq:sigmaSPAbroad} (solid lines) are displayed.}
\label{f:kernal2}
\end{center}
\end{figure}

\subsubsection{Broad spectrum case}

Let us now consider the broad-spectrum case when the width of the log-normal curvature perturbation is $\Delta\gtrsim1$. To evaluate Eq.~\eqref{sigmadelta1}, we define a dimensionless wave number $\kappa\equiv kR_s$, giving
\begin{align}\label{def:I(kappa)}
\sigma_\delta^2(R_s,\Delta)&\equiv \frac{16}{81}\frac{\mathcal{A}_{\mathcal{R}}}{\sqrt{2\pi}\Delta}\int\dif\kappa\exp\left(-f(\kappa)\right),\\
f(\kappa)&\equiv\frac{1}{2\Delta^2}\left(\ln\frac{\kappa}{\kappa_*}\right)^2+\kappa^2-3\ln\kappa.
\end{align}
The integral is dominated by contributions from the zero point of $f'(\kappa)$, which can be solved as
\be
\kappa_0=\frac{1}{2\Delta} W^{1/2}\left( 4e^{6\Delta^2}\Delta^2 \kappa_*^2 \right),
\label{eq:SPA}
\ee
where $W$ is the Lambert function. Then the variance of Eq.~\eqref{sigmadelta1} becomes
\be
\begin{aligned}
\sigma^2_\delta(R_s,\Delta)&\simeq\frac{16}{81}\frac{\AR}{\sqrt{2\pi}\Delta}\sqrt{\frac{2\pi}{|f''(\kappa_0)|}}e^{-f\left(\kappa_0\right)}\\\nn
&=\frac{16}{81}\frac{\AR}{\Delta}\kappa_*^3\left(\frac{4}{W\left(y\right)}+4\right)^{-1/2}
   \exp \left(\frac{9 \Delta ^2}{2}-\frac{W^2\left(y\right)+2W\left(y\right)}{8 \Delta ^2}\right),
\end{aligned}
\label{eq:sigmaSPAbroad}
\ee
where in the last step,
\be
y\equiv 4e^{6\Delta^2}\Delta^2 \kappa_*^2.
\ee
As shown in Fig.~\ref{f:kernal2},
these analytical results work well around the peak.

\begin{figure}[t]
\begin{center}
\includegraphics[width=0.65\textwidth]{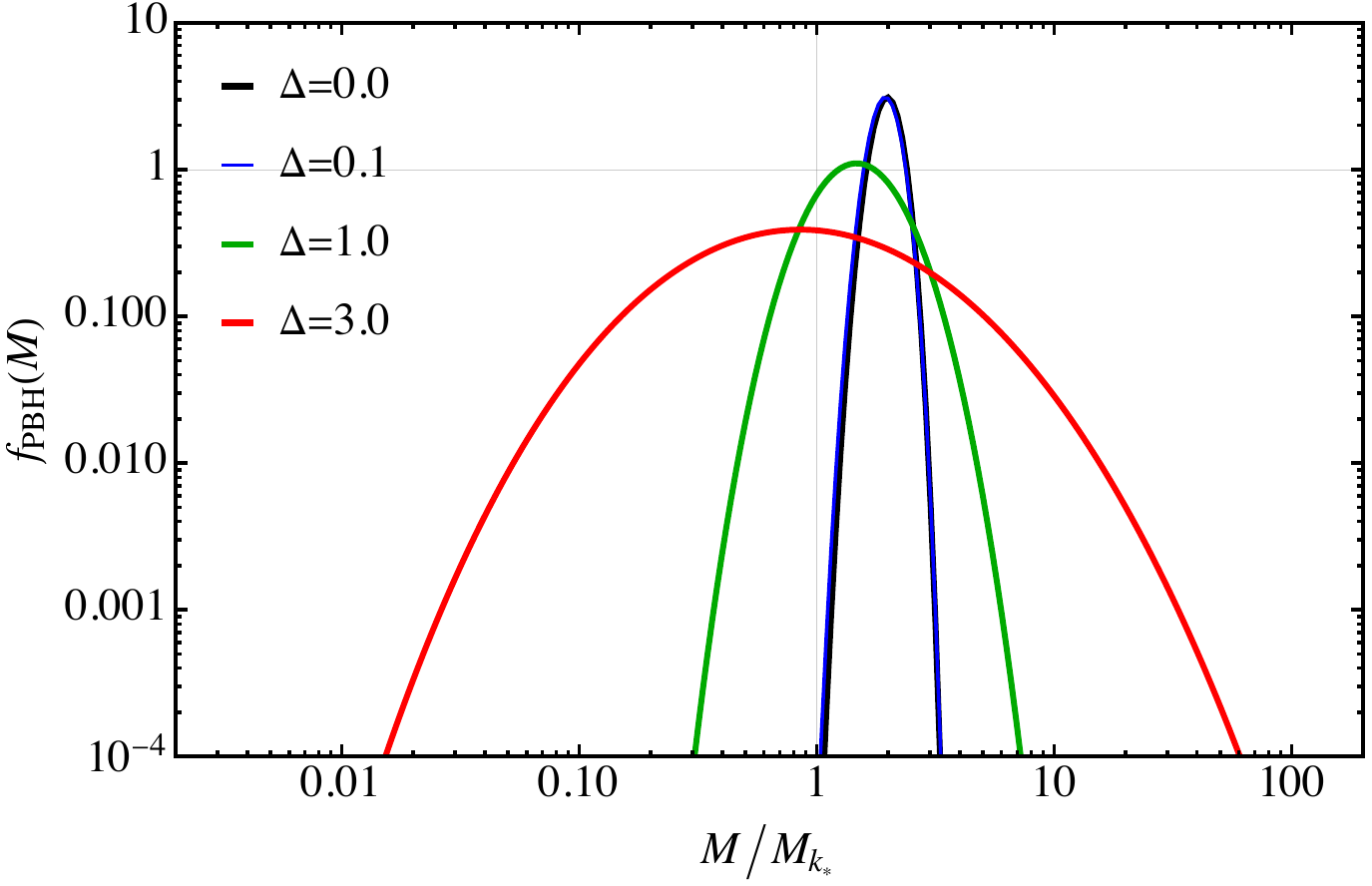}
\caption{The PBH mass function calculated from PS$\delta$ method using density contrast with a Gaussian window function, for different choices of log-normal curvature perturbation widths $\Delta$. All mass functions are normalized to $f_{\mathrm{PBH}}^{\mathrm{tot}} = 1$. The corresponding values of $\mathcal{A}_{\mathcal{R}}$ are shown in Tab.~\ref{tab:PSchart}.}
\label{f:fpbhwide}
\end{center}
\end{figure}

We can calculate the PBH mass function $f_\mathrm{PBH}(M)$ for broad power spectrum as
\begin{align}
f_\text{PBH}(M)&\simeq3.81\times10^{8}\alpha^{1/2}\left(\frac{g_{*i}}{106.75}\right)^{-1/4}\left(\frac{h}{0.67}\right)^{-2}
\beta_{\mathrm{PBH}}(M) \left(\frac{M}{M_\odot}\right)^{-1/2},
\label{eq:Redshift}
\end{align}
where  $\sigma_\delta(M)$ is given by Eq.~\eqref{sigmadelta1} or Eq.~\eqref{eq:sigmaSPAbroad}, with $M$ determined by Eq.~\eqref{R-MPBH}. The total PBH abundance can be calculated using Eq.~\eqref{def:ftot}, which depends on $\AR$ and $\Delta$. 
In Fig.~\ref{f:ftotPS} we display
the function $f_{\mathrm{PBH}}^{\mathrm{tot}}(\AR)$ for several values of width $\Delta$. 
We observe that for a given  $\AR$, PBH formation is suppressed as the power spectrum becomes broader. This is because a broader peak in the power spectrum corresponds to a smaller amplitude, reducing the number of perturbations that exceed the critical threshold. 

In Fig.~\ref{f:fpbhwide} we display the PBH mass function for different values of width $\Delta$ considering a Gaussian window function 
allowing to achieve $f^{\mathrm{tot}}_{\mathrm{PBH}} = 1$, corresponding to PBHs constituting all DM. To attain $f^{\mathrm{tot}}_{\mathrm{PBH}} = 1$, $\AR$ can be determined by fixing $M_{k_*}$ and solving the equations numerically. 
These results are summarized in Tab.~\ref{tab:PSchart}.

\begin{figure}[t]
    \centering
    \includegraphics[width=0.65\linewidth]{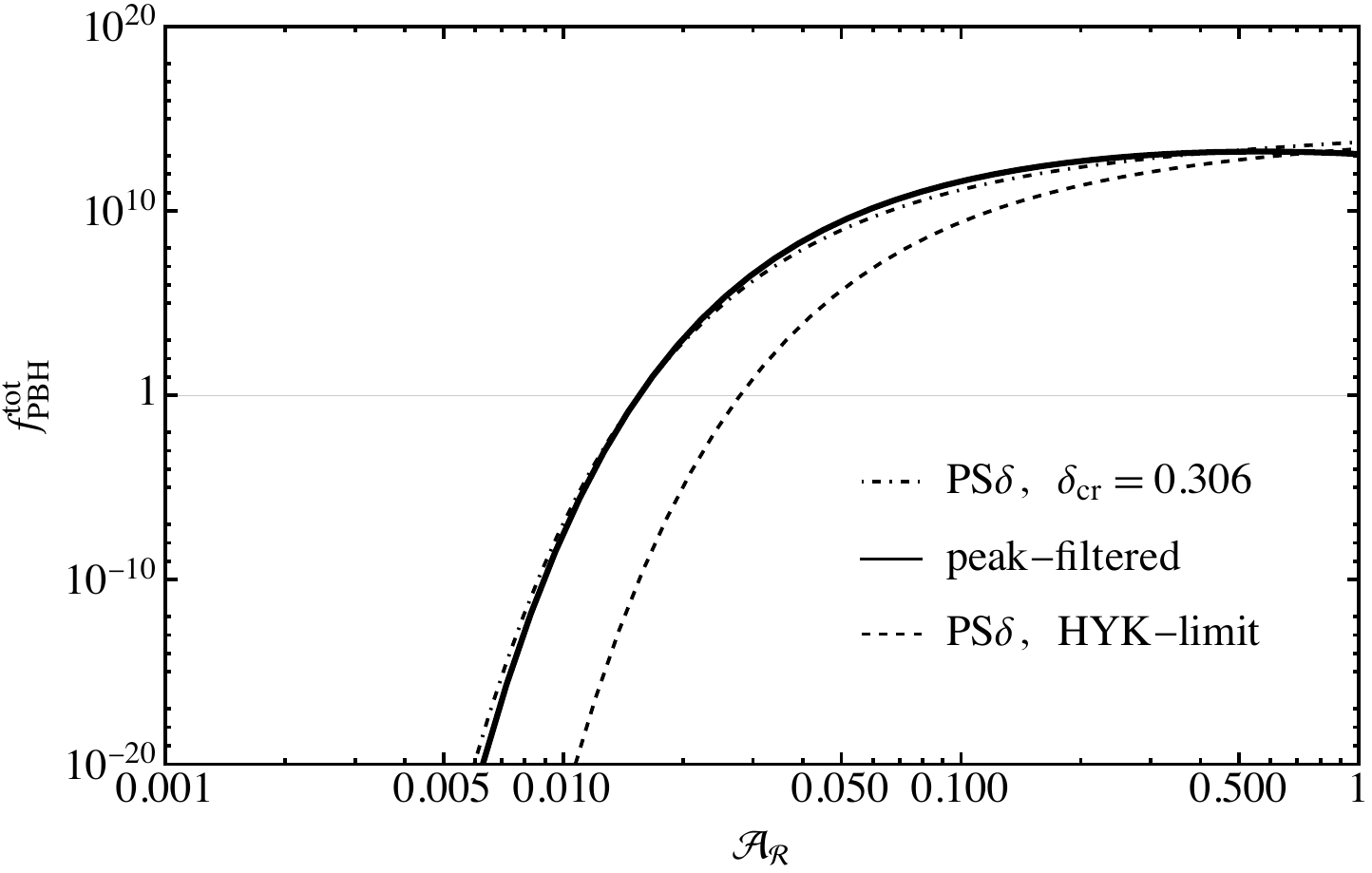}
    \caption{Total PBH abundance calculated considering peak-filtered method developed in the main text (solid line), PS$\delta$ method with threshold value of $\delta_{\rm cr}$ as in Eq.~\eqref{eq:HYKlimit} (dashed line, ``PS$\delta$, HYK-limit'')  and PS$\delta$ method with adjusted threshold of $\delta_{\mathrm{cr}}\simeq 0.306$ (dot-dashed line, ``PS$\delta$, $\delta_{\mathrm{cr}}\simeq 0.306$'') allowing to achieve enhanced PBH formation and comparable to peak-filtered method, considering perturbation width of $\Delta = 0$. Gaussian window function is assumed.}
    \label{fig:fitthinPS}
\end{figure}

The PBH formation efficiency and abundance in the PS$\delta$ method are highly sensitive to the threshold value $\delta_{\rm cr}$. By appropriately adjusting the input value of $\delta_{\rm cr}$, the PBH efficiency and abundance in the PS$\delta$ formalism can be enhanced and made comparable to those obtained with the method developed in the main text, based on peaks theory smoothed by a Gaussian window function and filtered by a filter function that we refer to as the ``peak-filtered" method. 
In Fig.~\ref{fig:fitthinPS}, for the illustrative case of $\Delta = 0$, we demonstrate that by adjusting the threshold in the PS$\delta$ method, its PBH formation efficiency and abundance can be significantly enhanced, reaching levels comparable to the peak-filtered method. In Tab.~\ref{tab:fitPS}, we provide the corresponding threshold values $\delta_{\mathrm{cr}}$ in the PS$\delta$ method  using density contrast  required to generate similar PBH abundance as the peak-filtered method, achieving the same {variance} $\mathcal{A_R}$ as listed in Tab.~\ref{tab:PTFullchart}.

\begin{table}[t]
\begin{center}
\begin{equation*}
\begin{array}{ccccccccc}
\hline\hline
\Delta & \text{color} & \mathcal{A}_{\mathcal{R}} & M_{c}/M_{k_*} &  &  \\
\hline 
0 & \text{Black} & 2.78\times 10^{-2}  & 2.00 &  & \\
0.1 & \text{Green} & 2.89\times 10^{-2}  & 1.91 &  & \\
0.4 & \text{Orange} & 4.10\times 10^{-2}  & 1.74 &  & \\
1.0 & \text{Blue} &  7.80\times 10^{-2}  & 1.45 &  & \\
2.0 & \text{Red} & 1.45\times 10^{-1}  & 1.20 &  & \\
3.0 & \text{Cyan} & 2.11\times 10^{-1}  & 0.87 &  & \\
\hline\hline
\end{array}
\end{equation*}
\caption{\label{tab:PSchart} Results from PS$\delta$ method for calculating PBH abundance using density contrast with a Gaussian window function allowing to achieve $f^{\mathrm{tot}}_{\mathrm{PBH}} = 1$.}
\end{center}
\end{table}
 
\begin{figure}[t]
    \centering   \includegraphics[width=0.49\linewidth]{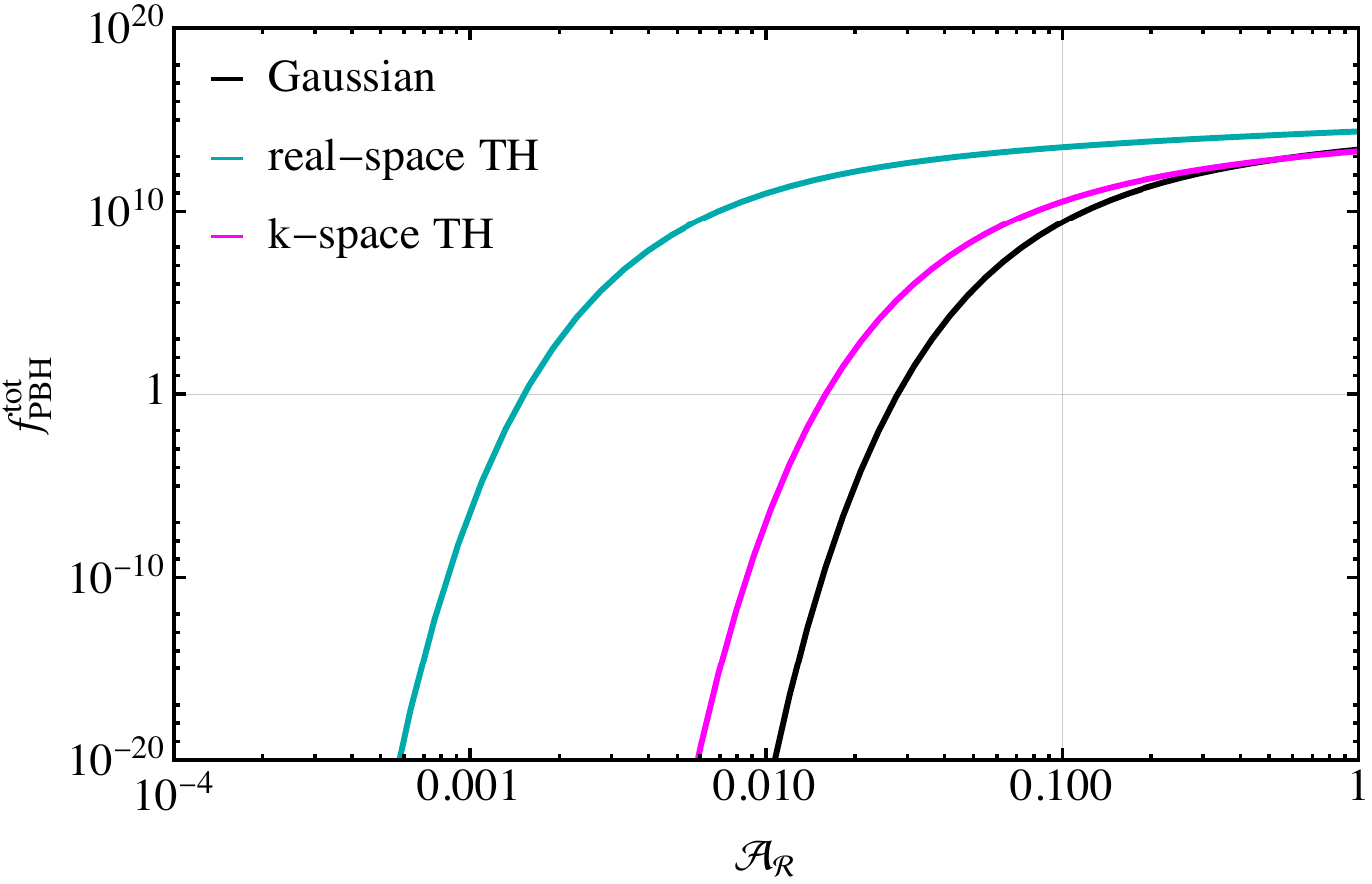}
\includegraphics[width=0.49\linewidth]{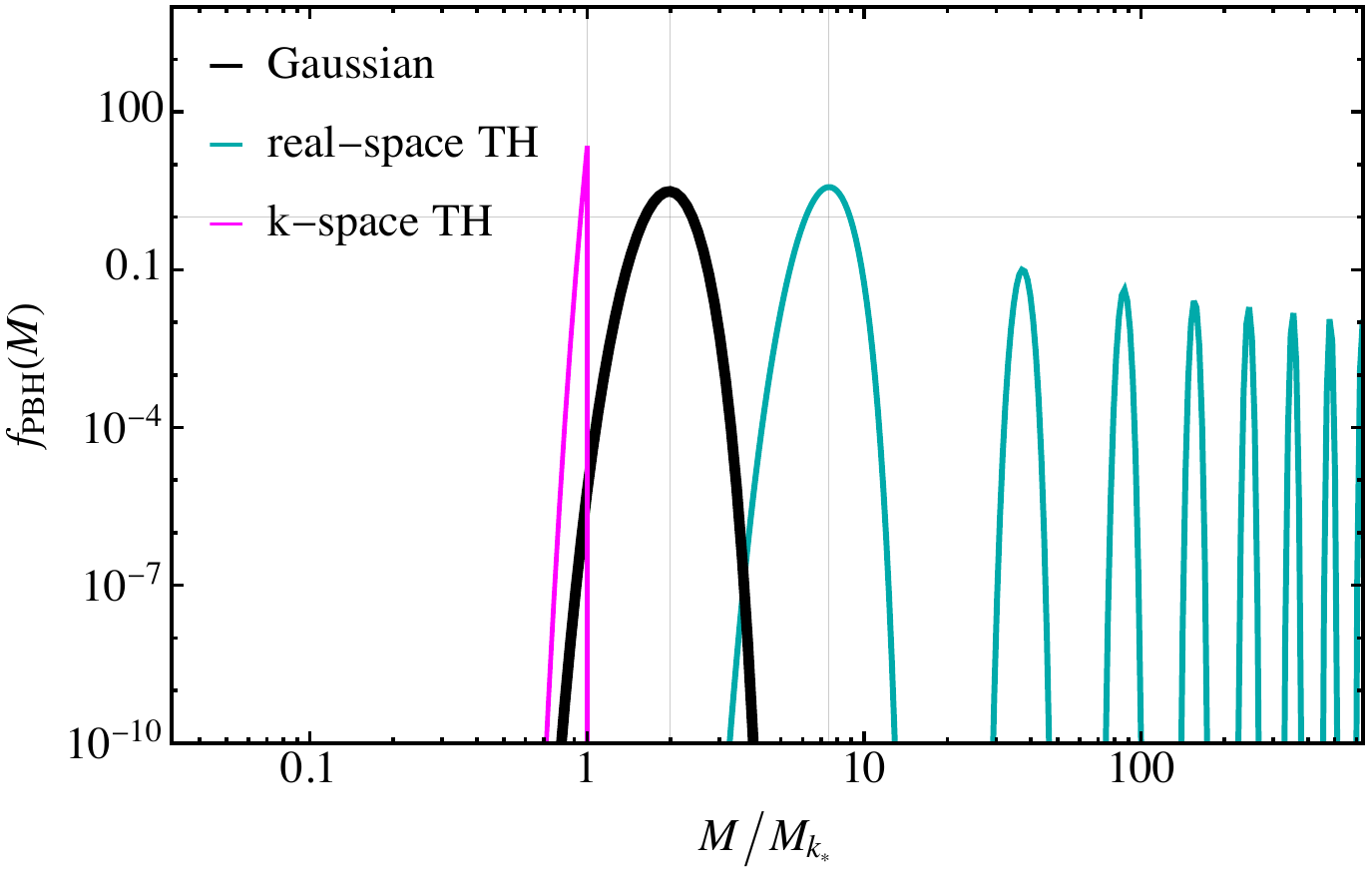}    
    \caption{[Left] Total PBH abundance calculated by the PS$\delta$ method using different types of window function, considering monochromatic power spectrum and normalized such that PBHs are all DM with $M_{k_*}= 10^{20}$g. The {variance} needed such that PBHs constitute all DM is $\mathcal{A}_\mathcal{R}=2.78\times 10^{-2}, 1.53\times 10^{-3},1.59\times 10^{-2}$ for the Gaussian, the real-space top-hat, and the $k$-space top-hat window function, respectively. [Right] PBH mass function calculated by the PS$\delta$ method using different types of window function, considering monochromatic power spectrum and normalized such that PBHs are all DM with $M_{k_*}= 10^{20}$g. The central mass is given by the $M_c / M_{k_*}=2.0, 7.41,1$ for the Gaussian, the real-space top-hat, and the $k$-space top-hat window function, respectively.}
    \label{fig:pSdiffWinftotM}
\end{figure}

In Fig.~\ref{fig:pSdiffWinftotM}, we present the total PBH abundance and mass function calculated using the PS$\delta$ method considering
different types of window functions. 
Note that our results for the total PBH abundance shown in Fig.~\ref{fig:pSdiffWinftotM} follow a different trend compared to \cite{Ando:2018qdb}, where the Gaussian window function produces an intermediate abundance relative to the other two types of window functions. This discrepancy arises because we consider different power spectra in our analysis.

\begin{table}[t]
\begin{equation*}
\begin{array}{cccccccccc}
\hline\hline
\Delta & 0 & 0.1 & 0.4 & 1 & 2 & 3 \\
\hline 
\delta_{\mathrm{cr}} & 0.306 & 0.3015 & 0.303 & 0.306 & 0.3075 &  0.31 \\
\hline\hline
\end{array}
\end{equation*}
\caption{\label{tab:fitPS} Adjusted PS$\delta$ threshold values $\delta_{\mathrm{cr}}$ allowing to achieve similar {variance} $\mathcal{A_R}$ and PBH abundance as peak-filtered method results of Tab.~\ref{tab:PTFullchart}.}
\end{table}

\subsection{PSC Method Using Compaction Function}\label{s:EXPS}

The PS$\delta$ method estimates the formation rate of collapsed objects by calculating the probability that a coarse-grained field exceeds a critical threshold value. The PSC uses the linear compaction function as the random field, defined as the linear approximation of the density contrast given in Eq.~\eqref{def:densitycontrast}
\begin{equation}
\mathcal{C}_\ell(r)=\frac{2}{R(r, t)} \rho_b(t) \int_0^{R(r, t)} \mathrm{d} R\left[4 \pi R(r, t)^2\right] \delta_\ell (r, t)=-\frac{4}{3} r \mathcal{R}^{\prime}(r).
\label{def:Cell}
\end{equation}
The full compaction function $\mathcal{C}(r)$ of Eq.~\eqref{eq:coarse-grained} can be expressed as a quadratic function of $\mathcal{C}_\ell(r)$~\cite{Kawasaki:2019mbl,Young:2019yug,DeLuca:2019qsy}
\begin{equation}
\mathcal{C}(r)=\mathcal{C}_{\ell}(r)-\frac{3}{8 } \mathcal{C}_{\ell}(r)^2.
\label{eq:CfullCell}
\end{equation}
In the PSC method using the compaction function, the main statistical variable is $\mathcal{C}_\ell$. Type I and Type II fluctuations can be distinguished by studying the monotonicity of compaction function,
\begin{equation}
\frac{\mathrm{d}\mathcal{C}}{\mathrm{d}\mathcal{C}_{\ell}}=1-\frac{3}{4} \mathcal{C}_{\ell}=0 \quad \Rightarrow \quad
\mathcal{C}_{\ell,\text{II}}=\frac{4}{3}.
\label{eq:ClII}
\end{equation}
Here, $\mathcal{C}_{\ell} > \mathcal{C}_{\ell,\text{II}}$ corresponds to Type II perturbations, where $\mathcal{C}$ decreases while $\mathcal{C}_{\ell}$ increases. To focus exclusively on Type I fluctuations, we can consider the upper limit of $\mathcal{C}_{\ell}$ to $\mathcal{C}_{\ell,\text{II}}$.

The definition of $\mathcal{C}_\ell$ in Eq.~\eqref{def:Cell} is equivalent to applying a real-space top-hat window function to the linear density contrast $\delta_\ell \sim -\nabla^2 \mathcal{R}$ \cite{Young:2022phe},
\begin{equation}
\mathcal{C}_{\ell}(\boldsymbol{r}_0,r) 
 =-\frac{4}{9} r^2 \int \mathrm{d}^3 \boldsymbol{r} \nabla^2 \mathcal{R}(\boldsymbol{r}) W(\boldsymbol{r}_0-\boldsymbol{r}, r),
 \label{eq:Cltophat}
\end{equation}
where 
\begin{equation}
W(\boldsymbol{r}_0-\boldsymbol{r}, r)\equiv\frac{3}{4 \pi r^3} \Theta(r-\left| \boldsymbol{r}_0-\boldsymbol{r} \right|)
\label{def:Tophat}
\end{equation}
is the real-space top-hat window function and $\boldsymbol{r}_0$ denotes the center of a peak.
The Fourier transform of this window function is
\begin{equation}
\widetilde{W}(k, r)=3 \frac{\sin (k r)-(k r) \cos (k r)}{(k r)^3}.
\label{eq:kspaceTophat}
\end{equation}
If the compactness of a given region exceeds a certain threshold value, $\mathcal{C}_{\mathrm{th}}$, a PBH will form when the region re-enters the horizon.

The PBH mass, $M_{\mathrm{PBH}}$, is related to the full compaction function, $\mathcal{C}$, through the critical collapse relation \cite{Choptuik:1992jv,Evans:1994pj,Koike:1995jm,Niemeyer:1997mt,Hawke:2002rf,Musco:2008hv}:
\begin{equation}
M_{\mathrm{PBH}} = \mathcal{K} M_H\left(\mathcal{C} - \mathcal{C}_{\mathrm{th}}\right)^\gamma,
\label{eq:massEXPS}
\end{equation}
where $\mathcal{K}$, $\mathcal{C}_{\mathrm{th}}$, and $\gamma$ are parameters that characterize the collapse, depending on the shape of the perturbations.
Using the threshold value $\mathcal{C}_{\mathrm{th}} = 0.587$ from Fig.~\ref{fig:Mono-Gau-ACF}, the corresponding threshold for $\mathcal{C}_\ell$ is found to be $\mathcal{C}_{\ell,\mathrm{th}} = 0.872$. The value of $\mathcal{C}_{\mathrm{th}}$ is derived from peaks theory and is valid only for a monochromatic power spectrum. While this may be invalid for a broad power spectrum, here we follow the method outlined in Ref.~\cite{Gow:2022jfb} for consistency.

\begin{figure}
\begin{center}
\includegraphics[width=0.65\textwidth]{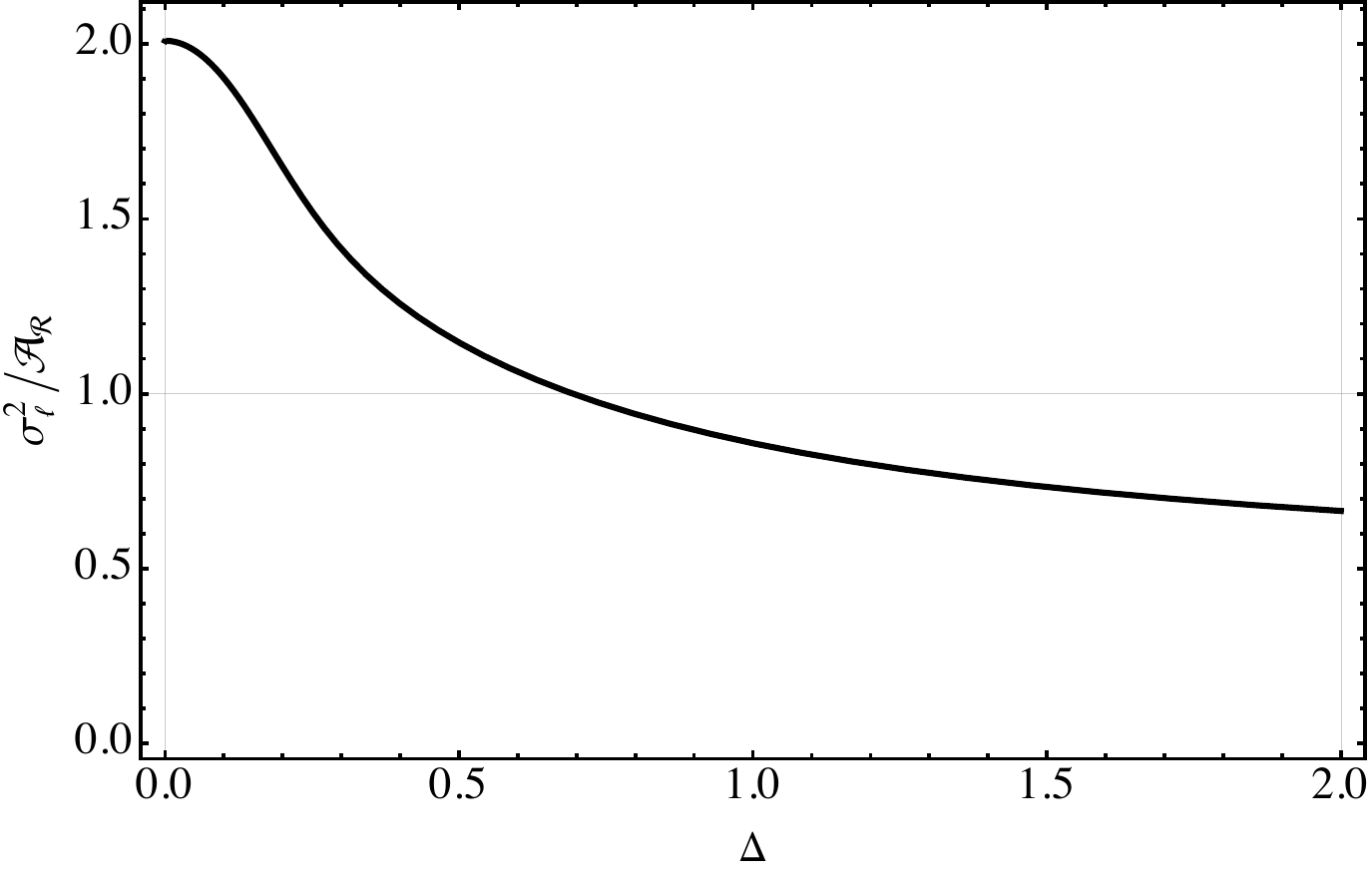} \\
\caption{The dependence of $\sigma_\ell^2 / \mathcal{A}_{\mathcal{R}}$ in Eq.~\eqref{eq:sigmaCell} on the power spectrum width $\Delta$. }
\label{fig:ftotEXPS}
\end{center}
\end{figure}

The number of over-dense regions for a given $\mathcal{C}_\ell$ that determines the PBH abundance for the corresponding mass  can be calculated by integrating the PDF of the relevant variables under the specified constraint conditions. The constraint is $\mathcal{C}_{\ell,\mathrm{th}} < \mathcal{C}_{\ell} < \mathcal{C}_{\ell,\mathrm{II}}$, and the PDF of the variable $\mathcal{C}_\ell$ is Gaussian, as it is derived from the gradient of a Gaussian field, as given in Eq.~\eqref{def:Cell}
\begin{equation}
\mathbb{P}_G (\mathcal{C}_\ell)=\frac{1}{\sqrt{2 \pi} \sigma_\ell} \exp \left(-\frac{1}{2} \frac{\mathcal{C}_\ell^2}{\sigma_\ell^2} \right).
\label{eq:PDFofCell}
\end{equation}
Next, we calculate the variance of this PDF, defined as $\sigma_\ell^2 \equiv \langle \mathcal{C}_\ell^2 \rangle$. Using the expansion of $\mathcal{R} \left(\boldsymbol{r}\right)$ in momentum space, we obtain 
\begin{equation}
\sigma_\ell^2 
=\frac{16}{9} \int \frac{\mathrm{d} k}{k}  \mathcal{P}_{\mathcal{R}}(k)  (k r)^4   \left(\frac{\sin (k r)-(k r) \cos (k r)}{(k r)^3}\right)^2.
\label{eq:sigmaCell}
\end{equation}
For a monochromatic power spectrum case, Eq.~\eqref{eq:sigmaCell} gives
\begin{equation}
\sigma_\ell^2 
=\mathcal{A}_{\mathcal{R}} \frac{16}{9}(k_* r)^4   \left(\frac{\sin (k_* r)-(k_* r) \cos (k_* r)}{(k_* r)^3}\right)^2 .
\end{equation}
In Fig.~\ref{fig:ftotEXPS}, we analyze the dependence of $\sigma_\ell^2 / \mathcal{A}_{\mathcal{R}}$ in Eq.~\eqref{eq:sigmaCell} on the width of the power spectrum, $\Delta$. A narrower power spectrum results in a broader PDF of $\mathcal{C}_\ell$. For a fixed threshold value, narrower power spectra produce more PBHs.

The PBH mass function at formation time $\beta_{\mathrm{PBH}}(M)$ is defined as
\begin{equation}
\mathrm{d}\ln M   \beta_{\mathrm{PBH}}(M)=\frac{M}{M_H} \mathbb{P}_G (\mathcal{C}_{\ell})\mathrm{d}\mathcal{C}_{\ell}.
\label{def:betaPBH}
\end{equation}
Thus,
\begin{equation}
\beta_{\mathrm{PBH}}(M)=\frac{ \mathcal{K} \left(\mathcal{C}-\mathcal{C}_{\mathrm{th}}\right)^{\gamma+1}}{\gamma \left(1-\frac{3}{4} \mathcal{C}_{\ell}\right)}  \mathbb{P}_G(\mathcal{C}_{\ell}). 
\label{eq:fPBH-EXPSM}
\end{equation}
Note that $M_H$ dependency is canceled in $ \beta_{\mathrm{PBH}}(M)$.
The denominator in Eq.~\eqref{eq:fPBH-EXPSM} originates from the Jacobian factor introduced when changing variables from $\mathcal{C}_\ell$ to the PBH mass $M$. This factor leads to a divergence at $\mathcal{C}_{\ell,\mathrm{II}} = 4/3$.

\begin{figure}
\begin{center}
\includegraphics[width=0.65\textwidth]{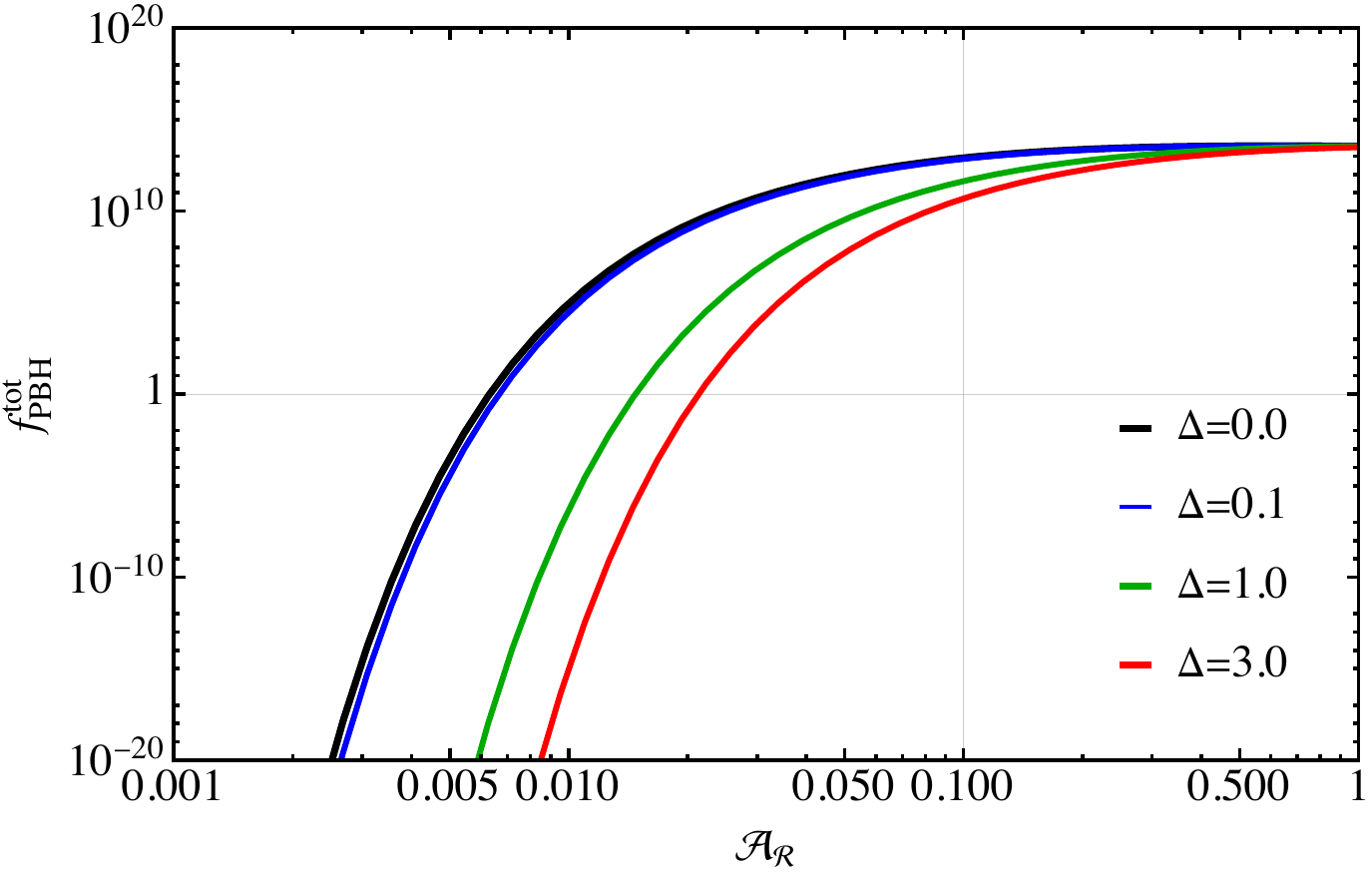} \\
\caption{Total PBH abundance calculated by PSC method using the compaction function considering different power spectrum widths $\Delta$, normalized such that PBHs constitute all DM and with $M_{k_*}= 10^{20}$g.
}
\label{fig:ftotEXPS1}
\end{center}
\end{figure}

In Fig.~\ref{fig:ftotEXPS1} we apply this method to calculate total PBH abundance, following Ref.~\cite{Gow:2022jfb}, to the log-normal power spectrum defined in Eq.~\eqref{def:PR} considering different widths $\Delta$. Here, we use a central frequency of $k_* = 1.56 \times 10^{13} \, \mathrm{Mpc}^{-1}$, with $M_{k_*}= 10^{20}$g. The scale $r$ is fixed such that $k_* r = 2.74$, which maximizes the compaction function for a monochromatic power spectrum.  Additionally, we adopt the same critical collapse parameters as in Ref.~\cite{Gow:2022jfb}, with $\mathcal{K} = 1$, $\mathcal{C}_{\mathrm{th}} = 0.587$, and $\gamma = 0.36$. Finally, we apply a redshift factor as given in Eq.~\eqref{eq:Redshift}. For PBHs to constitute all DM, we find that for PSC method with $\Delta = 0, 0.1, 0.4, 1, 2, 3$ the required {variance} is $\mathcal{A}_{\mathcal{R}} = 6.31\times 10^{-3}, 6.67\times 10^{-3}, 1.01\times 10^{-2}, 1.48\times 10^{-2}, 1.91\times 10^{-2},  2.14\times 10^{-2}$, respectively.

\begin{figure}
\begin{center}
\includegraphics[width=0.65\textwidth]{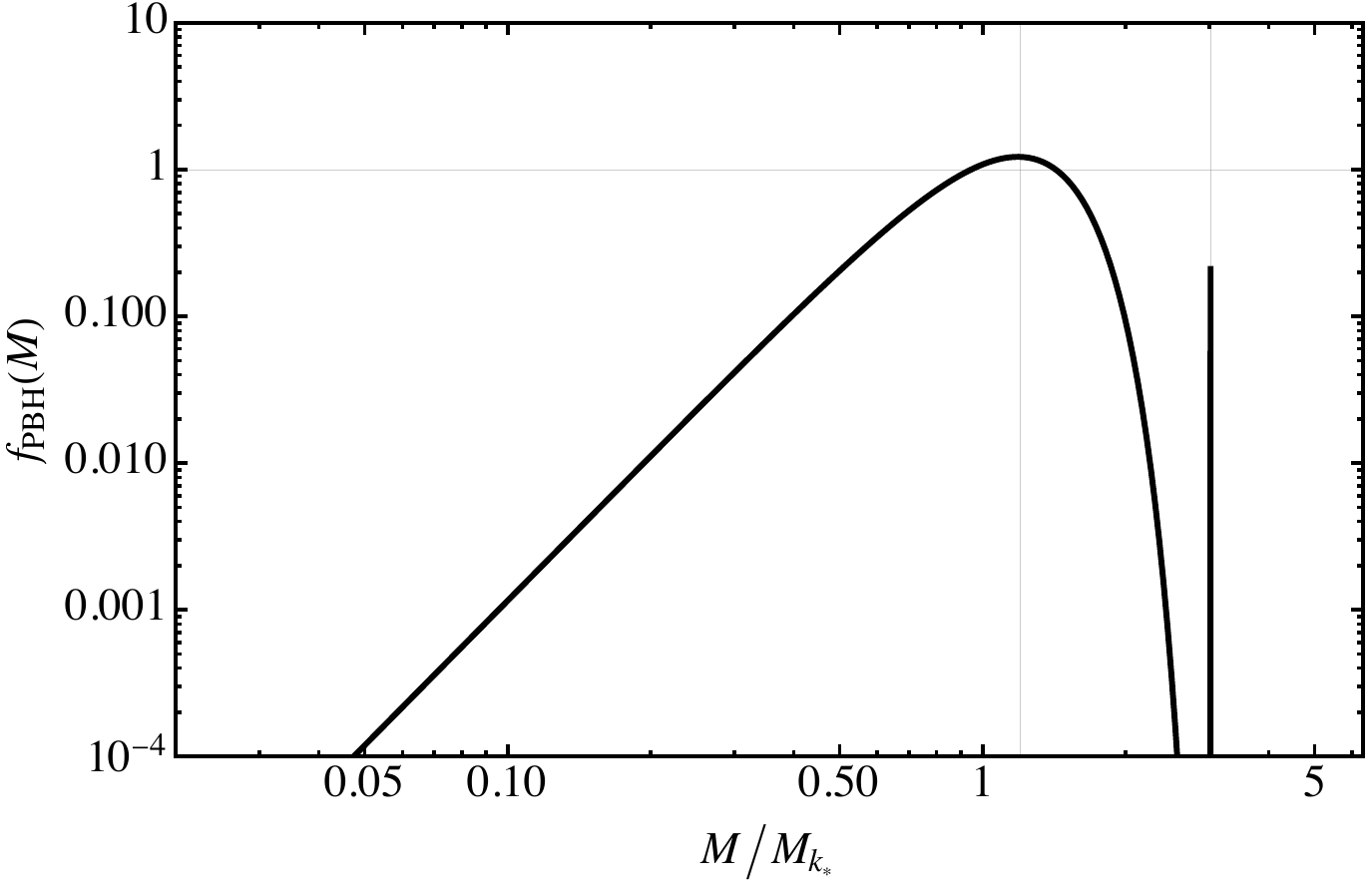} \\
\caption{PBH mass function calculated by PSC method using the
compaction function considering a monochromatic power spectrum, normalized such that PBHs are all DM with $M_{k_*}= 10^{20}$g. Power spectrum with different width has the same mass spectrum. There is a divergence at $M_{\mathrm{PBH,max}}\simeq 3.02 M_{k_*}$. At $M_c\simeq 1.2 M_{k_*}$ it reaches the maximum abundance.}
\label{fig:ftotEXPS2}
\end{center}
\end{figure}

In Fig.~\ref{fig:ftotEXPS2}, we present the PBH mass function calculated using the PSC method, assuming a monochromatic power spectrum and normalization such that PBHs constitute all DM, with $M_{k_*} = 10^{20} \, \mathrm{g}$. Notably, power spectra with different widths produce similar mass spectra. A divergence is observed in the mass function. While the total PBH abundance remains convergent, this divergence appears non-physical. It is expected to vanish once additional relevant variables describing the scenario are considered, resulting in the distribution being multivariate.

Our discussion highlights that dependence on smoothing scale $R_s$ is essential to consider, as $\mathrm{d}\ln M / \mathrm{d}R_s$ is never zero due to $M \propto R_s^2$ in Eq.~\eqref{eq:MH}. The divergence in Eq.~\eqref{eq:fPBH-EXPSM} arises because the so-called smoothing scale $r$ is fixed near the central wave number, which is a strong assumption. While this approach can provide a rough estimate of the total PBH abundance, it fails to yield an accurate mass spectrum, as perturbations away from the central wavenumber are not properly taken into account. Thus, we arrive at an important conclusion that the real PBH mass function should not exhibit a divergence if the $R_s$ dependence is properly accounted for. The divergent mass, as given by Eq.~\eqref{eq:ClII}, is 
\begin{equation}
\frac{M_{\mathrm{PBH,div}}}{M_{k_*}} = \mathcal{K} (k_* r)^2 \left(\mathcal{C}_{\ell,\mathrm{II}} - \frac{3}{8}\mathcal{C}_{\ell,\mathrm{II}}^2 - \mathcal{C}_{\mathrm{th}} \right)^\gamma \simeq 3.02.
\label{eq:Mmax}
\end{equation}

\section{Peaks Theory  Statistics Overview}\label{s:OverviewPT}

We review the fundamental concepts and derivations of peaks theory following BBKS Ref.~\cite{Bardeen:1985tr}, leading to the derivation of the comoving number density in Eq.~\eqref{eq:npeakk3} and the profile in Eq.~\eqref{eq:peakprofileg} of peaks. These serve as the starting points for calculating PBH abundance using peaks theory. 
The comoving number density of peaks is obtained as the probability average over variable sets that satisfy the specific conditions in Eq.~\eqref{eq:Npkvx}. The ``profile" refers to the mean value of the random field that also satisfies these conditions.

Peaks theory estimates the number of peaks using a point process approach. It assumes that the maxima in a random field are isolated from each other, and the density field at these points is represented by a delta function 
\begin{equation}
n_{\mathrm{pk}}(\boldsymbol{r})=\sum_p \delta^{(3)}_{\mathrm{D}}\left(\boldsymbol{r}-\boldsymbol{r}_p\right),
\label{eq:pointprocess}
\end{equation}
where, $\boldsymbol{r}_p$ represents the positions of the maxima, which satisfy certain physically motivated conditions, such as the requirement that the field heights exceed a specific threshold in our case.

To study the maxima of a field, it is necessary to expand the perturbation to at least second order. Here, the random field $F(\boldsymbol{r})$ is expanded around 
$\boldsymbol{r}_p$ in a Taylor series 
\begin{equation}
F(\boldsymbol{r}) \simeq F\left(\boldsymbol{r}_p\right)+\frac{1}{2} \sum_{i j} \zeta_{i j}\left(r-r_p\right)_i\left(r-r_p\right)_j,\quad 
\eta_i(\boldsymbol{r}) \simeq \sum_j \zeta_{i j}\left(r-r_p\right)_j.
\label{eq:expandofFgrad}
\end{equation}
In the neighborhood of a maxima point $\boldsymbol{r}_p$, $\eta_i\left(\boldsymbol{r}_p\right)=0$,
and $\boldsymbol{\zeta}$ must be negative definite at $\boldsymbol{r}_p$ to be a maxima.
If $\boldsymbol{\zeta}$ is non-singular at $\boldsymbol{r}_p$, the second formula in Eq.~\eqref{eq:expandofFgrad} gives $\boldsymbol{r}-\boldsymbol{r}_p \simeq \boldsymbol{\zeta}^{-1}\left(\boldsymbol{r}_p\right) \boldsymbol{\eta}(\boldsymbol{r})$. Then, one can transform the selection of points in real space Eq.~\eqref{eq:pointprocess} into the selection of derivative variables 
\begin{equation}
\delta^{(3)}_{\mathrm{D}}\left(\boldsymbol{r}-\boldsymbol{r}_p\right)=\left|\operatorname{det} \boldsymbol{\zeta}\left(\boldsymbol{r}_p\right)\right| \delta^{(3)}_{\mathrm{D}}\left(\boldsymbol{\eta}(\boldsymbol{r})-\boldsymbol{0}\right),
\label{eq:detzeta}
\end{equation}
using $\delta_{\mathrm{D}}( f(x))=\sum_k \left| f' ( x_k)\right|^{-1}\delta_{\mathrm{D}}( x-x_k)$.
The expression for the
number density of extreme points thus becomes
\begin{equation}
n_{\text {ext }}(\boldsymbol{r})=|\operatorname{det} \boldsymbol{\zeta}(\boldsymbol{r}_p)| \delta^{(3)}_{\mathrm{D}}\left(\boldsymbol{\eta}(\boldsymbol{r})\right).
\label{eq:next1}
\end{equation}
The ensemble average of Eq.~\eqref{eq:next1}  is 
\begin{equation}
\left\langle n_{\mathrm{ext}}(\boldsymbol{r})\right\rangle 
\equiv\int|\operatorname{det} \boldsymbol{\zeta}|  \delta^{(3)}_{\mathrm{D}} \left(\boldsymbol{\eta}(\boldsymbol{r})\right) P(F, \boldsymbol{\eta}, \boldsymbol{\zeta}) \mathrm{d} F \mathrm{d}^3 \boldsymbol{\eta} \mathrm{d}^6 \boldsymbol{\zeta}.
\label{eq:ensemblenext}
\end{equation}
The homogeneity guarantees that $\left\langle n_{\mathrm{ext}}(\boldsymbol{r})\right\rangle$ will be independent of $\boldsymbol{r}$. 

Since the random field $F(\boldsymbol{r})$ is Gaussian, its derivatives, integrals, and any linear functions of them are also Gaussian. Consequently, the multivariable probability distribution $P(F, \boldsymbol{\eta}, \boldsymbol{\zeta})$ in Eq.~\eqref{eq:ensemblenext} is a multivariate Gaussian. Here, we denote $\boldsymbol{V}_0$ as the vector of variables and $\boldsymbol{M}$ as the covariance matrix of $\boldsymbol{V}_0$. 
Considering Eq.~\eqref{eq:expandofFgrad}, in our case, there are 10 independent variables: 1 from the scalar field $F$, 3 from $\boldsymbol{\eta}$, and 6 from $\boldsymbol{\zeta}$. For a scalar Gaussian field with zero mean, knowledge of the two-point correlation function is sufficient to fully characterize the field
\begin{equation}
\xi\left(\boldsymbol{r}_1, \boldsymbol{r}_2\right)\equiv\left\langle F\left(\boldsymbol{r}_1\right) F\left(\boldsymbol{r}_2\right)\right\rangle~.
\label{def:xi}
\end{equation}
Together with its derivatives, this is sufficient to compute all relevant statistical properties. By using the Fourier expansion of $F(\boldsymbol{r})$, all the elements of the covariance matrix can be readily determined
\begin{equation}
\begin{array}{ll}
\langle F F\rangle=\sigma_0^2, & ~~~~~\left\langle\eta_i \eta_j\right\rangle=\dfrac{\sigma_1^2}{3} \delta_{i j}, \\
\left\langle F \zeta_{i j}\right\rangle=-\dfrac{\sigma_1^2}{3} \delta_{i j}, & ~~~~~\left\langle\zeta_{i j} \zeta_{k l}\right\rangle=\dfrac{\sigma_2^2}{15}\left(\delta_{i j} \delta_{k l}+\delta_{i k} \delta_{j l}+\delta_{i l} \delta_{j k}\right), \\
\left\langle F \eta_i\right\rangle=0, & ~~~~~\left\langle\eta_i \zeta_{j k}\right\rangle=0 .
\end{array}
\label{eq:variance}
\end{equation}
The ensemble averages are computed using the correlation functions at the same point, which introduce multipole moments
\begin{equation}
\sigma_n^2 \equiv \int \frac{\mathrm{d} k}{k} k^{2 n} \mathcal{P}_F(k).
\label{def:Multipole moment}
\end{equation}

It is evident that $(\eta_1, \eta_2, \eta_3, \zeta_{23}, \zeta_{13}, \zeta_{12})$ has already been diagonalized, as shown in Eq.~\eqref{eq:variance}. 
To diagonalize the remaining four dimensions, $(F, \zeta_{11}, \zeta_{22}, \zeta_{33})$, we define new variables $\nu, x, y, z$ as follows
\begin{equation}
\begin{aligned}
\sigma_0 \nu = F, \quad &
~~~~~\sigma_2 x=-\nabla^2 F=-\left(\zeta_{11}+\zeta_{22}+\zeta_{33}\right),\\
\sigma_2 y=-\left(\zeta_{11}-\zeta_{33}\right) / 2,\quad &
~~~~~\sigma_2 z=-\left(\zeta_{11}-2 \zeta_{22}+\zeta_{33}\right) / 2.
\end{aligned}
\label{eq:lambda123toxyz}
\end{equation}
Using Eq.~\eqref{eq:lambda123toxyz} and Eq.~\eqref{eq:variance}, one can obtain 
the covariance matrix elements of new variables
\begin{equation}
\left\langle \nu^2\right\rangle=1, \quad\left\langle x^2\right\rangle=1, \quad\langle x \nu \rangle=\gamma_1=\frac{\sigma_1^2}{\sigma_0 \sigma_2}, \quad\left\langle y^2\right\rangle=\frac{1}{15}, \quad\left\langle z^2\right\rangle=\frac{1}{5},
\label{def:gamma}
\end{equation}
the matrix is now also diagonal in $y$ and $z$. Then the quadratic form appearing in the multi-Gaussian joint probability distribution function can be written as 
\begin{equation}
\boldsymbol{V}_0^{\mathrm{T}} \boldsymbol{M}^{-1} \boldsymbol{V}_0=
\left(\begin{array}{l}
\boldsymbol{V}_1 \\
\boldsymbol{V}_2 
\end{array}\right)^{\mathrm{T}}\left(\begin{array}{cc}
\boldsymbol{M}_1^{-1} &  0 \\
 0 & \boldsymbol{M}_2^{-1}
\end{array}\right)\left(\begin{array}{l}
\boldsymbol{V}_1 \\
\boldsymbol{V}_2 
\end{array}\right),
\label{def:M1M2}
\end{equation}
where
\begin{equation}
\boldsymbol{V}_0=\left( \boldsymbol{V}_1,\boldsymbol{V}_2 \right), \quad \boldsymbol{V}_1=\left(\eta_1, \eta_2, \eta_3, \zeta_{23}, \zeta_{13}, \zeta_{12} \right), \quad \boldsymbol{V}_2=\left(\nu, x,y,z \right),
\label{eq:V2}
\end{equation}
and 
\begin{equation}
\boldsymbol{M}_1=\mathrm{diag}\left(\frac{\sigma_1^2}{3},\frac{\sigma_1^2}{3},\frac{\sigma_1^2}{3},\frac{\sigma_2^2}{15},\frac{\sigma_2^2}{15},\frac{\sigma_2^2}{15} \right), \quad 
\boldsymbol{M}_2=\left(\begin{array}{cccc}
1 & \gamma_1 & 0 & 0 \\
\gamma_1 & 1 & 0 & 0 \\
0 & 0 & 1 / 15 & 0 \\
0 & 0 & 0 & 1 / 5
\end{array}\right).
\label{eq:M1M2}
\end{equation}

Now one has an exponential term in the multi-Gaussian joint probability distribution function 
\begin{equation}
\begin{aligned}
 2 Q\equiv&~\boldsymbol{V}_0^{\mathrm{T}} \boldsymbol{M}^{-1} \boldsymbol{V}_0\\
=&~\nu^2+\frac{(x-x_\star)^2}{1-\gamma_1^2}+15 y^2+5 z^2+\frac{3}{\sigma_1^2}\left( \eta_1^2 +\eta_2^2 +\eta_3^2  \right)+\frac{15}{\sigma_2^2}\left( \zeta_{23}^2 +\zeta_{13}^2 +\zeta_{12}^2  \right),
\end{aligned}
\label{eq:Q}
\end{equation}
where we have defined $x_\star\equiv \gamma_1   \nu$.  
The six independent $\zeta_{ij}$ components in Eq.~\eqref{eq:expandofFgrad} can be labeled as $\zeta_{A}$, with $A=1,2...6$ such that
\begin{equation}
(\zeta_1,\zeta_2,\zeta_3,\zeta_4,\zeta_5,\zeta_6)\equiv (\zeta_{11},\zeta_{22},\zeta_{33},\zeta_{23},\zeta_{13},\zeta_{12}).
\label{def:zetaA}
\end{equation}
Since $\boldsymbol{\zeta}$ is symmetric, it can be diagonalized using a rotation matrix $\hat{S}$ 
\begin{equation}
\boldsymbol{\lambda} =\mathrm{diag}(\lambda_1,\lambda_2,\lambda_3)=-\hat{S} \boldsymbol{\zeta} \hat{S}^\dagger
.
\label{def:hatS}
\end{equation}
Note that here the hat denotes an operator and not the peak profile as in the main text.

The six independent variables in $\boldsymbol{\zeta}$, as defined in Eq.~\eqref{def:zetaA}, can be transformed into three eigenvalues and three Euler angles representing the principal directions. We now focus on the principal axes and introduce the eigenvalues of $-\boldsymbol{\zeta}$ as
\begin{equation}
\lambda_A=-\zeta_A, \quad A=1,2,3. 
\label{def:lambdaA}
\end{equation}
The remaining three degrees of freedom, $\zeta_A$ for $A = 4, 5, 6$, can be expressed in terms of the Euler angles. 
To determine the volume element $\mathrm{d}^6\boldsymbol{\zeta}$ in Eq.~\eqref{eq:ensemblenext}, we start by defining the inner product of two arbitrary matrices in the space of all third-order symmetric matrices (such as $\boldsymbol{\zeta}$) as the trace of their matrix multiplication. The metric in this space is then defined as the self-inner product of perturbation elements within the matrices (such as $\mathrm{d}^6\boldsymbol{\zeta}$).
Using the above definitions and Eq.~\eqref{def:hatS}, the metric element for $\boldsymbol{\zeta}$ is given by
\begin{equation}
\mathrm{Tr}\left( (\mathrm{d}\boldsymbol{\zeta})^2 \right)=\mathrm{Tr}\left( \left( \mathrm{d}\boldsymbol{\lambda} \right)^2 \right)+\mathrm{Tr}\left( \left[ \boldsymbol{\lambda}, \hat{S}^\dagger (\mathrm{d}\hat{S}) \right]^2 \right).
\label{eq:ds2}
\end{equation}
Since $\hat{S}$ and its transpose $\hat{S}^\dagger$ always have opposite symmetries, $\hat{S}^\dagger (\mathrm{d}\hat{S})$ is anti-symmetric. 
The elements of the anti-symmetric matrix $\hat{S}^\dagger \mathrm{d}\hat{S}$ and the symmetric matrix $\boldsymbol{\lambda}$ can be expressed, respectively, as
\begin{equation}
\left( \hat{S}^\dagger \mathrm{d}\hat{S} \right)_{\alpha_1 \alpha_2}= \varepsilon_{\alpha_1 \alpha_2 \alpha_3} \omega_{\alpha_3} , \quad \lambda_{\beta_1 \beta_2}=\delta_{\beta_1 \beta_2}\lambda_{\beta_2}~.
\label{eq:wepsilon}
\end{equation}
Here, $\boldsymbol{\omega}$ is an infinitesimal vector, and $\boldsymbol{\varepsilon}$ represents the Levi-Civita symbol. The indices $\alpha_{1,2,3}$ and $\beta_{1,2}$ denote tensor components. 
Using Eq.~\eqref{eq:wepsilon}, the line element in Eq.~\eqref{eq:ds2} can be written as 
\begin{equation}
\mathrm{Tr}\left( (\mathrm{d}\boldsymbol{\zeta})^2 \right) =\sum_{A=1}^3 (\mathrm{d}\lambda_A)^2 +  \left(\lambda_{2} -\lambda_{3} \right)^2 \omega_{1}^2 + \left(\lambda_{3} -\lambda_{1} \right)^2 \omega_{2}^2 + \left(\lambda_{1} -\lambda_{2} \right)^2 \omega_{3}^2.
\end{equation}

By comparing it with the line element in a Cartesian coordinate system, we obtain the ``orthogonal basis" in the $\boldsymbol{\zeta}$-space
\begin{equation}
\left\{ \mathrm{d}\lambda_1,\mathrm{d}\lambda_2,\mathrm{d}\lambda_3,  \left|\lambda_{2} -\lambda_{3} \right|\omega_{1},  \left|\lambda_{3} -\lambda_{1} \right| \omega_{2} , \left|\lambda_{1} -\lambda_{2} \right| \omega_{3}  \right\}.
\end{equation}
Hence, the volume element is the wedge product of orthogonal basis elements  
\begin{equation}
\mathrm{d}^6 \boldsymbol{\zeta}\equiv \prod_{A=1}^6 \mathrm{d}\zeta_A = \mathrm{d}\lambda_1 \wedge \mathrm{d}\lambda_2 \wedge \mathrm{d}\lambda_3 \wedge \left|\lambda_{2} -\lambda_{3} \right| \omega_{1} \wedge \left|\lambda_{3} -\lambda_{1} \right| \omega_{2} \wedge \left|\lambda_{1} -\lambda_{2} \right| \omega_{3} .
\end{equation}
The Lie algebra of the SO(3) group is the space formed by all $3 \times 3$ anti-symmetric matrices, such as $\left( \hat{S}^\dagger \mathrm{d}\hat{S} \right)_{\alpha_1 \alpha_2} = \varepsilon_{\alpha_1 \alpha_2 \alpha_3} \omega_{\alpha_3}$, where $\omega_i$, $i=1,2,3$, form an orthonormal basis. Consequently, the volume element in Eq.~\eqref{eq:ensemblenext} is
\begin{equation}
\mathrm{d}^6 \boldsymbol{\zeta}
= \left|\left( \lambda_1 -
\lambda_2 \right) \left( \lambda_2 -
\lambda_3 \right) \left( \lambda_3 -
\lambda_1 \right)\right| \mathrm{d}\lambda_1  \mathrm{d}\lambda_2 \mathrm{d}\lambda_3    \mathrm{d}\mathrm{vol}(\mathrm{SO(3)}),
\label{eq:lambda}
\end{equation}
where the volume element of $\mathrm{SO(3)}$ is
given by 
\begin{equation}
\mathrm{d}\mathrm{vol}(\mathrm{SO(3)}) =\omega_{1} \wedge \omega_{2} \wedge \omega_{3} \equiv\mathrm{d}\Omega_{S^3}.
\label{def:dOmega3}
\end{equation}
This represents the volume element on the surface of the three-sphere  $S^3$. Considering the rotational symmetry of the three Euler angle axes, the total volume of $S^3$ is given by $\int \mathrm{d}\Omega_{S^3} / 3! = 2\pi^2 / 3!$.

The volume element transformation from $\left\{\lambda_1, \lambda_2, \lambda_3\right\}$ to $\left\{x, y, z\right\}$ coordinates can be performed using Eq.~\eqref{eq:lambda123toxyz}, Eq.~\eqref{def:zetaA}, and Eq.~\eqref{def:lambdaA}. 
The total differential element in Eq.~\eqref{eq:ensemblenext} then becomes 
\begin{equation}
\begin{aligned}
\mathrm{d}\boldsymbol{V}_0\equiv &~\mathrm{d}\nu   \mathrm{d}^3 \boldsymbol{\eta}   \mathrm{d}^6 \boldsymbol{\zeta} \\
=&~\mathrm{d}\nu   \mathrm{d}^3 \boldsymbol{\eta}   \left| \left( \lambda_1-\lambda_2 \right)\left( \lambda_2-\lambda_3 \right)\left( \lambda_3-\lambda_1 \right) \right| \mathrm{d}\lambda_1  \mathrm{d}\lambda_2 \mathrm{d}\lambda_3  \frac{\mathrm{d}\Omega_{S^3}}{6} \\
=&~\mathrm{d}\nu   \mathrm{d}^3 \boldsymbol{\eta}   \sigma_2^3 \left| 2y \left( y^2 -z^2 \right)\right| \frac{2}{3}\sigma_2^3 \mathrm{d} x \mathrm{d} y \mathrm{d} z  \frac{\mathrm{d}\Omega_{S^3}}{6} .
\end{aligned}
\label{def:dV}
\end{equation}
In the following, we derive the comoving number density of peaks corresponding to maxima points. 
By substituting Eq.~\eqref{eq:lambda123toxyz}, Eq.~\eqref{eq:M1M2}, and Eq.~\eqref{def:dV} into the joint multivariate Gaussian probability distribution function in Eq.~\eqref{eq:ensemblenext}, the differential element of the comoving number density ensemble average can be obtained
\begin{equation}
\begin{aligned}
  P(\nu, \boldsymbol{\eta}, x, y, z ) \mathrm{d}\nu \mathrm{d}^3 \boldsymbol{\eta} \mathrm{d} x \mathrm{d} y \mathrm{d} z 
=&~ \int_{\Omega_3} \frac{e^{-Q}}{\left[(2\pi)^{10}\mathrm{det}(\boldsymbol{M})\right]^{1/2}} \left| \frac{\mathrm{d} \lambda_1 \mathrm{d} \lambda_2 \mathrm{d} \lambda_3}{\mathrm{d} x \mathrm{d} y \mathrm{d} z}  \right|^{-1} \mathrm{d}\boldsymbol{V}_0 \\
=&~ \frac{15^{5/2}}{32\pi^3} \frac{1}{\sigma_1^3 \sqrt{1-\gamma^2}}   \left| 2y \left( y^2 -z^2 \right)\right| e^{-Q}  \mathrm{d}\nu \mathrm{d}^3 \boldsymbol{\eta} \mathrm{d} x \mathrm{d} y \mathrm{d} z ,
\end{aligned}
\label{eq:Pnuxyz}
\end{equation}
where $Q$ is defined in Eq.~\eqref{eq:Q}. In the second equality, we use the fact that the total integral over the three-sphere yields $2\pi^2 / 6$, accounting for rotational symmetry.

We now consider a specific condition where the peak height $\nu$, as defined in Eq.~\eqref{eq:lambda123toxyz}, exceeds a certain threshold value $\nu_0$. First, we impose an ordering on the eigenvalues of $\boldsymbol{\lambda}$
\begin{equation}
\lambda_1 \geq \lambda_2 \geq \lambda_3.
\label{eq:order}
\end{equation} 
Since there are five other possible orderings we could have chosen, and $P(F, \boldsymbol{\eta}, \boldsymbol{\zeta})$ is invariant under changes in ordering, we must multiply the probability expression by 6 to account for this. This compensates for the reduction in the available $S^3$ volume by a factor of $3!$, which arises due to the identical nature of the axes. Specifically, the entire volume of $S^3$ is available for the triad rotation to the principal axes. 
An additional constraint is introduced if we require all eigenvalues to be positive, which is necessary for maxima. Under our ordering convention, this condition is equivalent to $\lambda_3 > 0$. 
Using Eq.~\eqref{eq:detzeta}, the number density of maxima points with heights between $\nu_0$ and $\nu_0 + \mathrm{d}\nu_0$ is given by
\begin{equation}
n_{\mathrm{pk}}\left(\boldsymbol{r}, \nu_0\right) \mathrm{d}\nu =|\operatorname{det}(\boldsymbol{\zeta})| \delta^{(3)}(\boldsymbol{\eta})
\Theta\left(\lambda_3\right) \delta_{\mathrm{D}}\left(\nu-\nu_0\right) \mathrm{d} \nu.
\label{def:nu0}
\end{equation}
The average of $n_{\mathrm{pk}}$ in (\ref{def:nu0}) is
\begin{equation}
\mathscr{N}_{\mathrm{pk}}\left(\nu_0\right)  \equiv\left\langle n_{\mathrm{pk}}\left(\boldsymbol{r}, \nu_0\right)\right\rangle =\left\langle\left|\lambda_1 \lambda_2 \lambda_3\right| \Theta\left(\lambda_3\right) \delta_{\mathrm{D}}\left(\nu-\nu_0\right)\right\rangle ,
\label{def:curlyN}
\end{equation}
where to obtain the final equality we retain $\mathrm{det}(\boldsymbol{\zeta}) = \left|\lambda_1 \lambda_2 \lambda_3\right|$. 
Note that the mean density $\mathscr{N}_{\mathrm{pk}}\left(\nu_0\right)$ is position-independent due to the homogeneity of the underlying random density field.

Next, we further constrain the maxima points to those with parameters $\nu$, $x$, $y$, and $z$ within corresponding infinitesimal ranges. We define $\mathrm{d}\boldsymbol{V}_2 \equiv \mathrm{d}\nu \, \mathrm{d}x \, \mathrm{d}y \, \mathrm{d}z$, resulting in
\begin{equation}
\begin{aligned}
 \mathscr{N}_{\mathrm{pk}}(\nu, x, y, z) \mathrm{d}\boldsymbol{V}_2 
\equiv&~\iint_{\Omega^3,\boldsymbol{\eta}}  \left|\lambda_1 \lambda_2 \lambda_3\right|\delta^{(3)}(\boldsymbol{\eta})
\Theta\left(\lambda_3\right) \delta_{\mathrm{D}}\left(\nu-\nu_0\right) P(\nu, \boldsymbol{\eta}, x, y, z, ) \mathrm{d}^3 \boldsymbol{\eta} \mathrm{d}\boldsymbol{V}_2 \mathrm{d}\Omega_{S^3} \\
=&~\iint_{\Omega^3,\boldsymbol{\eta}}    \left|\lambda_1 \lambda_2 \lambda_3\right|\delta^{(3)}(\boldsymbol{\eta}) \Theta\left(\lambda_3\right) \delta_{\mathrm{D}}\left(\nu-\nu_0\right) P(\nu, \boldsymbol{\eta}, x, y, z, ) \mathrm{d}\boldsymbol{V}_0 \left| \frac{\mathrm{d} \lambda_1 \mathrm{d} \lambda_2 \mathrm{d} \lambda_3}{\mathrm{d} x \mathrm{d} y \mathrm{d} z}  \right|^{-1} \\
=&~2\pi^2 \left|\lambda_1 \lambda_2 \lambda_3\right| \boldsymbol{\chi} \frac{e^{-\widetilde{Q}}}{\left[(2 \pi)^{10} \operatorname{det}(\boldsymbol{M})\right]^{1 / 2}} \left| \left( \lambda_1-\lambda_2 \right)\left( \lambda_2-\lambda_3 \right)\left( \lambda_3-\lambda_1 \right) \right| \\
&\times \mathrm{d} \nu \mathrm{d}\lambda_1 \mathrm{d}\lambda_2 \mathrm{d}\lambda_3 \left| \frac{\mathrm{d} \lambda_1 \mathrm{d} \lambda_2 \mathrm{d} \lambda_3}{\mathrm{d} x \mathrm{d} y \mathrm{d} z}  \right|^{-1} \\
=&~2\pi^2 (2\pi)^{-5}\frac{3\times 15^{5/2}}{\sigma_1^3 \sigma_2^3 \sqrt{1-\gamma^2}} \left|\lambda_1 \lambda_2 \lambda_3\right| \left| \left( \lambda_1-\lambda_2 \right)\left( \lambda_2-\lambda_3 \right)\left( \lambda_3-\lambda_1 \right) \right| e^{-\widetilde{Q}} \boldsymbol{\chi} \mathrm{d}\boldsymbol{V}_2 \\
=&~2\pi^2 \frac{1}{(2\pi)^{3}}\frac{1}{4\pi^2}\frac{3\times 3^{5/2} \times 5^{5/2}}{\sigma_1^3 \sigma_2^3 \sqrt{1-\gamma^2}} \frac{2}{3^3} \sigma_2^6 F(x,y,z) e^{-\widetilde{Q}} \boldsymbol{\chi} \mathrm{d}\boldsymbol{V}_2 \\
=&~\frac{5^{5 / 2} 3^{1 / 2}}{(2 \pi)^3}\left(\frac{\sigma_2}{\sigma_1}\right)^3 \frac{1}{\sqrt{1-\gamma^2}} e^{-\widetilde{Q}} F(x, y, z) \boldsymbol{\chi} \mathrm{d}\boldsymbol{V}_2,
\end{aligned} 
\label{eq:Npk1}
\end{equation}
where the two integrals in the first line are performed over $\mathrm{d}^3 \boldsymbol{\eta}$ and $\mathrm{d}\Omega_{S^3}$. 
Here, the function $F(x, y, z)$ is defined as 
\begin{equation}
\begin{aligned}
F(x, y, z)\equiv&~\frac{3^3}{2} \sigma_2^{-6} \left|\lambda_1 \lambda_2 \lambda_3\right|\left(\lambda_1-\lambda_2\right)\left(\lambda_2-\lambda_3\right)\left(\lambda_1-\lambda_3\right)\\
=&~(x-2 z)\left[(x+z)^2-(3 y)^2\right] y\left(y^2-z^2\right),
\end{aligned}
\label{def:Fxyz}
\end{equation}
which originates from $\mathrm{d}\boldsymbol{V}_0$, and Eq.~\eqref{eq:order} was used for the first equality. The quantity $\widetilde{Q}$ can be intuitively obtained from $Q$ in Eq.~\eqref{eq:Q} as
\begin{equation}
\widetilde{Q}\equiv Q(\boldsymbol{\eta}=\boldsymbol{0}, \zeta_{4,5,6}=0)=\frac{\nu^2}{2}+\frac{\left(x-x_\star\right)^2}{2(1-\gamma_1^2)}+\frac{15}{2} y^2+\frac{5}{2} z^2.
\label{def:widetildeQ}
\end{equation}

The condition for selecting the maxima points is given by
$
\boldsymbol{\chi} \sim \Theta\left(\lambda_3\right) \delta_{\mathrm{D}}\left(\nu - \nu_0\right),
$
where $\Theta(\lambda_3)$ ensures that $\lambda_3 > 0$. Using Eq.~\eqref{def:widetildeQ}, $\mathscr{N}_{\mathrm{pk}}(\nu, x, y, z)$ in Eq.~\eqref{eq:Npk1} can be rewritten as
\begin{equation}
\begin{aligned}
 \mathscr{N}_{\mathrm{pk}}(\nu, x, y, z) \mathrm{d}\nu \mathrm{d}x \mathrm{d}y \mathrm{d}z  
=&~\frac{e^{-\frac{\nu^2}{2}} }{(2 \pi)^2 R_*^3} \frac{\exp\left[ -\left(x-x_\star\right)^2/2(1-\gamma_1^2) \right]}{\sqrt{2 \pi(1-\gamma_1^2)}} \mathrm{d}\nu  \mathrm{d}x  \boldsymbol{\chi} \notag\\
&\times \frac{3^2 5^{5 / 2}}{\sqrt
{2\pi} }  e^{-\frac{15}{2}y^2}\mathrm{d}y F(x, y, z)  e^{-\frac{5}{2}z^2}  \mathrm{d}z,
\end{aligned}
\label{eq:Npkvxyz}
\end{equation}
by introducing a new definition
\begin{equation}
R_* \equiv \sqrt{3} \frac{\sigma_1}{\sigma_2}.
\label{def:Rstar}
\end{equation}
Integrating $y$ and $z$ in Eq.~\eqref{eq:Npkvxyz}, one has
\begin{equation}
\mathscr{N}_{\mathrm{pk}}(\nu, x) \mathrm{d}\nu \mathrm{d}x =\frac{e^{-\frac{\nu^2}{2}} }{(2 \pi)^2 R_*^3} \frac{\exp\left[ -\left(x-x_\star\right)^2/2(1-\gamma_1^2) \right]}{\sqrt{2 \pi(1-\gamma_1^2)}} \mathrm{d}\nu  \mathrm{d}x   f(x),
\label{eq:Npkvx}
\end{equation}
where
\begin{equation}
\begin{aligned}
f(x)\equiv \frac{3^2 5^{5 / 2}}{\sqrt
{2\pi} } \iint \mathrm{d}y \mathrm{d}z~\boldsymbol{\chi}   e^{-\frac{15}{2}y^2}  F(x, y, z)  e^{-\frac{5}{2}z^2}    .
\end{aligned}
\label{def:f(x)}
\end{equation}
Given the condition $\lambda_1 \geq \lambda_2 \geq \lambda_3 \geq 0$, it follows from the first line of Eq.~\eqref{def:Fxyz} that $F(x, y, z) > 0$.  Consequently, the final integration intervals for $y$ and $z$ are determined as
\begin{equation}
\begin{aligned}
f(x)
=&~\frac{3^2 5^{5 / 2}}{\sqrt{2 \pi}}\left[\int_0^{x / 4} e^{-\frac{15}{2} y^2} \mathrm{d} y \int_{-y}^{y} F(x, y, z) e^{-\frac{5}{2} z^2} \mathrm{d} z\right]\\
&~+\frac{3^2 5^{5 / 2}}{\sqrt{2 \pi}}\left[\int_{x / 4}^{x / 2} e^{-\frac{15}{2} y^2} \mathrm{d} y \int_{3 y-x}^{y} F(x, y, z) e^{-\frac{5}{2} z^2} \mathrm{d} z\right].
\end{aligned}
\end{equation}
Employing numerical methods, the result in Eq.~\eqref{def:f(x)} can be obtained directly. Essentially, $\mathscr{N}_{\mathrm{pk}}(\nu, x)$ in Eq.~\eqref{eq:Npkvx} provides the comoving number density of peaks with a specific height $\nu$ and width $x$.

Next, we focus on the peak profiles. Any peak can be characterized by a parameter set $\boldsymbol{C}$, which contains all the necessary information to ensure that $\boldsymbol{r} = 0$ is a peak.
We aim to calculate the probability that, at a displacement $\boldsymbol{r}$ from the peak that is assumed to be located at $\boldsymbol{r} = \boldsymbol{0}$, the field takes on the value $F$. Specifically, we are interested in $P[F(\boldsymbol{r}) \mid \boldsymbol{C}] \, \mathrm{d}F(\boldsymbol{r})$.
Using the conditional probability theorem for Gaussian distributions, $P[F(\boldsymbol{r}) \mid \boldsymbol{C}]$ is also Gaussian distributed.
By translating $\zeta_A$ ($A = 1, 2, 3$) into $x, y, z$ and rotating to the principal axes system, we find that only $\nu$ and $x$ are required to specify the field. Since the field is orientation-averaged, it is therefore spherically symmetric such that
\begin{equation}
\boldsymbol{C}=\left.\mid \nu,x\right.\rangle,
\label{eq:CV2}
\end{equation}
which is a partial component of $\boldsymbol{V}_2$ as defined in Eq.~\eqref{eq:V2}. 

The mean value of the conditional variable $(F \mid \boldsymbol{C})$ is then given by
\begin{equation}
\begin{aligned}
\left\langle F \mid \boldsymbol{C} \right\rangle=&~\left\langle F \otimes \boldsymbol{C} \right\rangle\left\langle \boldsymbol{C} \otimes \boldsymbol{C} \right\rangle^{-1} \boldsymbol{C}^{\mathrm{T}} \\
=&~ \frac{\nu}{\left(1-\gamma_1^2\right)}(\langle F \nu \rangle-\gamma_1\langle F x\rangle)+\frac{x}{\left(1-\gamma_1^2\right)}(\langle F x\rangle-\gamma_1\langle F \nu \rangle),
\end{aligned}
\label{eq:FCprofile}
\end{equation}
where $\left\langle \boldsymbol{C} \otimes \boldsymbol{C} \right\rangle$ is given by $\boldsymbol{M}_2$ in Eq.~\eqref{eq:M1M2}. 
To derive $\langle F x \rangle$ and $\langle F \nu \rangle$, recall the definition of the two-point correlation function of
Eq.~\eqref{def:xi}, one can define
\begin{equation}
\xi_{ij}\equiv\xi\left(\boldsymbol{r}_{i j}\right)\equiv \left\langle F\left(\boldsymbol{r}_i\right) F\left(\boldsymbol{r}_j\right)\right\rangle, \quad 
\xi\left(\boldsymbol{r}\to\boldsymbol{0}\right)\equiv  \left\langle F\left(\boldsymbol{r}_i\right) F\left(\boldsymbol{r}_j \to \boldsymbol{r}_i \right)\right\rangle= \sigma_0^2.
\label{eq:recallxi}
\end{equation}
Here $\boldsymbol{r}_{i,j}$ represent distinct points in space, with $\boldsymbol{r}_{ij} \equiv \boldsymbol{r}_i - \boldsymbol{r}_j$. The value $\xi(\boldsymbol{r} \to \boldsymbol{0})$ corresponds to the correlation at the same point, which is also the multipole moment.
Additionally, the normalized two-point correlation function can be defined as
\begin{equation}
\psi_{ij}\equiv\psi\left(\boldsymbol{r}_{i j}\right) \equiv  \dfrac{\xi\left(\boldsymbol{r}_{i j}\right)}{ \xi(\boldsymbol{r}\to\boldsymbol{0})}, \quad \text{or} \quad \psi\left(\boldsymbol{r}\right) \equiv  \dfrac{\xi\left(\boldsymbol{r}\right)}{ \sigma_0^2 }
\label{def:psi}
\end{equation}
for short, neglecting the subscripts $ij$. For the points in the region around peak $\boldsymbol{r}=\boldsymbol{0}$, Eq.~\eqref{def:psi} gives
\begin{equation}
\xi\left(\boldsymbol{r}\right)=\left\langle F\left(\boldsymbol{r}\right) \left(\left.F\right|_{\boldsymbol{r}\to\boldsymbol{0}}\right)\right\rangle=\psi\left(\boldsymbol{r}\right)\sigma_0^2.
\end{equation}
Enforcing the condition that the peak is located at $\boldsymbol{r}=\boldsymbol{0}$, following Eq.~\eqref{eq:lambda123toxyz} one has
\begin{equation}
\nu = \dfrac{\left.F\right|_{\boldsymbol{r}\to\boldsymbol{0}}}{ \sigma_0}, \quad 
\sigma_2 x=-\left.\left(\nabla^2 F\right)\right|_{\boldsymbol{r}\to\boldsymbol{0}}.
\label{def:vx}
\end{equation}
We can readily define the profile width around peak as
\begin{equation}
K_1^2\equiv-\left.\frac{\nabla^2 F}{F}\right|_{\boldsymbol{r}\to\boldsymbol{0}}=\frac{x}{\nu}\frac{\sigma_2}{\sigma_0}.
\label{eq:profilepara}
\end{equation}
According to Eq.~\eqref{def:vx} we also have
\begin{equation}
\begin{aligned}
 \langle F \nu \rangle=&~ \dfrac{\left\langle F\left(\boldsymbol{r}\right) \left(\left.F\right|_{\boldsymbol{r}\to\boldsymbol{0}}\right)\right\rangle}{\sigma_0}=\psi\left(\boldsymbol{r}\right) \sigma_0 ,\\
\langle F x \rangle=& - \dfrac{\left\langle F\left(\boldsymbol{r}\right) \left.\left(\nabla^2 F\right)\right|_{\boldsymbol{r}\to\boldsymbol{0}}\right\rangle}{ \sigma_2} =- \frac{\sigma_0^2}{\sigma_2} \nabla^2 \psi\left(\boldsymbol{r}\right).
\end{aligned}
\label{eq:FvFxpsi}
\end{equation}
In the last equality of Eq.~\eqref{eq:FvFxpsi}, since we are analyzing the peak profile at a given point $\boldsymbol{r}$, the contribution to $\nabla^2$ at $\boldsymbol{r}$ primarily arises from the perturbation at the origin.

We can now define the mean profile of the arbitrary Gaussian field $\hat{F}(\boldsymbol{r})$ as  
\begin{equation}
\hat{F}(\boldsymbol{r})\equiv \left\langle F(\boldsymbol{r}) \mid \nu, x \right\rangle.
\label{def:profileBBKS}
\end{equation}
Substituting Eq.~\eqref{eq:FvFxpsi} into Eq.~\eqref{eq:FCprofile}, we obtain
\begin{equation}
 \frac{\left\langle F \mid \nu, x \right\rangle}{\sigma_0} \\
= \frac{\nu}{\left(1-\gamma_1^2\right)} \left(\psi\left(\boldsymbol{r}\right) +\gamma_1 \frac{\sigma_0}{\sigma_2} \nabla^2 \psi\left(\boldsymbol{r}\right) \right) -\frac{x/\gamma_1}{\left(1-\gamma_1^2\right)} \left( \gamma_1^2 \psi\left(\boldsymbol{r}\right) + \gamma_1 \frac{\sigma_0}{\sigma_2} \nabla^2 \psi\left(\boldsymbol{r}\right) \right). 
\label{eq:profile00}
\end{equation}
Along with Eq.~\eqref{def:gamma}, Eq.~\eqref{def:Rstar} and Eq.~\eqref{eq:profilepara},
we also have
\begin{equation}
\frac{R_*^2}{3} = \gamma_1 \frac{\sigma_0}{\sigma_2}, \quad x=\nu K_1^2 \frac{\sigma_0}{\sigma_2}.
\label{eq:checkvx}
\end{equation}
Using the first relation in Eq.~\eqref{eq:checkvx}, we find that Eq.~\eqref{eq:profile00} reproduces equation Eq.~(7.10) found in BBKS~\cite{Bardeen:1985tr}, provided that $r$ is measured in units of $R_*$. 

The profile defined in Eq.~\eqref{def:profileBBKS} then becomes
\begin{equation}
\begin{aligned}
\hat{F}(\boldsymbol{r})\equiv& \left\langle F \mid \nu, x \right\rangle\\
=&~\frac{\nu   \sigma_0}{ 1-\gamma_1^2 } \left\{ \left(\psi\left(\boldsymbol{r}\right) +\frac{R_*^2}{3}  \nabla^2 \psi\left(\boldsymbol{r}\right) \right) -\frac{K_1^2}{\gamma_1} \frac{\sigma_0}{\sigma_2} \left( \gamma_1^2 \psi\left(\boldsymbol{r}\right) + \frac{R_*^2}{3}  \nabla^2 \psi\left(\boldsymbol{r}\right) \right) \right\}.
\end{aligned}
\label{eq:profileNum}
\end{equation}
Replacing
\begin{equation}
\boxed{
\nu   \sigma_0\to \mu_0, \quad \psi \to \psi_0, \quad R_* \to R_1, 
}
\label{eq:newcoo}
\end{equation}
we obtain 
\be
\hat{F}(r)=\frac{\mu_0}{1-\gamma_1^2}\left[\psi_0(r)+\frac{R_1^2}{3} \nabla^2 \psi_0(r)-\frac{K_1^2}{\gamma_1} \frac{\sigma_0}{\sigma_2}\left(\gamma_1^2 \psi_0(r)+\frac{R_1^2}{3} \nabla^2 \psi_0(r)\right)\right].
\label{eq:peakprofileg}
\ee
Taking into account Eq.~\eqref{eq:Q}, Eq.~\eqref{def:Rstar}, Eq.~\eqref{eq:checkvx}, Eq.~\eqref{eq:newcoo}, and the variable transformations
\begin{equation}
\frac{\partial\left( \nu,x\right)}{\partial\left( \mu_0,K_1\right)}=\frac{2 \mu_0 K_1}{\sigma_0 \sigma_2},
\end{equation}
we can finally obtain the comoving peak number density for a fixed peak height and width from Eq.~\eqref{eq:Npkvx} as
\begin{equation}
\begin{aligned}
\mathscr{N}_{\mathrm{pk}}\left(\mu_0,K_1\right) \mathrm{d}\mu_0 \mathrm{d}K_1  
=&~2\frac{3^{-3/2}}{(2 \pi)^{5/2} } \mu_0 K_1 \frac{\sigma_2^2}{\sigma_0 \sigma_1^3} f\left(  \frac{\mu_0 K_1^2}{\sigma_2}\right) \frac{1}{\sqrt{1-\gamma_1^2}} \notag\\
&~\times \exp\left[ -\frac{1}{2}\left(\frac{\mu_0}{\sigma_0}\right)^2\left(1+\frac{\left(K_1^2 \frac{\sigma_0}{\sigma_2}-\gamma_1 \right)^2}{1-\gamma_1^2 
} \right)\right] \mathrm{d}\mu_0  \mathrm{d}K_1,
\end{aligned}
\label{eq:comNpkvx}
\end{equation}
which is exactly Eq.~\eqref{eq:npeakmuok1}.

\clearpage
\bibliography{PBH-peak}
\addcontentsline{toc}{section}{Bibliography}
\bibliographystyle{JHEP}
\end{document}